\documentclass[
    prx,
    twocolumn,
    superscriptaddress,
    floatfix,
    longbibliography
]{revtex4-2}

\usepackage{amsmath,amssymb,amsfonts,amsthm}
\usepackage{bm}
\usepackage{mathtools}
\usepackage{empheq}
\usepackage{graphicx}
\usepackage{subcaption}
\usepackage{float}
\usepackage{adjustbox}

\usepackage{algorithm}
\usepackage{algorithmic}
\usepackage{comment}
\usepackage{quantikz}
\usepackage{tikz}
\usetikzlibrary{arrows.meta, positioning}

\usepackage{placeins}
\usepackage{xcolor}
\usepackage[pdfencoding=auto, psdextra]{hyperref}
\usepackage[nameinlink,capitalize]{cleveref}
\crefname{appendix}{SM}{SM}
\Crefname{appendix}{SM}{SM}

\usepackage{bbm}
\usepackage{orcidlink}
\usepackage[normalem]{ulem}
\usepackage{booktabs}

\usepackage{acro}
\DeclareAcronym{NP}{
    long = Non-deterministic Polynomial time,
    short = NP
}

\DeclareAcronym{QSP}{ 
    long = Quantum State Preparation,
    short = QSP
}
\DeclareAcronym{MCX}{
    long = Multi-Controlled NOT,
    short = MCX
}
\DeclareAcronym{MCR}{
    long = Multi-Controlled Rotation,
    short = MCR
}
\DeclareAcronym{HHL}{
    long = Harrow-Hassidim-Lloyd,
    short = HHL
}
\DeclareAcronym{QFT}{ 
    long = Quantum Fourier Transform,
    short = QFT
}
\DeclareAcronym{QLS}{ 
    long = Quantum Loading Scheme,
    short = QLS
}
\DeclareAcronym{NISQ}{ 
    long = Noisy Intermediate-Scale Quantum,
    short = NISQ
}
\DeclareAcronym{QRAM}{ 
    long = Quantum Random-Qccess Memory,
    short = QRAM
}
\DeclareAcronym{TSP}{ 
    long = Travelling Salesperson Problem,
    short = TSP
}
\DeclareAcronym{TCI}{
    long = Tensor Cross Interpolation,
    short = TCI
}
\DeclareAcronym{TT}{
    long = Tensor Train,
    short = TT
}
\DeclareAcronym{MPS}{
    long = Matrix Product State,
    short = MPS
}
\DeclareAcronym{MPO}{
    long = Matrix Product Operator,
    short = MPO
}
\DeclareAcronym{SVD}{
    long = Singular Value Decomposition,
    short = SVD
}
\DeclareAcronym{DMRG}{
    long = density matrix renormalization group,
    short = DMRG
}
\DeclareAcronym{WHT}{
    long = Walsh-Hadamard transform,
    short = WHT
}
\DeclareAcronym{BT}{
    long = Binary-Tree,
    short = BT
}
\DeclareAcronym{BT-WHT}{
    long = Binary-Tree Walsh--Hadamard Transform,
    short = BT-WHT
}
\DeclareAcronym{BT-BE}{
    long = Binary-Tree Boolean Encoder,
    short = BT-BE
}

\DeclareAcronym{SM}{
    long = Supplemental Material,
    short = SM
}

\DeclareAcronym{DSOP}{
    long = Disjoint Sum of Products,
    short = DSOP
}
\DeclareAcronym{ESOP}{
    long = Exclusive Sum of Products,
    short = ESOP
}
\DeclareAcronym{SOP}{
    short = SOP,
    long = Sum of Products
}
\DeclareAcronym{TT-SVD}{
    short = TT-SVD,
    long = Tensor Train Singular Value Decomposition
}
\DeclareAcronym{PDE}{ 
    long = Partial Differential Equation,
    short = PDE
}

\newtheorem{theorem}{Theorem}

\newtheorem{lemma}[theorem]{Lemma}

\newcommand{\vc}{\bm}

\renewcommand\bra[1]{\langle #1 |}
\renewcommand\ket[1]{| #1 \rangle}

\newcommand{\I}{\mathrm{I}}

\AtBeginDocument{%
}
\AtBeginDocument{%
    \crefname{appendix}{SM}{SM}%
    \Crefname{appendix}{SM}{SM}%
}
\begin{document}

\title{
Scalable Unitary State-Encoding for Generic Large Data
}

\author{Vittorio Pagni\orcidlink{0009-0006-9753-3656}}
\email{vittorio.pagni@dlr.de}
\affiliation{German Aerospace Center (DLR), Institute of Software Technology, Sankt Augustin, Germany}
\affiliation{University of Cologne, Department of Mathematics and Computer Science, Cologne, Germany}
\author{Peter K. Schuhmacher\orcidlink{0000-0003-1232-4363}}
\affiliation{German Aerospace Center (DLR), Institute of Software Technology, Sankt Augustin, Germany}
\author{Michael Felderer\orcidlink{0000-0003-3818-4442}}
\affiliation{German Aerospace Center (DLR), Institute of Software Technology, Sankt Augustin, Germany}
\affiliation{University of Cologne, Department of Mathematics and Computer Science, Cologne, Germany}
\author{Michael Epping\orcidlink{0000-0003-0950-6801}}
\affiliation{German Aerospace Center (DLR), Institute of Software Technology, Sankt Augustin, Germany}
\author{Kevin Lively\orcidlink{0000-0003-2098-1494}}
\email{kevin.lively@dlr.de}
\affiliation{German Aerospace Center (DLR), Institute of Software Technology, Sankt Augustin, Germany}

\begin{abstract}
Efficient encoding of data into quantum states remains a fundamental bottleneck in many quantum computing protocols, which would otherwise offer advantages over their classical counterparts. Here, we introduce deterministic and scalable amplitude-encoding methods with low two-qubit gate costs and tunable trade-offs between circuit depth and encoding fidelity. Our approaches require no amplitude amplification or mid-circuit measurement and are capable of encoding arbitrary complex data or queryable functions by using tensor-networks to go beyond what can be held in classical memory. Our principle contributions are two complementary algorithms for simplifying fully controlled multiplexer circuits to one- and two-qubit gates. In the first, we formally map the elimination of redundant gate controls to a Boolean logic minimization problem, while in the second we utilize a single ancilla qubit to map the multi-controlled rotation gates to a sequence of one- and two-qubit operations via the Walsh-Hadamard spectrum of the gate angles. We benchmark these methods across a broad range of function classes, characterizing their total depth in an all-to-all connectivity graph against system size and target fidelity. Hybridizing our methods gives up to orders of magnitude improvement in circuit depth at equivalent state fidelity compared to the default state-preparation method in the Qiskit package. We furthermore demonstrate our method on IQM hardware with a square lattice connectivity, finding excellent fidelity, even with minimal noise mitigation. By significantly reducing gate depth for encoding, this work enables the experimental execution of applications on current and near-term hardware which would otherwise be reduced to noise before the first calculation step, shortening the time horizon to robust quantum utility.
\end{abstract}
\maketitle
\acresetall
\section{Introduction}
Quantum computing has attracted substantial scientific attention and economic investment in recent years because certain quantum algorithms promise substantial, in some cases exponential, speedups over the best currently known classical methods~\cite{Shor1994,Grover1996,Harrow2009,QuantumAlgorithmZoo}. However, even assuming the eventual availability of a fully fault-tolerant quantum computer, such algorithmic speedups do not necessarily translate into an end-to-end computational advantage when including the cost of classically pre-processing, preparing, and reading out the data. In particular, a nominal quantum advantage may be eliminated before the computation even begins if the classical data required by the algorithm cannot be encoded efficiently into the physical state of the quantum device.

A pertinent example is the \ac{HHL} algorithm~\cite{Harrow2009}, together with its subsequent refinements~\cite{Childs2017,Liu2022,Bravoprieto2026} which underpin many quantum machine learning methods \cite{Morales2026, Ji2026}. Regardless of all other conditions in the problem, if preparing the input state $\ket{b}$ scales as $O(N)$ or polynomially in $N$ elements, then the proposed speedup is already lost~\cite{Aaronson2015a}. Similar qualifications apply to many quantum algorithms requiring classical data loading \cite{Dalzell2025}. Consequently, demonstration of end-to-end applied quantum advantage for many non-trivial tasks of scientific, engineering or economic interest necessitates encoding methods which have efficient classical pre-processing, a low total gate-depth, high encoding fidelity, deterministic or high-probability success, and feasible qubit resource requirements.

While the problem of \ac{QSP} is a rich and diverse topic which has been worked on by various communities for decades \cite{Zhang2026}, when excluding variational approaches, we can broadly divide it into methods that exploit constraints on the type of input or generic methods, i.e. those agnostic to the input data-structure. In the former case, many schemes have been put forward for specific data classes such as efficiently integrable probability distributions, sparse and low-density states, states which admit low-rank tensor-network representations, or specific function families like Gaussians, piecewise-continuous polynomials, or power-law turbulence spectra \cite{GroverRudolph2002,Zhang2022, Pagni_2026,Schön2005,Jaderberg2026,Green2026,Huggins2025,Xie2025,OBrien2025,Manabe2025,Guseynov2026,Meng2026}. Data-agnostic schemes are generally based on fully controlled multiplexers (also known as multiplexors, or uniform controlled rotations)~\cite{Mottonen,Shende_2006,Carvalho2025}:
\begin{equation}
    \mathcal{M}_l(\vc{\theta}) = \sum_{j=1}^{2^l}|j\rangle\langle j|\otimes R(\theta_j),
\end{equation}
which are controlled by $l$ qubits, require $2^l$ angles $\vc{\theta}\in\mathbb{R}^{2^l}$, and affect the rotation gate $R(\theta)$ on the target qubit. Generic state preparation centered on this gate was first described by Möttönen et. al \cite{Mottonen}, and expanded via a recursive decomposition by Shende et. al \cite{Shende_2006}, giving a total circuit depth of $O(N)$. Sun et. al further optimized this, giving a tight-bound optimal depth of $\Theta(N/n)$ for $N=2^n$ \cite{Sun2023}. While further depth-optimizations are possible, they require the introduction of up to $O(2^n)$ ancillary qubits \cite{Araujo2021,Sun2023} which can be reduced at the cost of mid-circuit measurement and feedback \cite{Zhang2022,Nmaju2025}. Approximation methods have been put forward, but only for real, positive functions and require variational elements which are known to suffer from barren plateaus \cite{Marin-Sanchez2023}.

While these works provide a formal prescription for generic QSP, they are insufficient for practical implementation on hardware when considering how to calculate and store exponentially many angles as well as finding and eliminating redundant multi-controlled gates in a computationally efficient way. In this work, we address both of these problems, allowing us to construct encoding circuits for arbitrarily large queryable complex functions and data sets under well-controlled fidelity error bounds with orders of magnitude improvement in gate-depth compared to the Qiskit multiplexer based state preparation method \cite{Iten_2016,qiskit}.

We do so first by starting from the binary-tree graph often accompanying multiplexer based state preparation \cite{Cortese2018,Araujo2021,Araujo2023,Clader2023,Nmaju2025}. This binary-tree graph exposes the hierarchichal dependence of angles which enter the fully controlled multiplexers at different layers $l$, allowing us to systematically truncate branches corresponding to small probability amplitude weight of the final state. Finding qubit layouts where this elimination occurs near the top of the tree subsequently leads to an exponential reduction in the number of angles. 

Second, for the remaining angles, we introduce the \ac{BT-BE} and \ac{BT-WHT} algorithms for multiplexer simplification. \ac{BT-BE} builds on the previous work of some of the present authors which performs a finite-precision binary decomposition of the continuous rotation angles~\cite{Pagni2025}. Here, we formally map the $l-$qubit controlled gates onto a Boolean vector space of size $2^l$, and for the first time demonstrate how this problem is equivalent to the Boolean logic \ac{DSOP} simplification problem, which is known to be \ac{NP}-hard \cite{gimpel1965reduction,Bernasconi2013}. 

The \ac{BT-WHT}, is based on the decomposition of multiplexed controlled rotations using the \ac{WHT} of their angle vectors. Related approaches which have previously appeared in the literature~\cite{Mottonen,Shende_2006,Welch_2014} rely on specific ordering of the angles in order to efficiently address all $2^l$ angles. Here, we exploit both the truncation of the binary-tree and sparsity within the \ac{WHT} transform to address only the most important \ac{WHT} coefficients via the introduction of a single ancilla. We ensure the scalability of our approach to extremely large data sizes by formulating the entire \ac{BT-WHT} pipeline to be able to operate entirely on data which is in a \ac{TT} representation, also known as a \ac{MPS} \cite{Oseledets2011TensorTrain,Schollwock2011,Khoromskij2012TensorMethods,Waintal2026}. This drops our memory requirements from $O(2^l)\to O(2l\chi^2)$ for multiplexer angle vectors controlling $l$ qubits which admit a \ac{TT} bond dimension of $\chi$. Given the recent advances in \ac{TCI} for fitting arbitrary data, this allows our method to be applied to any queryable function or data-set which admits some \ac{TT} representation of acceptably large $\chi$ \cite{Ritter2024,N_ez_Fern_ndez_2025}. Finally, we combine the two methods into a BT-Hybrid approach, which uses a \ac{TT} for once $l$ exceeds a chosen value.

\begin{figure*}
\centering
\includegraphics[width=0.99\textwidth]{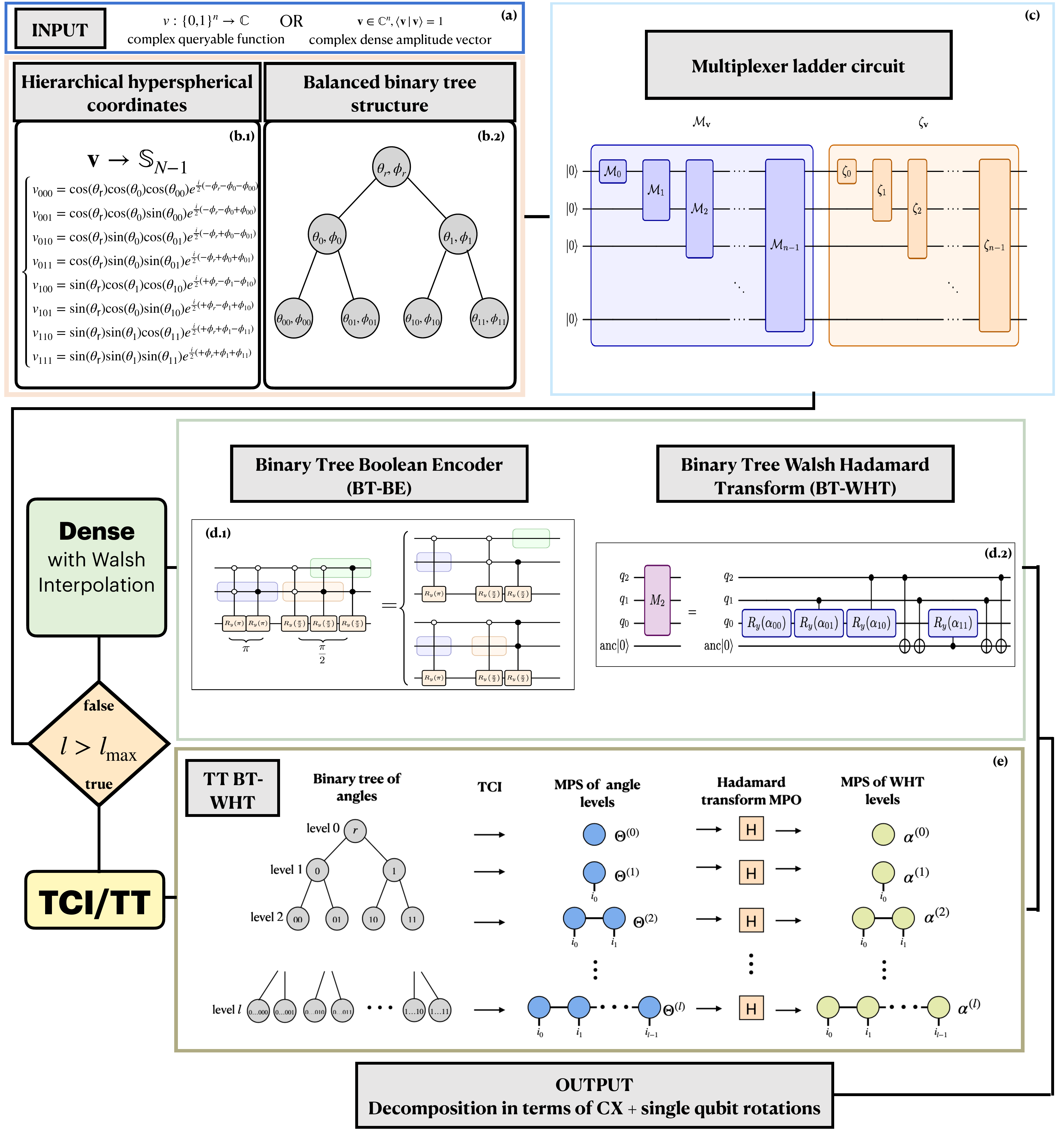}
\caption{Overview of the structure of our algorithms and this paper. The input, (a), of the procedure can be a queryable complex function, data-set or  dense amplitude vector. The entries are recursively parametrized using hierarchical hyperspherical  coordinates (b.1) organized into a binary tree (b.2), calculated from the top down. These angles enter the multiplexer-ladder circuit (c), with each layer of $2^{l}$ binary-tree angles corresponding to the $l$-th multiplexer $\mathcal{M}_l$. Different methods are used to simplify and decompose a multiplexer into a CX and one-qubit rotation gate basis, depending on the size and structure of the corresponding angle layer. For layers up to some $l_{\max}$, corresponding to $2^{l_{\max}}$ angles, we construct the full angle vectors and subsequently simplify and decompose them using \ac{BT-BE} (d.1), \ac{BT-WHT} (d.2), or a hybrid of the two methods which acts directly on dense angle vectors. For deeper layers, the required nodes can instead be queried using \ac{TCI} applied to \cref{eq:modulus_tensor_angles,eq: phase angle}, producing a \ac{TT} representation of each layer (e). Each TT can then be transformed efficiently into the Walsh basis by applying the Hadamard transform and the spectrum can be queried to retrieve the angles $\vc{\alpha}$ needed by \ac{BT-WHT}.}
\label{fig:paper_scheme_summary}
\end{figure*}

A visual overview of the conceptual structure of the algorithms is provided in \cref{fig:paper_scheme_summary}. The main body of the text is meant to provide an accessible overview of our methods, defering extensive and detailed derivations to the \ac{SM}. In \cref{sec:theory}, we lay out the binary-tree representation, its correspondence with the multiplexer-ladder circuit and explain how we utilize the \ac{TT} representation. In \cref{sec:algorithms}, we introduce the BT-BE and BT-WHT algorithms as well as how to synthesize them into a BT-Hybrid method. We evaluate their scaling and accuracy for a variety of input functions and data in \cref{sec:results}, and demonstrate the performance of our encoder on real hardware. We conclude in \cref{sec:conclusions} with the expected impact of this work and the subsequent research directions it inspires.

\section{Theory}
\label{sec:theory}

Here we introduce our two key ansätze: 
the recursive angle parametrization of the input amplitude vector and the associated encoding circuit.
First in \cref{sec:binary-tree} we utilize a hierarchical angle parameterization to represent the target state. We do so following a binary tree architecture, whose nodes contain the angles required by our encoding circuit. This recursive structure allows for efficient storage and truncation of angle data.
In \cref{sec:circuit-ansatz} we define our circuit ansatz, the so-called multiplexer ladder, ultimately descending from the work of Möttönen and Shende et. al \cite{Mottonen,Shende_2006}. Our algorithms, defined in \cref{sec:algorithms} are concerned with efficiently calculating, storing, and encoding  the up-to $2(2^n-1)$ free parameters of the circuit as one- and two-qubit gates.

\subsection{The binary-tree representation}
\label{sec:binary-tree}
Let us consider a quantum state of n qubits of the form 
\begin{equation}
\ket{v}:=\sum_{i=0}^{2^n-1} v_i \ket{i}
\label{encoding_state}
\end{equation} whose amplitudes are normalized with respect to the $L_2$-norm inner product. The ket state $\ket{i}$ is written in the computational basis, with the index $i\in\{0,\ldots,2^n-1\}$ expanded as a binary integer. We may trivially reshape the amplitude vector $\vc{v}\in \mathbb{C}^{2^n} $ into a tensor $\mathbf{v}\in\mathbb{C}^{2,2,\dots,2}$ as follows:
\begin{equation}\label{eq:tensor-amplitude}
\begin{split}
    \ket{v}&=\sum_{i_0,\dots,i_{n-1}=0,\ldots, 0}^{1,\ldots, 1}\text{v}_{i_0,i_1,\dots,i_{n-1}}\ket{i_0}\ket{i_1}\dots \ket{i_{n-1}}\\
    &=\sum_{\vc{i} = \vc{0}}^{\vc{1}}\text{v}_{\vc{i}}\ket{\vc{i}},
\end{split}
\end{equation}
where we use the bold character to indicate tuple notation $\vc{i} = (i_0,\ldots, i_{n-1})\in\mathbb{Z}_2^n$ to avoid notational clutter. This bold tuple notation will also be used for sub-spaces of the $n$-qubit Hilbert space by indicating $\vc{i}[l] = (i_0,\ldots,i_{l-1})\in\mathbb{Z}_2^l$. Each qubit index $i_k\in\mathbb{Z}_2=\{0,1\}$ and $\vc{0} = (0,\ldots, 0)$.
We can then use the polar representation of the complex entries of $\mathbf{v} $ to write
\begin{equation}
     \text{v}_{\vc{j}}=M_{\vc{j}}e^{i \zeta_{\vc{j}}},
\end{equation}
with the modulus tensor $M$, and the phase tensor $\zeta $ both being elements of $\mathbb{R}^{2,\dots,2}$. Since the amplitude vector has unit norm and it is defined up to a global phase, we have the constraint
\begin{equation}\label{normalization_contraints}
\sum_{\vc{j}} M^2_{\vc{j}}=1.
\end{equation}
After imposing normalization and fixing one real degree of freedom corresponding to the global phase, the state is characterized by $2(2^n-1)$ independent real parameters.

The derivation of the modulus and phase tensors in terms of angles which enter the multi-controlled $R_y$ and $R_z$ gates in the multiplexer circuit in \cref{sec:circuit-ansatz}, are pedagogically well explained in other works  \cite{Araujo2021,Clader2023,Araujo2023,Zhang2026}, here we simply reproduce them out of space considerations:
\begin{equation}\label{eq:modulus_tensor_angles}
     M_{\vc{j}}
     = \prod_{l=0}^{n-1}G_{j_l}(\theta_{\vc{j}[l]}),
\end{equation}
where 
\begin{equation}
       G_{j_l}(x)\coloneqq \begin{cases}
           \cos(x)\null\quad&\text{if } j_l= 0\\
           \sin(x)\null\quad&\text{if } j_l= 1.
       \end{cases}
\end{equation}

Since the modulus tensor represents a point on the nonnegative orthant of the $N-1$ dimensional hypersphere, $\mathcal{S}_{N-1}$, we refer the above parameterization as hierarchical hyperspherical coordinates.
The corresponding angles are calculated from $\ket{v}$ via
\begin{equation}\label{eq:binarytreepolarangles}
\begin{aligned}
\theta_{\vc{j}[l]}&\coloneqq\arctan\left(\sqrt{\frac{
\bra{v}
\proj{\vc{j}[l]}\otimes\proj{1}
\otimes  {I}_{n-l-1}
\ket{v}
}{
\bra{v}
\proj{\vc{j}[l]}\otimes\proj{0}\otimes  {I}_{n-l-1}
\ket{v}
}}\right)\\
&=\arctan\left(\sqrt{\frac{\sum_{j_{l+1}\dots j_{n-1}}|v_{\vc{j}[l]1j_{l+1}\dots j_{n-1} }|^2}{\sum_{j_{l+1}\dots j_{n-1}}|v_{\vc{j}[l]0j_{l+1}\dots j_{n-1} }|^2}}\right),
\end{aligned}
\end{equation}
which has value $\frac{\pi}{2}$ if the denominator is zero and where $I_{k}$ is the identity operator on the $k$-qubit subspace. 
\begin{figure}%
\centering%
\resizebox{\columnwidth}{!}{%
\begin{tikzpicture}[
    every node/.style={
        draw,
        circle,
        align=center,
        minimum size=11mm,
        inner sep=1.5pt,
        font=\small
    },
    root/.style={fill=gray!15, font=\small\bfseries},
    internal/.style={fill=gray!8},
    leaf/.style={fill=white},
    placeholder/.style={
        fill=white,
        minimum size=7mm,
        inner sep=0.5pt,
        font=\scriptsize,
        line width=0.6pt
    },
    highlight/.style={
        fill=blue!20,
        draw=blue!70!black,
        line width=1pt
    }
]

\node[root,highlight] (root) at (0,0)
    {$\bigl(\theta,\phi\bigr)_{\mathrm{root}}$};

\node[internal,highlight] (n0) at (-1.8,-1.5)
    {$\bigl(\theta,\phi\bigr)_0$};

\node[internal] (n1) at (1.8,-1.5)
    {$\bigl(\theta,\phi\bigr)_1$};

\node[leaf] (n00) at (-2.8,-3.0)
    {$\bigl(\theta,\phi\bigr)_{00}$};

\node[leaf,highlight] (n01) at (-0.8,-3.0)
    {$\bigl(\theta,\phi\bigr)_{01}$};

\node[placeholder] (v000) at (-3.3,-4.4)
    {$v_{000}$};

\node[placeholder] (v001) at (-2.3,-4.4)
    {$v_{001}$};

\node[placeholder] (v010) at (-1.3,-4.4)
    {$v_{010}$};

\node[placeholder,highlight] (v011) at (-0.3,-4.4)
    {$v_{011}$};

\draw[line width=0.8pt] (root) -- (n1);
\draw[line width=0.8pt] (n0) -- (n00);

\draw[line width=0.8pt]
    (n1) -- (0.9,-2.8);

\draw[line width=0.8pt]
    (n1) -- (2.7,-2.8);

\draw[blue!70!black,line width=1.2pt]
    (root) -- (n0);

\draw[blue!70!black,line width=1.2pt]
    (n0) -- (n01);

\draw[blue!70!black,line width=1.2pt]
    (n01) -- (v011);

\draw[dashed,line width=0.7pt]
    (n00) -- (v000);

\draw[dashed,line width=0.7pt]
    (n00) -- (v001);

\draw[dashed,line width=0.7pt]
    (n01) -- (v010);


\end{tikzpicture}%
}%
\caption{ Binary-tree organization of the state-preparation angles for $n=3$ qubits. The $2^n-1$ internal nodes are labeled by a modulus angle $\theta$ and a phase angle $\phi$. The $2^l$ nodes at level $l$ define the angle vectors required by the $l$-th multiplexer, while the $2^n$ terminal paths correspond to the entries of the target state. On the tree we highlight in blue the example of the path $\phi$(011) from the root node to the leaf node $011$, and drop the children of node $(\theta,\phi)_1$ for clarity.}
\label{fig:angles_tree}%
\end{figure}
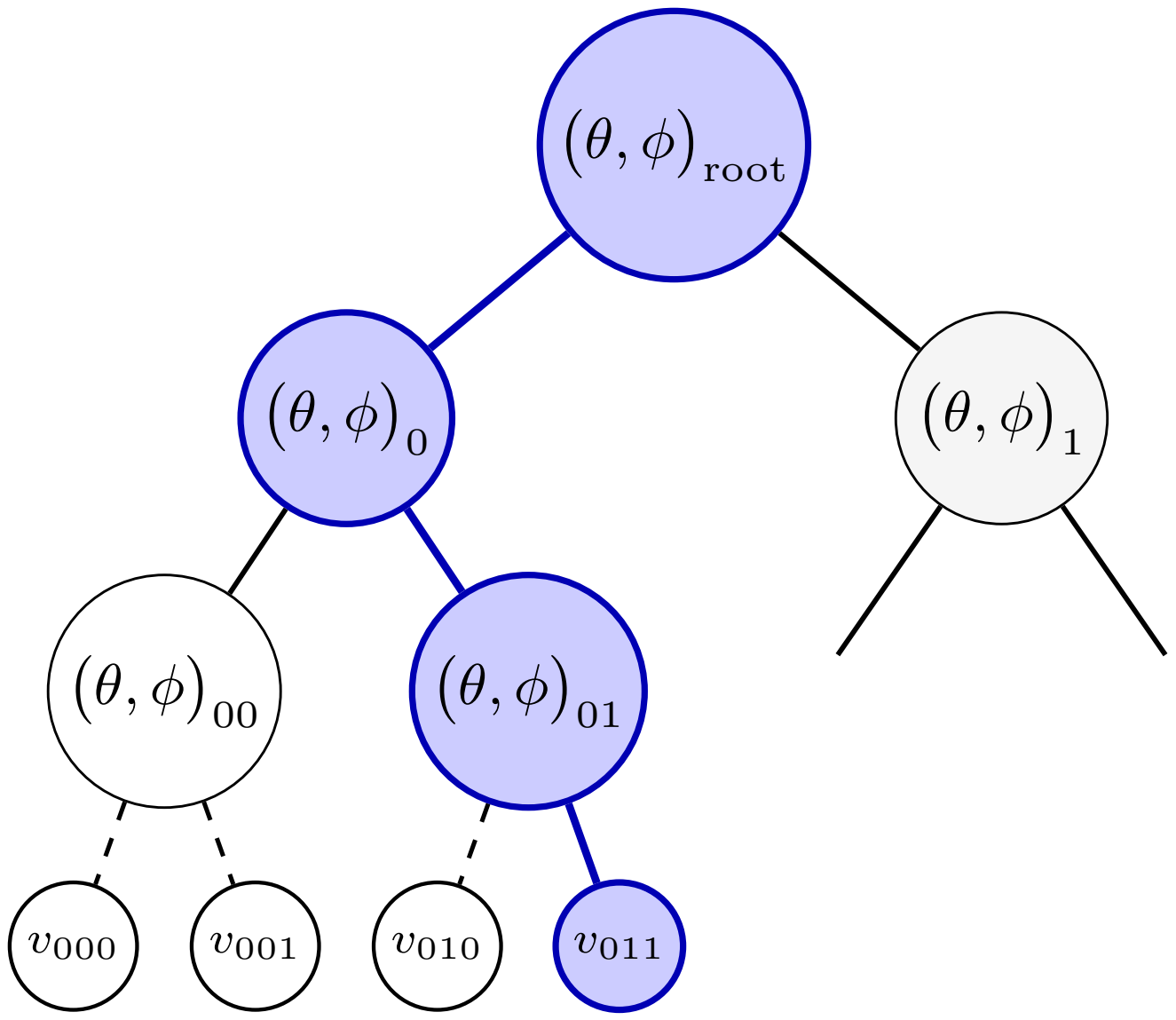%
\begin{figure*}
\centering
\input{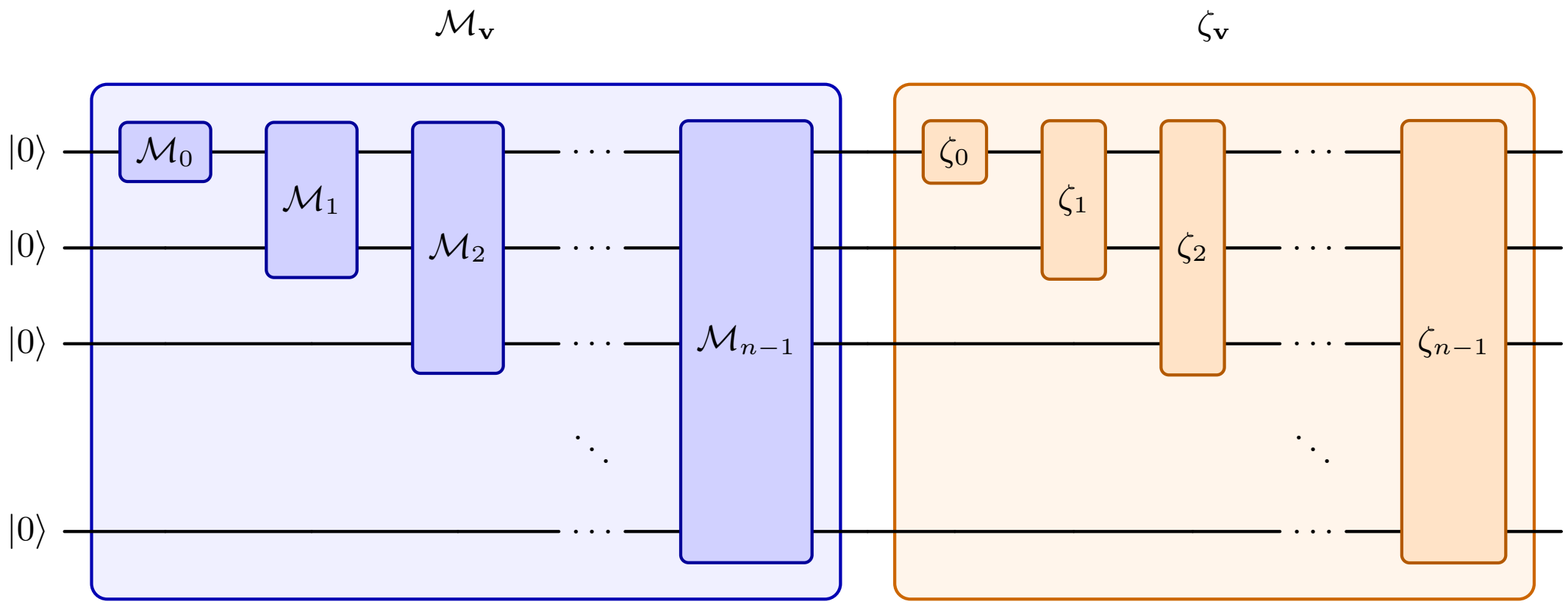}
\caption{Structure of the recursive state-preparation circuit. The circuit consists of $n$ multiplexers of $C^{l}R_y$ rotations, denoted by $\mathcal{M}_l$, which encode the moduli of the amplitudes, followed by $n$ multiplexers of $C^{l}R_z$ rotations, denoted by $\zeta_l$, which encode their phases. The structure of an individual $\mathcal{M}_l$ subcircuit is shown in \cref{fig:multiplexer}.}
\label{fig:multiplexer_ladder}
\end{figure*}
\begin{figure}
\centering
\input{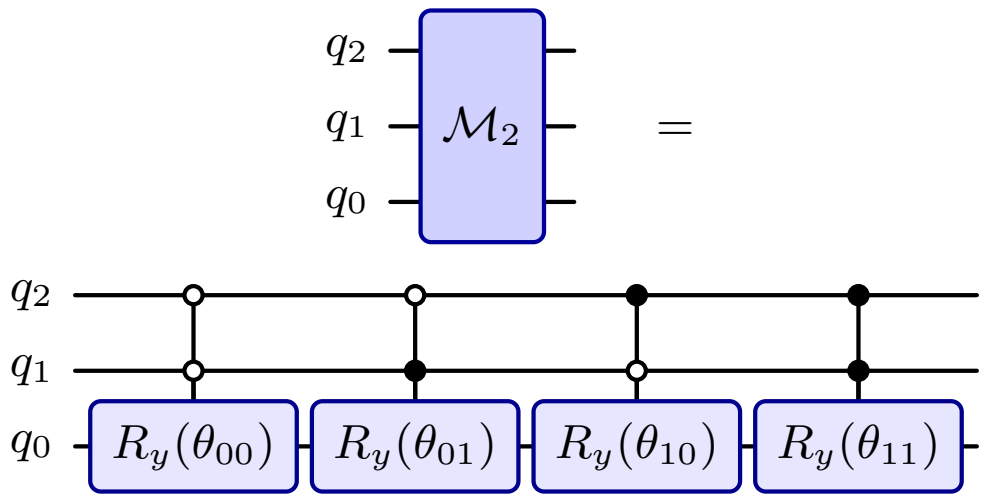}
\caption{Example of a multiplexer at layer $l=2$, consisting of $2^{2}=4$ multi-controlled $R_y$ rotations.}
\label{fig:multiplexer}
\end{figure}

We express the complex amplitudes as 
\begin{align}\label{generic_parametrization}
     v_{\vc{j}}=\prod_{k=0}^{n-1}G_{j_k}(\theta_{j_k})\exp\left((-1)^{j_k+1}\frac{i}{2} \vc{\phi}_{j_k}\right),
\end{align}
where $j_k$ is the $k^{\text{th}}$ entry of $\vc{j}$. The angles $\vc{\phi}_{\vc{j}[k]}$ giving the phase are obtained from the phase tensor
\begin{equation}\label{eq:zeta}
        \zeta_{\vc{j}} :=\arg (v_{\vc{j}}),
\end{equation}
as \cite{Mottonen}
\begin{equation}\label{eq: phase angle}
   \phi_{\vc{j}[l]}=\frac{1}{2^{n-l-1}}\sum_{\vc{k} \in\{0,1\}^{n-l-1}} (\zeta_{\vc{j}_{[l]} 1  \vc{k}}-\zeta_{\vc{j}_{[l]} 0 \vc{k}});
\end{equation}
see \cref{fig:paper_scheme_summary}, box (b.1) for an  example with three qubits. Note that for $l=0$ we have
\begin{align}\label{eq:theta_root}
    \theta_{\text{root}}=&\arctan\left(\sqrt{\frac{
    \bra{v}
    \ket{1}\bra{1}\otimes  {I}_{n-1}
    \ket{v}
    }{
    \bra{v}
    \ket{0}\bra{0}\otimes  {I}_{n-1}
    \ket{v}
    }}\right)
\end{align} 
and
\begin{align}\label{eq:phi_root}
    \phi_{\text{root}}=&\frac{1}{2^{n-1}}
    \sum_{\vc{k}\in\{0,1\}^{n-1}} \zeta_{1\vc{k}}- \zeta_{0\vc{k}}.
\end{align}
A critical observation to note is that starting from \cref{eq:theta_root} and \cref{eq:phi_root}, the partial traces over the state and phase tensors advances in a hierarchical manner as one advances $l\to l+1$ in \cref{eq:binarytreepolarangles} and \cref{eq: phase angle}. Thus we are free to organize this recursive bifurcation of the state into left, i.e. 0 projected, and right, i.e. 1 projected, portions via a balanced binary tree, as seen in \cref{fig:angles_tree}.
The modulus and phase angle vectors at level $l$ each contain $2^l$ entries. 

The binary-tree representation provides a visual interpretation of the parameterization in \cref{generic_parametrization}, which will elucidate our first approximation. 
To recover the amplitude $v_{\vc{j}}$, one begins at the root with an initial value of one and follows the path specified by the bit string $\vc{j}$. 
Going left one multiplies the amplitude with $\cos(\theta_{\vc{j}[l]})\exp(-\frac{i}{2}\phi_{\vc{j}[l]})$ and going right, with $\sin(\theta_{\vc{j}[l]})\exp(\frac{i}{2}\phi_{\vc{j}[l]})$.
The product accumulated along the complete path is $v_{\vc{j}}$. 
Therefore $\theta_{\vc{j}[l]}=\frac{\pi}{2}$ and $\theta_{\vc{j}[l]}=0$ implies
$v_{\vc{j}[l]0\vc{k}}=0$ and $v_{\vc{j}[l]1\vc{k}}=0$
for all $\vc{k}\in\{0,1\}^{n-l-1}$,
because the cosine and sine factor vanishes, respectively.
Thus if $\theta_{\vc{j}[l]}=\frac{\pi}{2}$, the left subtree below $\vc{j}[l]$ contains only zero amplitudes and need not be constructed while analogously, if $\theta_{\vc{j}[l]}=0$, the right subtree can be omitted.
In practice, we choose a numerical tolerance to decide whether $\theta_{\vc{j}[l]}$ is approximately $0$ or $\frac{\pi}{2}$. 
For a normalized state, discarding subtrees with conditional probability weight $\delta$ and renormalizing the remaining state produces a squared overlap of $1-\delta$ with the original state. 
Pruning the tree this way at high levels leads to exponential compression in storing the state.
Since these sub-trees do not contribute to the state amplitudes, but some of our encoding sub-steps require complete $2^l$ sized levels, we are free to interpolate their values such that they retain a convenient structure, see \cref{sec:BT-WHT}. 

Operationally, the binary tree is the data structure underlying our encoding algorithms. We differ from previous works by relying on the tangent function definition of the angle, as it provides a more balanced bisection between the cosine and sine in the binary-tree decomposition, allowing subsequent layers to be generated recursively. At each level $l$, the active nodes define the angle vectors $\vc{\theta}^{(l)},\vc{\phi}^{(l)}\in\mathbb{R}^{2^l}$ required by the corresponding circuit layer $l$ defined in the next section. For sufficiently small $l$, these vectors can be directly constructed, while at deeper levels, we can construct a \ac{TT} which encodes them using \ac{TCI}, see \cref{sec:tensor-network-theory}. 

\subsubsection{Special Case of Real Valued Functions}
In case the function we want to encode is real valued, we can represent it using the tree of $\theta$ angles alone. This is achieved by extending the range of $\theta$ of the nodes in the last layer of the tree from $[0,\frac{\pi}{2}]$ to $[-\pi,\pi]$. The $\theta$ values of each node of the layers $0$ to $n-2$ are still calculated as \cref{eq:binarytreepolarangles}, while those on the last layer are determined by the formula 
\begin{equation}
    \theta_{\vc j [n-1]}=\operatorname{atan2}\left(v_{\vc j[l]0},v_{\vc j[l]1) }\right),
\end{equation}
for $l=n-2$, where atan2 can be written as
\begin{equation}
\operatorname{atan2}(y, x) = 
2 \arctan\left( \frac{y}{\sqrt{x^2 + y^2} + x} \right),
\end{equation}
if $x>0$ or $y\neq 0$ and $\pi$ otherwise.

\subsection{The Circuit Ansatz}
\label{sec:circuit-ansatz}

The parametrization of a generic quantum state we gave in the previous subsection has a natural correspondence in terms of a state preparation circuit: 
each one of the $n$ layers of angles $\vc{\theta}^{(l)}$ and $\vc{\phi}^{(l)}$ in the binary-tree graph corresponds to a fully controlled quantum multiplexer block in the multiplexer ladder circuit shown in \cref{fig:multiplexer_ladder,fig:multiplexer} \cite{Araujo2021,Clader2023,Araujo2023,Zhang2026}. 
For any input vector $\vc{v} \in \mathbb{C}^N$, the two circuit blocks 
\begin{equation}
\label{eq:modulus_multiplexer_ladder_as_product}
    \mathcal{M}_{\vc{v}}=\prod_{l=0}^{n-1} \mathcal{M}_l\left(\vc{\theta}^{(l)}\right)
\end{equation}
and 
\begin{equation}
\label{eq:phase_multiplexer_ladder_as_product}
    \zeta_{\vc{v}}=\prod_{l=0}^{n-1} \zeta_l\left(\vc{\phi}^{(l)}\right)
\end{equation}
compose the unitary $U_{\vc{v}}=\zeta_{\vc{v}}\mathcal{M}_{\vc{v}}$, such that
\begin{equation}
    U_{\vc{v}}\ket{0}=\ket{v},
\end{equation}
where $\ket{v}$ defined in \cref{encoding_state}
is the target state whose amplitudes encode the normalized vector $\vc{v}$. 
Every element of the products in \cref{eq:modulus_multiplexer_ladder_as_product} and in \cref{eq:phase_multiplexer_ladder_as_product} corresponds to a given uniform multiplexer level $l$, with $l=0$ corresponding to the single qubit operator, $l=1$ to the two-qubit multiplexer and so on. 
We use multi-controlled $R_y$ gates for the modulus and multi-controlled $R_z$ for the phase respectively. Given that our modulus angles are constructed as $\theta\in[0,\pi/2]$, given the standard definition of rotation gates halving the angles, it is in fact $2\vc{\theta}^{(l)}$ which enters the multi-controlled $R_y$ gates.
We use the notation 
\begin{equation}
    C^lR(\vc{c},\theta)= \I + P_{\vc{c}} \otimes (R-I)
\end{equation}
to denote a controlled-$R$ gate with $l$ control qubits
where the binary string $\vc{c} \in \{0,1\}^l$ associated with the projector $P_{\vc{c}}=\proj{c}$ is referred to as the \textit{control string}.
The multiplexer ladder circuit has full expressivity, meaning it can prepare any quantum state in depth $\Theta(2^n/n)$ \cite{Sun2023}.  
Our core algorithmic developments exploit the underlying structure of the angles in each layer in order to simplify the multiplexer blocks and thereby reduce the total gate depth.

\subsection{Tensor-network compression}
\label{sec:tensor-network-theory}
\begin{figure*}
\centering
\includegraphics[width=\textwidth]{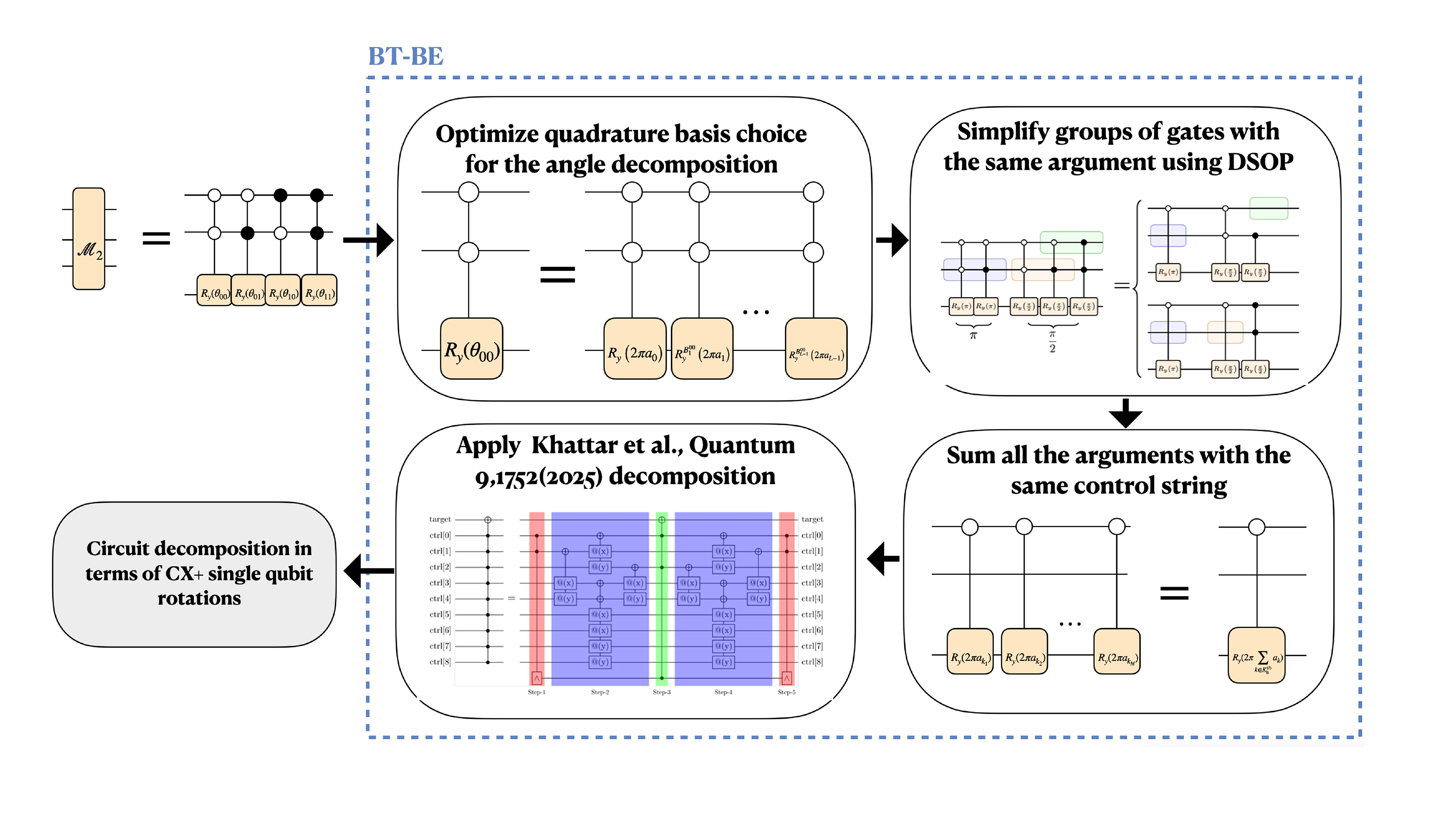}
\caption{Overview of the \ac{BT-BE} algorithm. A continuous multiplexer angle layer is first approximated in a finite basis, producing a collection of fixed-angle Boolean masks. Each mask is simplified independently using Disjoint Sum of Products (\ac{DSOP}) synthesis, after which rotations with identical generalized controls are collected. The resulting multi-controlled rotations are finally decomposed into the chosen elementary gate basis via the KG24 method from \cite{KhattarGidney2025ConditionallyCleanAncillae}.}
\label{fig:BT_BE_scheme}
\end{figure*}

When the data we wish to encode is too large to be held directly in classical memory, as may be expected for many quantum algorithms of interest, we can still evaluate \cref{eq:binarytreepolarangles,eq: phase angle} by first representing it as a tensor network, the simplest of which is the \ac{TT}. A \ac{TT}, also known in physics as an \ac{MPS}, is a tensor network with a one-dimensional topology which represents a high-order tensor as an ordered product of lower-order core tensors~\cite{Oseledets2011TensorTrain,Schollwock2011,Khoromskij2012TensorMethods,Waintal2026}. \ac{TT} representations can substantially reduce the memory and computational cost when the corresponding bond dimensions remain small relative to the Hilbert space size. For the present case of qubits, this binary tensorization is commonly referred to as the quantics tensor-train representation~\cite{Khoromskij2011QTT}

The state amplitude tensor from \cref{eq:tensor-amplitude}, admits a \ac{TT} representation of the form 
\begin{equation}
\label{eq:tt-decomposition}
\begin{aligned}
v_{\vc{i}}
=& \sum_{\vc{\alpha}}A^{(0)}_{i_0,\alpha_0}
A^{(1)}_{\alpha_0,i_1,\alpha_1}
\cdots
A^{(n-1)}_{\alpha_{n-2},i_{n-1}}
\end{aligned}
\end{equation}
with the internal `link' or `bond' indices 
\begin{equation}
\alpha_q\in\{1,\ldots,\chi_q\},
\end{equation}
and $\chi_q$ being the bond dimension across the $q$-th bipartition. For convenience, we define the boundary bond dimensions as $\chi_{-1}=\chi_{n-1}=1$, and thus each tensor-core has dimensions
\begin{equation}
A^{(q)}
\in
\mathbb{C}^{\chi_{q-1}\times 2\times\chi_q}.
\end{equation}

A dense vector requires the storage of $2^n$ complex coefficients. By contrast, the number of coefficients in its \ac{TT} representation is

\begin{equation}N_{\mathrm{TT}}=
\sum_{q=0}^{n-1}2\chi_{q-1}\chi_q\leq2n\chi^2,\qquad\chi\coloneqq\max_q\chi_q.
\end{equation}

Thus, the representation is efficient when the maximum bond dimension $\chi$ grows sufficiently slowly with $n$, which is not generally the case for random, unstructured data. However, this is in practice not a limitation for many applications of interest, including quantum many-body physics, quantum chemistry, or \ac{PDE} simulation \cite{ChanSharma2011DMRG,Wouters2014DMRG,Lively2025,Gourianov2025,Gross2026,Waintal2026}.

When all $2^n$ entries of $\mathbf v$ are available, the cores can be constructed using the \ac{TT-SVD} algorithm~\cite{Oseledets2011TensorTrain}. The tensor is first reshaped into a matrix separating the first physical index from all remaining indices, and a \ac{SVD} is applied. The left singular vectors form the first core, while the singular values and right singular vectors are reshaped and passed to the next step. 

Compression is obtained by retaining only the $\chi_q$ largest singular values at each bipartition. Given the squared norm of the discarded singular values $\vc{\sigma}^(q)$ at for core $q$,
\begin{equation}
\epsilon_q^2=\sum_{r>\chi_q}\left(\sigma_r^{(q)}\right)^2,
\end{equation}
the tensor produced by the truncated \ac{TT-SVD} the satisfies Frobenius norm error bound~\cite{Oseledets2011TensorTrain}
\begin{equation}
\left|
\mathbf v-\widetilde{\mathbf v}
\right|_{\mathrm F}^2
\leq
\sum_{q=0}^{n-2}\epsilon_q^2.
\end{equation}

The \ac{TT-SVD} construction nevertheless requires access to the complete dense tensor and is therefore unsuitable once $2^n$ becomes too large to store or feasibly perform SVDs over. In this regime, we instead employ \ac{TCI}, also known as the \ac{TT}-cross approximation~\cite{Oseledets2010TTCross,N_ez_Fern_ndez_2022,N_ez_Fern_ndez_2025}. Rather than evaluating every tensor entry, \ac{TCI} works by sampling entries of the $\mathbf{v}$, which are then used as pivots to construct the \ac{TT} cores from a generalized cross approximation. The pivots are then refined through successive sweeps until the prescribed convergence criterion is reached. If the queryable function or data set admits a low-rank TT approximation then TCI will converge to it. Qin \emph{et al.}~\cite{qin2022error} derived error bounds for the TCI, and found that there exists a cross approximation $\widetilde{\mathbf v}$ whose Frobenius norm error obeys
\begin{equation}
  |\mathbf{v}-\widetilde{\mathbf v}|_F\leq \frac{(3\kappa)^{\lceil \log_2 n \rceil - 1}}{3\kappa - 1}\,
  (\chi+1)\,\epsilon ,
  \label{eq:ttcross-exact}
\end{equation}
where $\epsilon = \max_{k} \lVert \mathbf{v}^{\langle k\rangle} - \mathbf{v}^{\langle k\rangle}_{\chi_k} \rVert_F$ 
is the largest rank $\chi_k$ truncation error over matrix reshapings of $\mathbf{v}$, namely, $\mathbf{v}^{\langle k\rangle}\in\mathbb{C}^{2^{k}\times 2^{n-k}}$, for $k\in\{1,\ldots,n\}$, while $\chi=\text{max}_k\chi_k$ is the largest tensor core rank, and $\kappa$ measures the conditioning of the selected pivot submatrices. The error thus grows only linearly in $\epsilon$, and since the order enters solely through the exponent $\lceil \log_2 n \rceil$, the amplification factor in \cref{eq:ttcross-exact} scales as $O\!\bigl(n^{\log_2 3\kappa}\bigr)$, which is algebraic rather than exponential in $n$, provided the pivots are well conditioned. This can be guaranteed for instance by maximum-volume selection \cite{Goreinov2010GoodSubmatrix}.
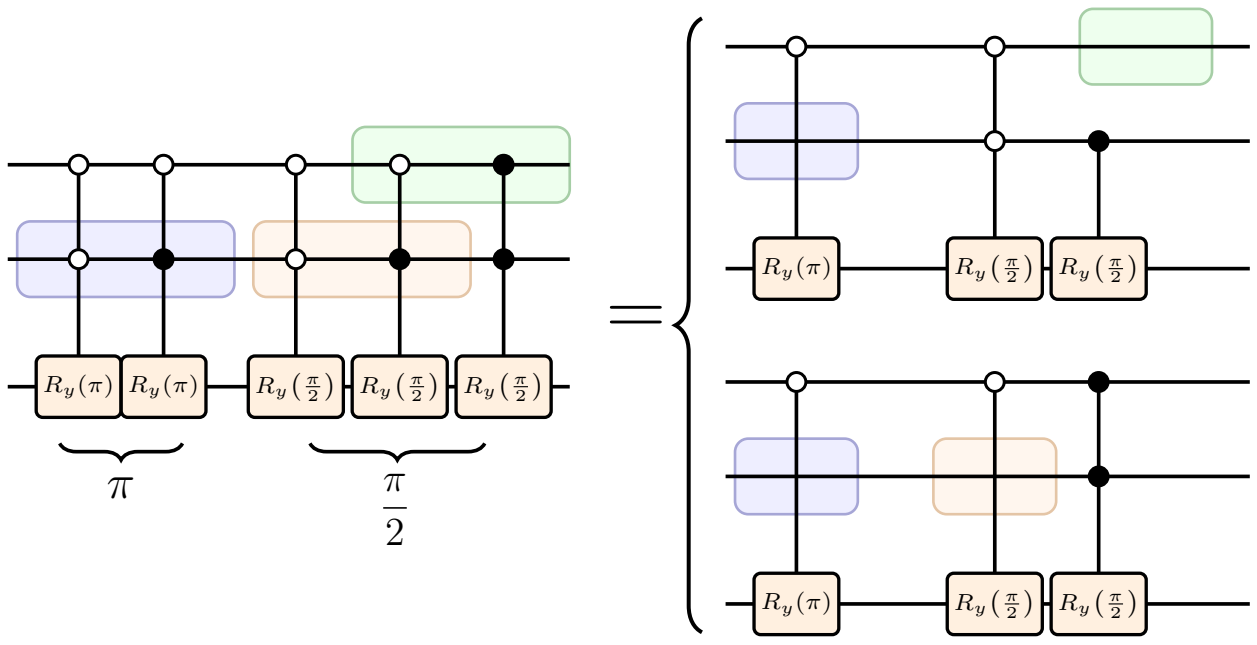
\begin{figure}
\centering
\resizebox{\columnwidth}{!}{\usetikzlibrary{
    backgrounds,
    decorations.pathreplacing,
    positioning,
    calc
}

\begin{tikzpicture}[
    >=Latex,
    font=\small,
    wire/.style={line width=1.1pt, black},
    vline/.style={line width=1.1pt, black},
    openctrl/.style={
        circle,
        draw=black,
        fill=white,
        line width=0.9pt,
        minimum size=2.0mm,
        inner sep=0pt
    },
    closedctrl/.style={
        circle,
        draw=black,
        fill=black,
        line width=0.9pt,
        minimum size=2.0mm,
        inner sep=0pt
    },
    rygate/.style={
        draw=black,
        fill=orange!12,
        line width=1.0pt,
        rounded corners=2pt,
        minimum width=0pt,
        minimum height=6.5mm,
        inner xsep=2pt,
        inner ysep=1.5pt,
        font=\scriptsize
    },
    bluebox/.style={
        draw=blue!55!black,
        fill=blue!18,
        opacity=0.35,
        rounded corners=4pt,
        line width=0.8pt
    },
    orangebox/.style={
        draw=orange!70!black,
        fill=orange!18,
        opacity=0.35,
        rounded corners=4pt,
        line width=0.8pt
    },
    greenbox/.style={
        draw=green!45!black,
        fill=green!18,
        opacity=0.35,
        rounded corners=4pt,
        line width=0.8pt
    },
    brace/.style={
        decorate,
        decoration={
            brace,
            mirror,
            amplitude=5pt
        },
        line width=1.1pt
    },
    casebrace/.style={
        decorate,
        decoration={
            brace,
            mirror,
            amplitude=8pt
        },
        line width=1.2pt
    }
]


\def\ytopL{2.2}
\def\ymidL{1.2}
\def\ybotL{-0.15}

\def\ytopRU{3.45}
\def\ymidRU{2.45}
\def\ybotRU{1.10}

\def\ytopRL{-0.10}
\def\ymidRL{-1.10}
\def\ybotRL{-2.45}


\begin{scope}[on background layer]
    \node[bluebox, minimum width=23mm, minimum height=8mm]
        at (0.75,\ymidL) {};
    \node[orangebox, minimum width=23mm, minimum height=8mm]
        at (3.25,\ymidL) {};
    \node[greenbox, minimum width=23mm, minimum height=8mm]
        at (4.30,\ytopL) {};
\end{scope}

\draw[wire] (-0.5,\ytopL) -- (5.45,\ytopL);
\draw[wire] (-0.5,\ymidL) -- (5.45,\ymidL);
\draw[wire] (-0.5,\ybotL) -- (5.45,\ybotL);

\coordinate (LxA) at (0.25,0);
\coordinate (LxB) at (1.15,0);
\coordinate (LxC) at (2.55,0);
\coordinate (LxD) at (3.65,0);
\coordinate (LxE) at (4.75,0);

\draw[vline] (0.25,\ytopL) -- (0.25,\ybotL);
\node[openctrl] at (0.25,\ytopL) {};
\node[openctrl] at (0.25,\ymidL) {};
\node[rygate] at (0.25,\ybotL) {$R_y(\pi)$};

\draw[vline] (1.15,\ytopL) -- (1.15,\ybotL);
\node[openctrl] at (1.15,\ytopL) {};
\node[closedctrl] at (1.15,\ymidL) {};
\node[rygate] at (1.15,\ybotL) {$R_y(\pi)$};

\draw[vline] (2.55,\ytopL) -- (2.55,\ybotL);
\node[openctrl] at (2.55,\ytopL) {};
\node[openctrl] at (2.55,\ymidL) {};
\node[rygate] at (2.55,\ybotL) {$R_y\!\left(\frac{\pi}{2}\right)$};

\draw[vline] (3.65,\ytopL) -- (3.65,\ybotL);
\node[openctrl] at (3.65,\ytopL) {};
\node[closedctrl] at (3.65,\ymidL) {};
\node[rygate] at (3.65,\ybotL) {$R_y\!\left(\frac{\pi}{2}\right)$};

\draw[vline] (4.75,\ytopL) -- (4.75,\ybotL);
\node[closedctrl] at (4.75,\ytopL) {};
\node[closedctrl] at (4.75,\ymidL) {};
\node[rygate] at (4.75,\ybotL) {$R_y\!\left(\frac{\pi}{2}\right)$};

\draw[brace] (0.05,-0.75) -- (1.35,-0.75);
\node[font=\Large] at (0.70,-1.22) {$\pi$};

\draw[brace] (2.70,-0.75) -- (4.55,-0.75);
\node[font=\large] at (3.60,-1.45) {$\dfrac{\pi}{2}$};

\node[font=\Huge] at (6.15,0.55) {$=$};

\draw[casebrace] (6.85,3.75) -- (6.85,-2.75);


\begin{scope}[on background layer]
    \node[bluebox, minimum width=13mm, minimum height=8mm]
        at (7.85,\ymidRU) {};
    \node[greenbox, minimum width=14mm, minimum height=8mm]
        at (11.55,\ytopRU) {};
\end{scope}

\draw[wire] (7.1,\ytopRU) -- (12.65,\ytopRU);
\draw[wire] (7.1,\ymidRU) -- (12.65,\ymidRU);
\draw[wire] (7.1,\ybotRU) -- (12.65,\ybotRU);

\draw[vline] (7.85,\ytopRU) -- (7.85,\ybotRU);
\node[openctrl] at (7.85,\ytopRU) {};
\node[rygate] at (7.85,\ybotRU) {$R_y(\pi)$};

\draw[vline] (9.95,\ytopRU) -- (9.95,\ybotRU);
\node[openctrl] at (9.95,\ytopRU) {};
\node[openctrl] at (9.95,\ymidRU) {};
\node[rygate] at (9.95,\ybotRU) {$R_y\!\left(\frac{\pi}{2}\right)$};

\draw[vline] (11.05,\ymidRU) -- (11.05,\ybotRU);
\node[closedctrl] at (11.05,\ymidRU) {};
\node[rygate] at (11.05,\ybotRU) {$R_y\!\left(\frac{\pi}{2}\right)$};


\begin{scope}[on background layer]
    \node[bluebox, minimum width=13mm, minimum height=8mm]
        at (7.85,\ymidRL) {};
    \node[orangebox, minimum width=13mm, minimum height=8mm]
        at (9.95,\ymidRL) {};
\end{scope}

\draw[wire] (7.1,\ytopRL) -- (12.65,\ytopRL);
\draw[wire] (7.1,\ymidRL) -- (12.65,\ymidRL);
\draw[wire] (7.1,\ybotRL) -- (12.65,\ybotRL);

\draw[vline] (7.85,\ytopRL) -- (7.85,\ybotRL);
\node[openctrl] at (7.85,\ytopRL) {};
\node[rygate] at (7.85,\ybotRL) {$R_y(\pi)$};

\draw[vline] (9.95,\ytopRL) -- (9.95,\ybotRL);
\node[openctrl] at (9.95,\ytopRL) {};
\node[rygate] at (9.95,\ybotRL) {$R_y\!\left(\frac{\pi}{2}\right)$};

\draw[vline] (11.05,\ytopRL) -- (11.05,\ybotRL);
\node[closedctrl] at (11.05,\ytopRL) {};
\node[closedctrl] at (11.05,\ymidRL) {};
\node[rygate] at (11.05,\ybotRL) {$R_y\!\left(\frac{\pi}{2}\right)$};

\end{tikzpicture}}
\caption{Boolean simplification of controlled rotations with identical angles. Fully controlled rotations are grouped into disjoint subcubes and replaced by fewer rotations with fewer controls. The complete correspondence between generalized control strings, Boolean cubes, projectors, and controlled gates is given in \cref{sec:properties_of_the_multiplexer}.}
\label{fig:boolean_simplification_colored}
\end{figure}

\begin{figure}
\centering
\input{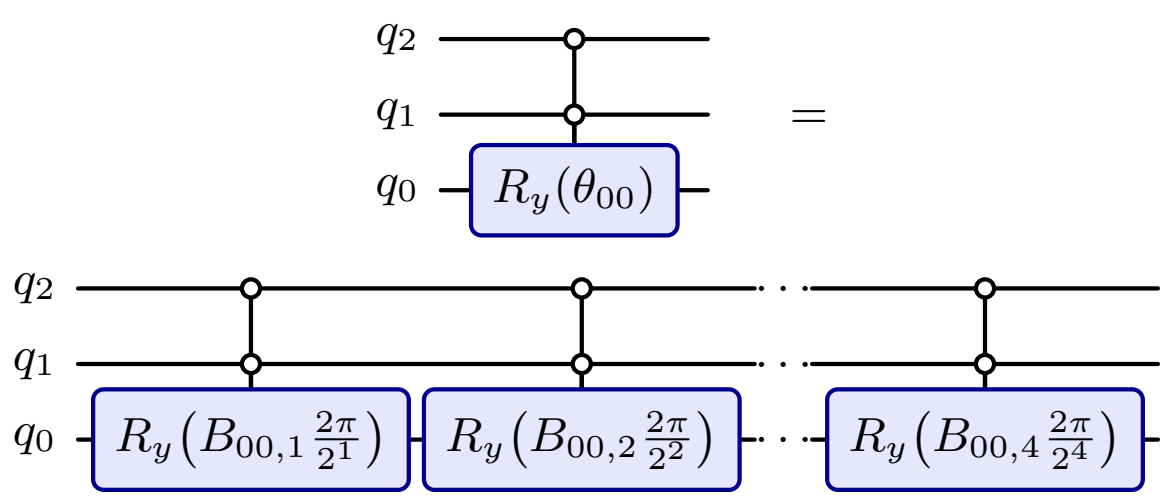}
\caption{Example of the decomposition of a $\text{MCR}_y$ gate at layer $l=2$ in terms of a product of $L$ $\text{MCR}_y$ gates with fixed angles. Here the basis are the negative integer powers of two times $2\pi$.}
\label{fig:binary_decomposition_of_angles}
\end{figure}

\begin{figure*}
    \centering
    \includegraphics[width=\textwidth]{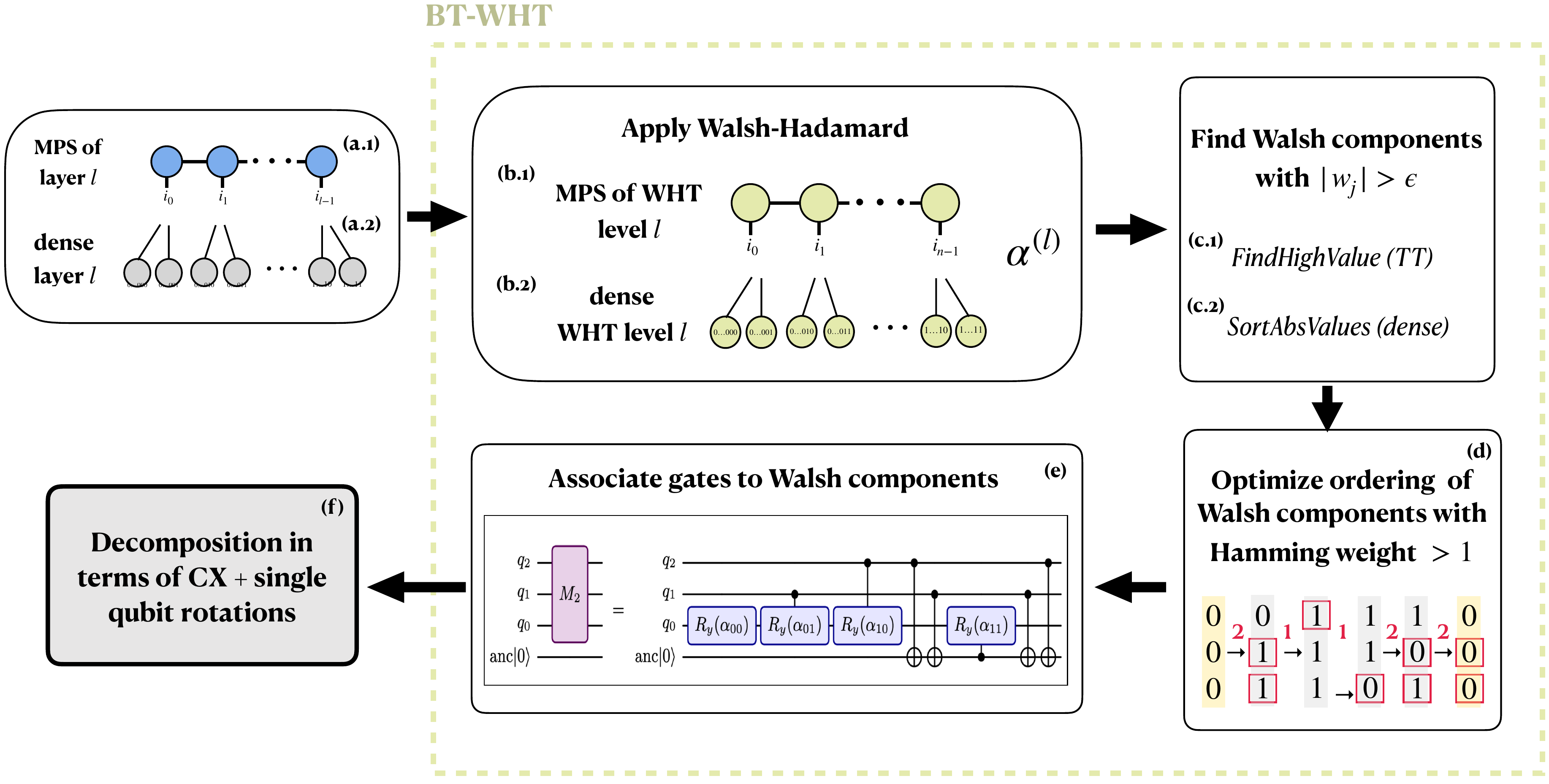}
    \caption{Scheme of the BT-WHT algorithm with or without \acfp{TT}. The input is either the \acf{MPS} of the angle vector of level $l$, see (a.1), or its dense version, see (a.2). The first step is to apply the Walsh-Hadamard transform and compute the \ac{MPS} of the Walsh spectrum (b.1) or its dense version (b.2). Then the components of the spectrum with absolute value greater than the threshold $\epsilon$ are selected, while the rest is set to zero. In the \ac{TT} case this step is done via the bisection algorithm \cref{alg:findHighValue} (c.1), while in the dense case we simply sort the absolute values of the array (c.2). After truncation, the selected components are divided into groups based on the Hamming weight of their binary address. Those with a Hamming weight that is greater or equal to $2$ are then sorted in order to minimize the length of the Hamming path starting and ending with the zero vector and visiting each binary address (see \cref{app:BT-WHT} for more details) (d). Each component in the given order is assigned to a specific sequence of single and two- qubit gates (e) (see \cref{tab:j_cases}), giving the final decomposition (f).}
    \label{fig:bt_wht_overview}
\end{figure*}

In our application individual entries of the angle tensors can be evaluated directly using \cref{eq:binarytreepolarangles,eq: phase angle}. When $\ket{v}$ is itself represented as a \ac{TT}, \cref{eq:binarytreepolarangles} is simply the ratio of inner products over projectors, while \cref{eq:zeta} can itself be constructed via \ac{TCI}, and subsequently \cref{eq: phase angle} evaluated by contracting the $\vc{j}[l]$ sites with their appropriate $\ket{0},\ \ket{1}$ states, site $l-1$ with the state $(\ket{1}-\ket{0})$, and the remaining states with vectors $(\ket{0}+\ket{1})$ over the site indices. Consequently, \ac{TCI} can construct a compressed representation of an angle layer without explicitly forming all $2^l$ entries. We can further keep track of the binary-tree nodes which have been truncated at previous levels when choosing \ac{TCI} sample pivots, reducing the number of samples required. The efficiency of this construction depends on the angle tensor admitting an accurate low-bond-dimension representation. In the absence of such structure, the required number of samples and the bond dimensions may still become large.

The \ac{TT} representation is also particularly convenient for the \ac{BT-WHT} method. The multidimensional \ac{WHT} is a product of local two-dimensional Hadamard transforms and can therefore be applied independently to the physical index of each \ac{TT} core. For $H$, the normalized $2\times2$ Hadamard matrix, the transformed cores are
\begin{equation}
\widetilde{A}^{(q)}_{\alpha_{q-1},s_q,\alpha_q}=\sum_{i_q=0}^{1}
H_{s_qi_q}
A^{(q)}_{\alpha_{q-1},i_q,\alpha_q}.
\end{equation}
A representation of the Walsh coefficients of an angle layer can therefore be obtained without reconstructing the corresponding dense vector or increasing the bond-dimension of the \ac{TT}. There is nothing special about this particular tensor-network, except that it provides simple and computationally efficient contraction. Extensions to other tensor-networks such as tensor trees which are known to have better bond-dimension scaling for different types of entangled states are also possible, and similarly admit \ac{TCI} for calculation of angles \cite{Tindall2026}. 

\section{Algorithms}
\label{sec:algorithms}

As shown in \cref{fig:multiplexer}, a naive implementation for a given level $l$ would require $2^l$ gates corresponding to every angle $\vc{\theta}^{(l)}$, and $\vc{\phi}^{(l)}$. 
In this section we introduce two complementary methods, 
namely the \ac{BT-BE} and the \ac{BT-WHT} based decompositions,
that simplify a given multiplexer inside the state preparation circuit in \cref{fig:multiplexer_ladder}, reducing gate count. 
These approaches can also be combined in a hybrid strategy which we later show in \cref{sec:results} generally outperforms either one in isolation. 

In the first method we perform a binary decomposition of the angles of the multi-controlled rotations inside each multiplexer up to a given binary precision $L$.
We then exploit the commutativity of the \ac{MCR} gates inside each layer to rearrange and group them according to the fixed angle value and simplify their controls, 
hence reducing the problem to classical Boolean logic simplifcation \cite{brayton1984espresso, Bernasconi2013, ESOP_article}.  
The second approach leverages the Walsh Hadamard Transform (WHT) to decompose the multiplexer into one-qubit and single-qubit-controlled rotations via a single ancilla, allowing a reduction in gate count as long as the Walsh spectrum remains sufficiently sparse. 
\subsection{The Binary-Tree Boolean-Encoder Method}
\label{sec:BT-BE}

The \ac{BT-BE} method acts on a single angle layer of the binary-tree representation. Once the gate angles corresponding to the modulus $\vc{\theta}^{(l)}$ or phase $\vc{\phi}^{(l)}$ at level $l$ have been calculated the method reduces the corresponding multiplexer in four steps: finite-precision angle decomposition, Boolean simplification of the resulting fixed-angle multiplexers, merging of the \ac{MCR} gates with the same control and synthesis of the simplified \ac{MCR} gates into the selected elementary gate basis. \cref{fig:BT_BE_scheme} summarizes the procedure.

We first approximate each continuous angle using a finite basis of $L$ fixed angles. The most trivial decomposition is
\begin{equation}
\label{eq:btbe-binary-angle-decomposition}
\theta_{\vc{j}[l]}
\approx
\sum_{i=1}^{L}
B_{\vc{j}[l],i}a_i,
\qquad
a_i=\frac{2\pi}{2^i},
\qquad
B_{\vc{j}[l],i}\in\{0,1\}.
\end{equation}

However, we are free to optimize the choice of basis in order to minimize the approximation error and the final circuit depth as explained in \cref{sec:BT-BE quadrature}. The basis' coefficients form a binary matrix
\begin{equation}
\mathbf B^{(l)}
\in
\mathbb F_2^{2^l\times L}.
\label{level_binary_matrix_BT_BE}
\end{equation}

The binary integer expansion of the indices corresponding to the non-zero entries in the $i$-th column, $\vc{b}_i^{(l)}\in\mathbb{F}_2^{2^l}$, identifies the control strings to which the fixed rotation $R(a_i)$ must be applied, where $R$ is $R_y$ for the modulus or $R_z$ for the phase. The approximation in \cref{eq:btbe-binary-angle-decomposition} has a per-angle truncation error of order $O\left(\pi/2^L\right)$. Because rotations about the same axis commute and the angles add upon operator multiplication, the original multiplexer can be approximated as

\begin{equation}
\label{eq:btbe-fixed-angle-product}
\mathcal M_l
\bigl(\vc{\theta}^{(l)}\bigr)
=
\prod_{i=1}^{L}
\mathcal M_l
\bigl(a_i\vc{b}_i^{(l)}\bigr) + O\left(\pi/2^L\right).
\end{equation}
The error remains upper bounded by the binary approximation due to the MCR gates acting on mutually exclusive control subspaces. Thus, BT-BE replaces one multiplexer with $2^l$ arbitrary continuous angles by at most $L$ multiplexers whose nonzero rotations have fixed angles, and thus can be merged into a smaller total number of operations. This simplification is based on the fundamental substitution rule that merges any pair of $\text{MCR}$ gates whose \textit{generalized} control strings (those which include identity $\I$) differ in only one position with a single $\text{MCR}$ that does not control (has a $\I$) in that position. An example of the application of this rule is given in \cref{fig:boolean_simplification_colored}. The order of this simplification matters and can lead to significantly different final circuit depths, hence the necessity to map this onto a Boolean logic simplification problem. We do so via interpreting generalized control strings as sub-graphs of a binary Hamming graph. Here we focus on the results necessary to understand our implementation, for details see \cref{sec:appendix geometric interp}.

For a given column $\vc{b}_i^{(l)}$, the trivial implementation assigns one fully controlled rotation to every nonzero entry. BT-BE reduces this number by combining compatible control strings such that the same operation is applied with fewer, partially controlled rotations. We represent a partially controlled rotation by a generalized control string $\vc{c}\in\{0,1,\I\}^l$, where $\I$ denotes an uncontrolled qubit. The generalized string identifies a set of binary strings:
\begin{equation}
C_{\vc{c}}
\subseteq
\{0,1\}^l
\end{equation}
containing all binary control strings which are covered by $\vc{c}$. For example, $\vc{c}=\I0\I$ represents the set of indices $C_{\vc{c}}=\{000,001,100,101\}$ and therefore replaces four fully controlled rotations by a single rotation controlled only on the middle qubit. 

Exact simplification is ensured by finding a set $\mathcal G_j$ of generalized control strings, which we call a decomposition. The collection of binary indicator vectors associated to its generalized control strings form a disjoint partition of the support
\begin{equation}
\label{eq:btbe-main-dsop}
\mathcal S_i
:=
\Big\{
\vc{j}[l]\in\{0,1\}^l\ \big|\ b^{(l)}_{\vc{j}[l],i}=1
\Big\}
=
\bigsqcup_{\vc{c}\in\mathcal G_i} C_{\vc{c}}.
\end{equation}
\Cref{eq:btbe-main-dsop} is a \ac{DSOP} decomposition of the Boolean function represented by $\vc{b}_i^{(l)}$. Each term $\vc{c}\in\mathcal G_i$ is associated to a single multi-controlled rotation $C^{m(\vc{c})}R(a_i)$, where
\begin{equation}
\label{eq:amount_of_controls}
m(\vc{c})
:=
l- \#_{\I}(\vc{c}),
\end{equation}
is the number of remaining controls calculated as $l$ minus the amount of $\I $ symbols in $\vc{c}$. Consequently, minimizing the number of generalized control strings reduces the number of rotations. A related case is when the multi-controlled operation is self-inverse, for example the multi-controlled X (MCX) gate. Here the problem becomes an Exclusive Sum Of Products (\ac{ESOP}) decomposition. A theorem linking the simplification of the multiplexer to these two Boolean logic problems is derived in \cref{sec:properties_of_the_multiplexer}. For the fixed-angle rotations considered in BT-BE, we use the disjoint formulation in \cref{eq:btbe-main-dsop}.

Finding a minimum \ac{DSOP} decomposition is \ac{NP}-hard~\cite{gimpel1965reduction,Bernasconi2013}. We therefore employ a heuristic synthesis method. Based on the benchmark comparison discussed in \cref{app:multiplexer_simplification_DSOP}, we use the method of Ref.~\cite{Balasubramanian} as the default \ac{DSOP} solver. The optimization is applied independently to every column of $\mathbf B^{(l)}$. 
Afterwards, rotations associated with the same generalized control string are collected and their angles are added.

The resulting multi-controlled rotations can either be retained as native operations or decomposed into a universal gate basis. In the resource estimates reported below, we decompose each $C^mR_y$ and $C^mR_z$ operation into two $C^mX$ gates and single-qubit rotations, and synthesize the multi-controlled $X$ gates using the construction of Khattar and Gidney~\cite{KhattarGidney2025ConditionallyCleanAncillae}. 
The explicit decomposition and the resulting CX-count and depth bounds are given in \cref{app:mcr-synthesis}. 

\subsection{The Binary-Tree Walsh-Hadamard-Transform Method}
\label{sec:BT-WHT}
\begin{figure*}
\centering

\begin{subfigure}[b]{\textwidth}
\centering
\input{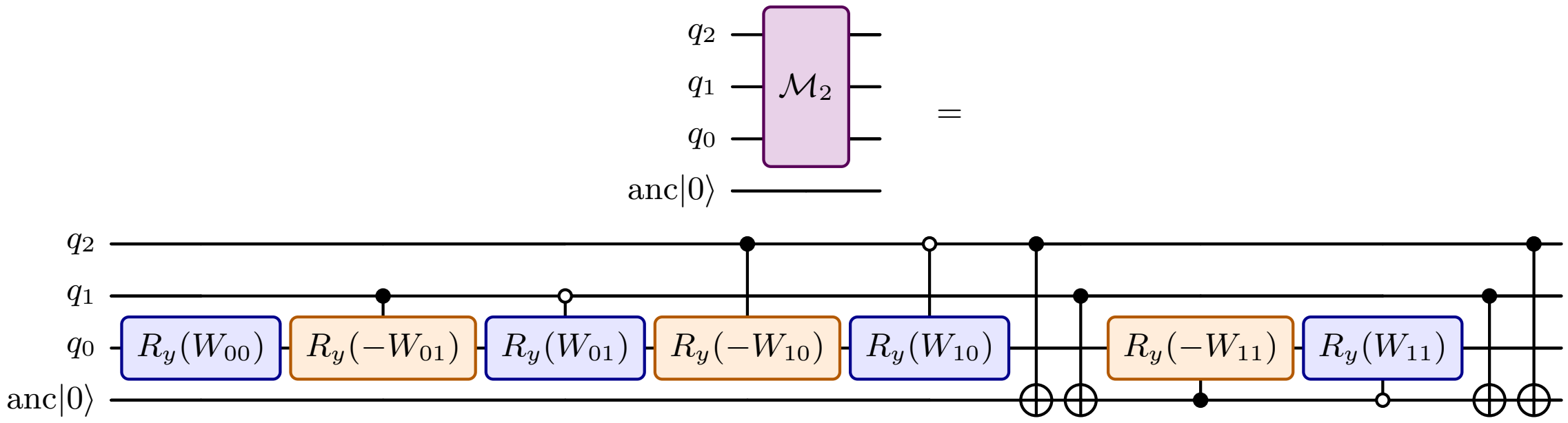}
\caption{Unsimplified decomposition of the $l=2$ multiplexer into
single-qubit rotations and ancilla-assisted controlled operations.}
\label{fig:walsh_decomposition_k2_full}
\end{subfigure}

\medskip

\begin{subfigure}[b]{\textwidth}
\centering
\input{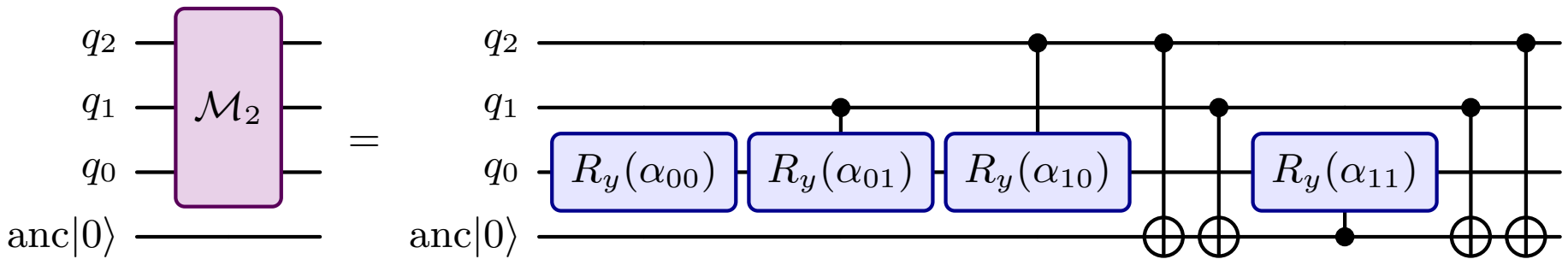}
\caption{Simplified decomposition obtained by absorbing the
zero-controlled rotations into the initial single-qubit rotation and
doubling the angles of the remaining one-controlled rotations.}
\label{fig:walsh_decomposition_k2_simplified}
\end{subfigure}

\caption{Walsh decomposition of the $l=2$ multiplexer. In
\cref{fig:walsh_decomposition_k2_full}, each nonzero Walsh component
produces zero- and one-controlled rotations with opposite angles. The
weight-two component is additionally preceded and followed by CX
ladders that compute and uncompute its parity on the ancilla. In
\cref{fig:walsh_decomposition_k2_simplified}, the zero-controlled
rotations have been absorbed into the initial single-qubit rotation.}
\label{fig:walsh_decomposition_k2}
\end{figure*}
An alternative strategy to simplify the multiplexer ladder circuit in \cref{fig:multiplexer_ladder} makes use of the \ac{WHT} of the angles in each layer of the binary-tree graph. \cref{fig:bt_wht_overview} gives an overview of this approach. Each component of the Walsh spectrum can be encoded with a single- or two-qubit gate, giving an exact decomposition of each multiplexer as a product of commuting $R_y$ and $CR_y$ operations for the moduli and $R_z$ and $CR_z$ for the phases. 
When most of the spectral weight is accounted for by a small number of Walsh coefficients, the remaining coefficients can be truncated in order to reduce the total depth.

The Walsh spectrum is computed as
\begin{equation}
\label{eq:main-WHT}
\vc{w}^{(l)}
=
H^{\otimes l}\boldsymbol{\theta}^{(l)},
\end{equation}
where $H^{\otimes l}$ is the $2^l\times2^l$ normalized Hadamard matrix. 
Its entries are
\begin{equation}
H^{\otimes l}_{\vc{i},\vc{j}}
=
\frac{(-1)^{\langle\vc{i},\vc{j}\rangle_2}}{\sqrt{2^l}},
\end{equation}
where, $\langle\vc{i}, \vc{j}\rangle_2=\sum_k \vc{i}_k\vc{j}_k\ \text{mod}\ 2$, denotes the binary inner product. The inverse transform gives
\begin{equation}
\begin{split}
\theta_{\vc{i}}^{(l)}&=\sum_{\vc{j}\in\{0,1\}^l}(-1)^{\langle\vc{i},\vc{j}\rangle_2}\frac{w_{\vc{j}}^{(l)}}{\sqrt{2^l}}\\
&=\sum_{\vc{j}\in\{0,1\}^l}
(-1)^{\langle\vc{i},\vc{j}\rangle_2}W_{\vc{j}}^{(l)},
\end{split}
\end{equation}
implicitly defining the normalized Walsh coefficient $W_{\vc{j}}^{(l)}$. We introduce a threshold $\mathcal{E}_{\mathrm{WHT}}>0$, we sort the Walsh coefficient according to a decreasing order of  their absolute value and we discard those whose index $ j$ is greater than $j_\mathrm{\tau}$ defined by the equation
\begin{equation}
\frac{\sum_{j> j_\mathrm{\tau}}
\left|w_{j}^{(l)}\right|^2}{\sum_{j}
\left|w_{j}^{(l)}\right|^2}\leq \mathcal{E}_{\mathrm{WHT}}^2.
\end{equation} 
Since $H^{\otimes l}$ is orthonormal, the resulting
Euclidean error in the angle vector is equal to the norm of the
discarded Walsh coefficients,

\begin{equation}\label{eq:walsh-truncation-error}
\left\|\vc\theta^{(l)}-\widetilde{\vc\theta}^{(l)}\right\|_2^2
=
\sum_{j>j_\mathrm{\tau}}
\left|w_{j}^{(l)}\right|^2.
\end{equation}
This implies that the relative error over the angle layer is the same at every level, namely
\begin{equation}
\label{eq:relative_error_def}
\frac{\left\|\vc\theta^{(l)}-\widetilde{\vc\theta}^{(l)}\right\|_2
}{\left\|\vc\theta^{(l)}\right\|_2}\leq \mathcal{E}_{\mathrm{WHT}},\ \forall \ l \in [n].
\end{equation} 

The resulting error on the encoded vector from this discarded spectral weight is given for complex data in a pointwise manner in \cref{eq:upperbound_pointwise_amplitude_abs_error} while the infidelity for real data is given by
\begin{equation}
\label{eq:relative_error_upperbound_main}
    \mathcal{I}\leq \mathcal{E}_{\mathrm{WHT}}^2\sum_{l=0}^{n-1}\left\|\vc \theta^{(l)}\right\|^2_2 (2l+1)+o\left(\mathcal{E}^2_{\mathrm{WHT}}\right),
\end{equation}
for the strict upper bound $o\left(\mathcal{E}^2_{\mathrm{WHT}}\right)$, which is derived in detail in \cref{app:minimization_of_BTWHT_error}.

Despite \cref{eq:relative_error_upperbound_main} being a loose upper bound for all of the functions in our analysis it is efficiently computable even within the TN formalism, where the norms of the angle layers can be obtained from their TT representations in $O(n\chi ^3)$.
For layers that can be constructed densely, the Walsh spectrum is
computed directly from \cref{eq:main-WHT} via the Fast WHT \cite{Fino1976}. 

\begin{algorithm}[H]
\caption{Wrapper of the High Value Finder}
\label{alg:findHighValueWrapper}
\begin{algorithmic}[1]

\REQUIRE
 MPS $\mathbf m $ of $l\in [n]$  sites, WHT threshold $\epsilon\in [0,1]$
\STATE $\text{values} \gets\left[\ \right]$
\WHILE{$|\text{values}[\text{end}]| > \epsilon$}
    \STATE $\text{val},\text{position} \gets \text{findHighValue}(\mathbf m, \{\})$\\
   \STATE $\mathbf m\ -= \text{Proj}(\text{position})\mathbf m$\\
   \STATE $\text{values}=\text{append}(\text{values},\text{val})$
\ENDWHILE

\RETURN $\text{values}$ 

\end{algorithmic}
\end{algorithm}
\begin{algorithm}[H]
\caption{findHighValue}
\label{alg:findHighValue}
\begin{algorithmic}[1]

\REQUIRE
MPS $\mathbf m$ of length $f\in [n]$, binary address $\mathbf b$
\IF {f=0}
\RETURN $\mathbf m, \vc b$\null\quad where $\mathbf{m}$ is a scalar
\ENDIF
\STATE $\mathbf m_0 \gets \text{fixMPSSiteToValue}(\mathbf m ,f,0)$
\STATE $\vc m_1\gets \text{fixMPSSiteToValue}(\mathbf m ,f,1)$
\STATE $v_0 = \|\vc m_0\|_2$
\STATE $v_1 = \|\vc m_1\|_2$
\STATE $\vc b_0 = \text{append}(\vc b,0)$
\STATE $\vc b_1 = \text{append}(\vc b,1)$

\STATE $p_0=\frac{v_0^2}{v_0^2+v_1^2}$
\STATE $p_1=\frac{v_1^2}{v_0^2+v_1^2}$
\STATE $r\gets\text{ rand}([0,1])$
\IF {$r \leq p_0$}
    \RETURN $\text{findHighValue}(\mathbf{m}_0,\vc b_0)$
\ELSE 
   \RETURN  $\text{findHighValue}(\mathbf{m}_1,\vc b_1)$
\ENDIF

\end{algorithmic}
\end{algorithm}

Given that we truncate many branches of the tree, at a given level we generally do not have a complete angle vector of size $2^l$. For small $l$, where we can still work with dense data, we construct a heuristic interpolation of the missing values, \cref{alg:walsh-interpolation}, in such a way that the Walsh spectrum remains sparse. This is described in detail in \cref{app:BT-WHT}. At deeper levels, the angle
vectors are instead sampled using \ac{TCI} and stored in \ac{TT} form. We choose the sample pivots based on the surviving binary-tree paths. To find the high magnitude coefficients from the TT, we utilize the recursive algorithm \cref{alg:findHighValue} inside its wrapper \cref{alg:findHighValueWrapper}. This algorithm recursively bisects the TT representation of the layer by sequentially fixing the value of one of its sites either to $0$ or $1$. It then chooses to inspect branch zero or branch one with probabilities
\begin{equation}
    p_j=\frac{v_j^2}{v_0^2+v_1^2}\quad j \in \{0,1\},
\end{equation}
where $v_j$ is the norm of branch $j \in \{0,1\}$. When the algorithm reaches the current maximum angle it outputs its value and binary address. The wrapper function collects these values one by one and each time removes them from the TT via projectors associated to their address. The process is repeated until we find a value that is lower in modulus than a given WHT truncation threshold.

\begin{figure*}
\centering
\input{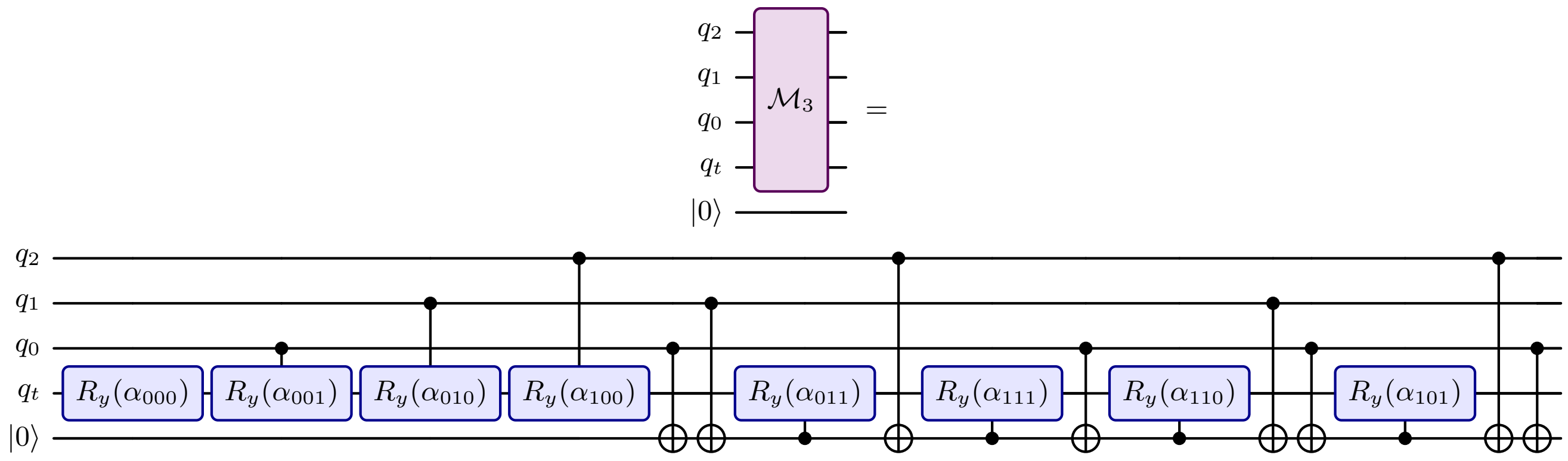}
\caption{Walsh-decomposed implementation of an $l=3$ multiplexer with
three controls, one target qubit, and one clean ancilla. The
weight-zero and weight-one Walsh components are implemented directly.
The higher-weight components are ordered along the cycle
$000\to011\to111\to110\to101\to000$, which requires eight CX gates and
returns the ancilla to $\ket{0}$.}
\label{fig:walsh_decomposition_k3}
\end{figure*}

As a first step in the decomposition, which we refer to as the
unsimplified version, each Walsh spectrum coefficient for a given level $l$, $W_{\vc{j}}^{(l)}$, is associated either
with a single-qubit rotation, a pair of singly controlled rotations, or a sequence consisting of a CX ladder, a pair of singly controlled
rotations, and a second CX ladder. The appropriate case is determined
by the Hamming weight $|\vc{j}|_{\mathrm H}$ and by the positions of the ones in the binary index $\vc{j}$, as summarized in
\cref{tab:j_cases}, and diagrammed in \cref{fig:walsh_decomposition_k2_full}.

The first component $W_{\vc{0}}^{(l)}$ with Hamming weight
zero corresponds to a single-qubit $R_y$ rotation of angle
$W_{\vc{0}}^{(l)}$ acting on the multiplexer target.
Every coefficient
$W_{\vc{j}}^{(l)}$ whose index has Hamming weight one corresponds to
two singly controlled rotations, one zero-controlled and one
one-controlled, with respective angles $W_{\vc{j}}^{(l)}$ and
$-W_{\vc{j}}^{(l)}$. The control is the unique qubit corresponding to a nonzero entry of $\vc{j}$.

Every coefficient whose index has Hamming weight greater than one is
associated with the same pair of opposite controlled rotations.
However, these rotations use the ancilla as their control and are
preceded and followed by CX ladders that compute and uncompute the
binary inner product on the ancilla. For a Walsh index $\vc{j}$, the
corresponding ladder is
\begin{equation}
\label{eq:main-Walsh-ladder}
\mathcal{L}_{\vc{j}}^{(l)}
=
\prod_{q=0}^{l-1}
\mathrm{CX}(q,a)^{j_q},
\end{equation}
where $a$ denotes the clean ancilla. Thus, the ladder contains one CX
gate for every nonzero entry of $\vc{j}$. 
\begin{table}
\centering
\caption{Operations associated with a Walsh coefficient
$W_{\vc{j}}^{(l)}$ before simplification.}
\label{tab:j_cases}
\begin{tabular}{|c|l|}
\hline
\textbf{$|\vc{j}|_{\mathrm H}$} &
\textbf{Associated operations}\\
\hline
$0$ &
One $R_y$ or $R_z$ rotation\\
\hline
$1$ &
Two $CR_y$ or $CR_z$ rotations\\
\hline
$\geq2$ &
$2|\vc{j}|_{\mathrm H}$ CX gates\\
&
Two ancilla-controlled $R_y$ or $R_z$ rotations\\
\hline
\end{tabular}
\end{table}

\begin{table*}[!t]
\centering%
\caption{
Function families used in the benchmark, their default parameters, and common figure legend.
}%
\label{tab:results-function-families}%
\begin{tabular}{llllc}%
\hline
Family
&
Parameters
&
Default Parameters
&
Legend
\\
\hline
Gaussian
&
Center $\mu$, width $\sigma$
&
$\mu=0,\ \sigma=1$
&
\adjustbox{valign=c}{\includegraphics[height=2.5ex]{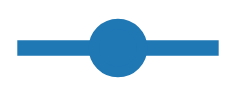}}
\\
Lorentzian
&
Width $\gamma$, center $x_0$
&
$\gamma=0.1,\ x_0=0$
&
\adjustbox{valign=c}{\includegraphics[height=2.5ex]{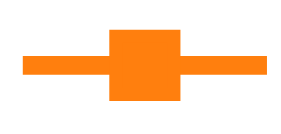}}
\\
Sinusoidal
&
Frequency parameter $k$
&
$k=10$
&
\adjustbox{valign=c}{\includegraphics[height=2.5ex]{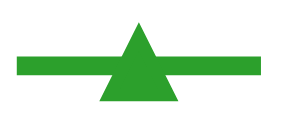}}
\\
Chebyshev polynomial
&
Degree $m$
&
$m=2$
&
\adjustbox{valign=c}{\includegraphics[height=2.5ex]{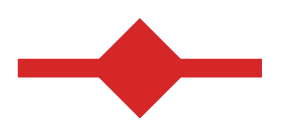}}
\\
Hermite Polynomial
&
degree $\ell$, width $\alpha$, center $x_0$
&
$\ell=16,\ \alpha=0.1,\ x_0=0$
&
\adjustbox{valign=c}{\includegraphics[height=2.5ex]{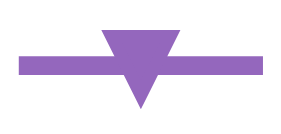}}
\\
Polynomial
&
Order $m$
&
$m=2$
&
\adjustbox{valign=c}{\includegraphics[height=2.5ex]{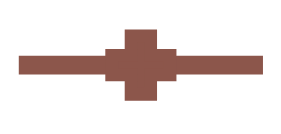}}
\\
RSRA
&
Sparsity $|\mathcal{S}|$, Distribution
&
$|\mathcal{S}|=32$, Normal
&
\adjustbox{valign=c}{\includegraphics[height=2.5ex]{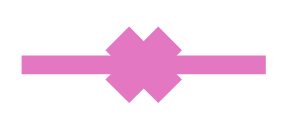}}
\\
RSUA
&
Sparsity $|\mathcal{S}|$, Distribution
&
$|\mathcal{S}|=32$, Uniform
&
\adjustbox{valign=c}{\includegraphics[height=2.5ex]{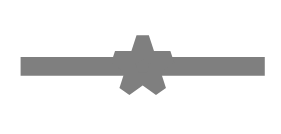}}
\\
Merton
&
$\lambda$, $\mu_{\mathrm{jump}}$, $\sigma_{\mathrm{jump}}$
&
$\lambda=0.25,\ \mu_{\mathrm{jump}}=-0.1,\ \sigma_{\mathrm{jump}}=0$
&
\adjustbox{valign=c}{\includegraphics[height=2.5ex]{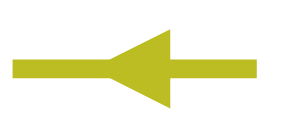}}
\\
Correlated Merton
&
$\lambda$, $\rho_{\mathrm{diff}}$, $\rho_{\mathrm{jump}}$
&
$\lambda=0.5,\ \rho_{\mathrm{diff}}=0.3,\ \rho_{\mathrm{jump}}=0$
&
\adjustbox{valign=c}{\includegraphics[height=2.5ex]{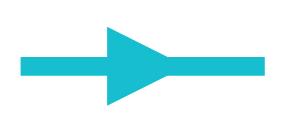}}
\\
Weierstrass type
&
$M$, $a$, $b$
&
$M=30,\ a=0.5,\ b=3$
&
\adjustbox{valign=c}{\includegraphics[height=2.5ex]{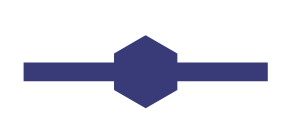}}
\\
\hline
\end{tabular}%
\end{table*}%

It turns out that the circuit depth in \cref{fig:walsh_decomposition_k2_full} can be cut almost in half through an effective `DC offset', namely adding all the positive angles associated with zero-controlled rotations. This gives a new argument of the initial single-qubit rotation:
\begin{equation}
\label{eq:main-alpha-zero}
\alpha_{\vc{0}}^{(l)}
=
\sum_{\vc{j}\in\{0,1\}^l}
W_{\vc{j}}^{(l)}.
\end{equation}
The angles of the remaining one-controlled rotations are doubled and
change sign,
\begin{equation}
\alpha_{\vc{j}}
=
-2W_{\vc{j}},
\qquad
\vc{j}\neq\vc{0}.
\end{equation}
This gives the simplified circuit shown in
\cref{fig:walsh_decomposition_k2_simplified}. The complete derivation of this decomposition, including the
weight-zero, weight-one, and higher-weight cases, is provided in \cref{sec:Walsh-spectral-circuit-proof} using the general multiplexer identities derived in
\cref{sec:properties_of_the_multiplexer}.

The CX ladders associated with the higher-weight Walsh components can
be simplified further by finding a specific sequence of CX ladders to move between non-zero Walsh coefficient addresses. As explained in our previous work
\cite{Pagni2025}, the action of a CX ladder on the ancilla can be
represented by the corresponding binary Walsh address. A transition
between two addresses $\vc{j}$ and $\vc{k}$ requires CX gates only at the positions in which the two addresses differ, and therefore has cost
$|\vc{j}\oplus\vc{k}|_{\mathrm H}$.

Since the logical operations associated with the different Walsh
coefficients commute, they can be ordered so as to minimize the total
Hamming distance between consecutive addresses. The sequence begins
and ends at $\vc{0}$, ensuring that the ancilla is returned to
$\ket{0}$ for the next multiplexer level block.
For example, the $l=3$ circuit in
\cref{fig:walsh_decomposition_k3} orders its retained higher-weight
addresses along the cycle
\begin{equation}
000
\to
011
\to
111
\to
110
\to
101
\to
000,
\end{equation}
whose corresponding transitions require a total of eight CX gates. When every
Walsh component is present, a Gray-code ordering traverses the full set of binary controls~\cite{Mottonen}. In the
truncated case, however, only a subset of the Walsh addresses is
retained, so the problem becomes that of finding a short cycle through
the retained indices. To reduce the number of CNOT gates required for parity-network updates, we reorder the selected Walsh components using a greedy nearest-neighbor heuristic in Hamming space. Starting from an initial binary configuration, each subsequent component is chosen to minimize the Hamming distance to the previously selected configuration. The circuit is then constructed using the XOR differences between consecutive configurations, thereby avoiding redundant parity computations.

\subsection{The Binary Tree Hybrid approach}
The circuit depth of the \ac{BT-WHT} decomposition is strongly influenced by the sparsity of the Walsh spectra of the modulus and phase angle vectors at each tree level. Localized irregularities in these vectors may generate a large number of Walsh coefficients, increasing the cost of the corresponding decomposition. The BT-Hybrid method addresses this problem by identifying and modifying selected angles to improve the compressibility of their Walsh spectra. The resulting angle vector is synthesized using \ac{BT-WHT}, while the differences from the original angles are restored through multi-controlled rotations decomposed using \ac{BT-BE}.

To identify suitable interpolation points, we determine those points which most strongly induce non-sparsity in the spectrum. Specifically, we set to zero all Walsh coefficients whose absolute values exceed a prescribed threshold, retaining only the small-amplitude spectral tail, and apply the inverse Walsh–Hadamard transform. Up to normalization, this yields the gradient of the squared spectral-tail energy with respect to the original angle vector, provided the spectral mask is held fixed. Entries with large absolute values indicate angles whose modification can strongly affect this energy and therefore serve as candidates for interpolation.

If we apply a mask to those points and interpolate them using \cref{alg:walsh-interpolation}, we aim to sparsify the Walsh spectrum, thereby reducing the gate count for that level.
The local variations associated with the removed points may correspond to cusps or discontinuities in the otherwise smooth angle vector.
These local ``corrections" must be applied after the one- and two-qubit-gate decomposition based on the \ac{WHT} in each level. 
Due to the element-wise flooring operation, this procedure requires a dense realization of the data. This could in principle be worked around via searching out WHT coefficients below the threshold, and resolving the specific angle values by querying the TT, but we do not pursue that path in this work. At deep levels, we are free to switch back to exclusively BT-WHT, allowing the use of tensor-networks for high dimensional data.

\section{Results}
\label{sec:results}
In order to benchmark our methods, we compare the performance of \ac{BT-BE},
\ac{BT-WHT}, and the BT-Hybrid method both against each other and  with respect to the \texttt{StatePreparation} routine in Qiskit, which implements the multiplexer based algorithm from Iten et. al \cite{Iten_2016}. We adjust the transpilation of the resulting unitary to match our method's encoding infidelity for a fair comparison of depth to accuracy, see \cref{sec:encoding_results}. 
The comparison is performed over several families of
one- and two-dimensional functions. Each family is chosen to probe a
different structural feature of the input state, including smoothness,
localization, heavy tails, oscillatory behavior, sparsity, and structure over multiple length scales. The legend showing the colors and line style used for the representative of each family of functions is in \cref{tab:results-function-families}.

We assess how the gate depth of each method scales with the size of the data for a few representative parameter configurations. We show  the dependence of the total gate depth, the CX depth and state fidelity with varying precision $L$ for \ac{BT-BE} and hybrid methods as well as the impact of the \ac{WHT} threshold $\mathcal{E}_{\mathrm{WHT}}$ for \ac{BT-WHT}. Here we consider the depth, i.e. non-parallelizable circuit execution path, after using the KG24 method to decompose the simplified \ac{MCR} gates into the $\{\text{CX},R_y,R_z\}$ basis. 
Finally, the circuit is compiled with Qiskit~\cite{qiskit} at the highest optimization level for all-to-all connectivity.

\subsection{Benchmark protocol}
\label{sec:results-protocol}
\begin{figure*}
\centering
\begingroup
\setlength{\tabcolsep}{0pt}
\renewcommand{\arraystretch}{1.0}

\newcommand{\panelplot}[1]{%
    \includegraphics[
        width=0.49\textwidth,
        height=0.205\textheight,
        keepaspectratio
    ]{#1}%
}

\begin{tabular}{@{}c@{\hspace{0.01\textwidth}}c@{}}

\multicolumn{1}{c}{\textbf{\small Native Circuit Depth}}
&
\multicolumn{1}{c}{\textbf{\small State-Preparation Infidelity}}
\\[0.3em]

\panelplot{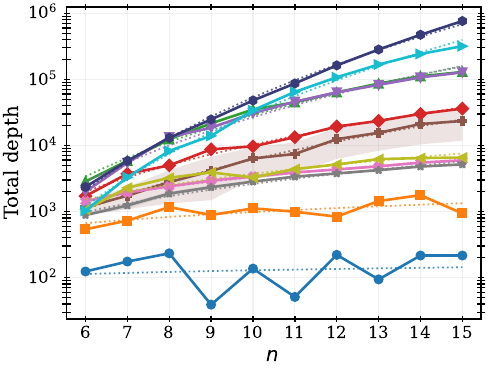}
&
\panelplot{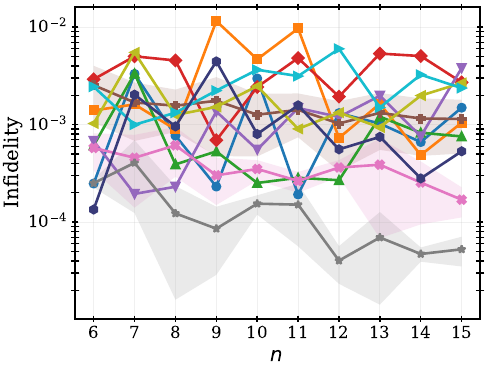}
\\[-0.5em]
{\small (a) BT-BE}
&
{\small (b) BT-BE}
\\[0.3em]

\panelplot{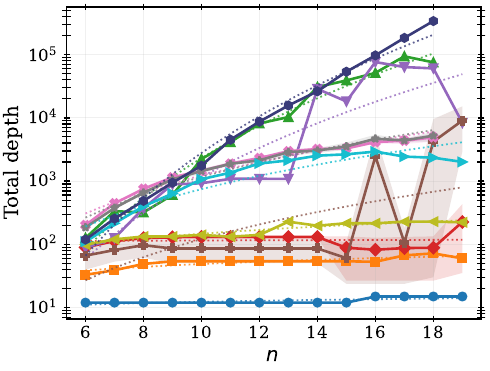}
&
\panelplot{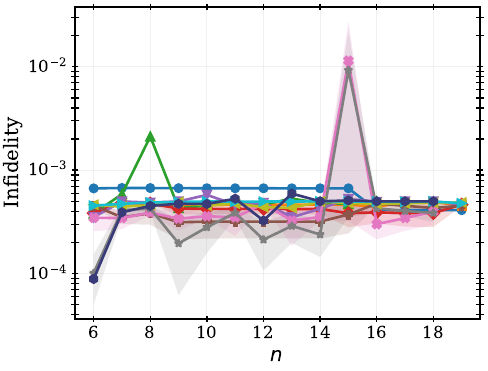}
\\[-0.5em]
{\small (c) BT-WHT}
&
{\small (d) BT-WHT}
\\[0.3em]

\panelplot{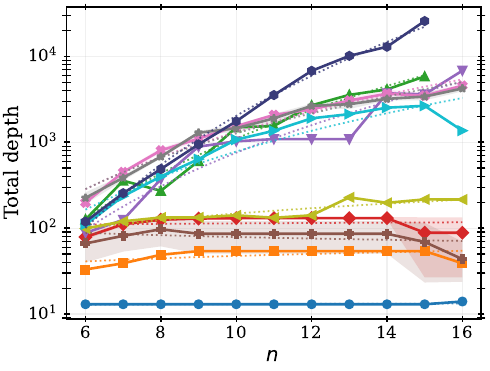}
&
\panelplot{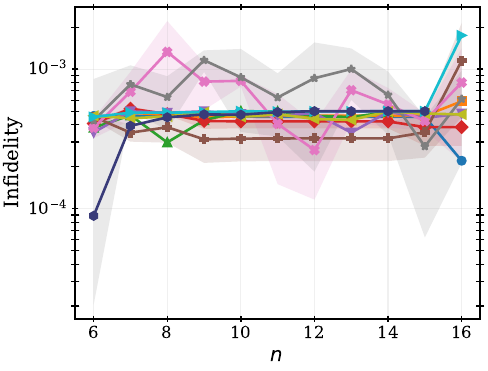}
\\[-0.5em]
{\small (e) BT-Hybrid}
&
{\small (f) BT-Hybrid}
\\[0.3em]

\panelplot{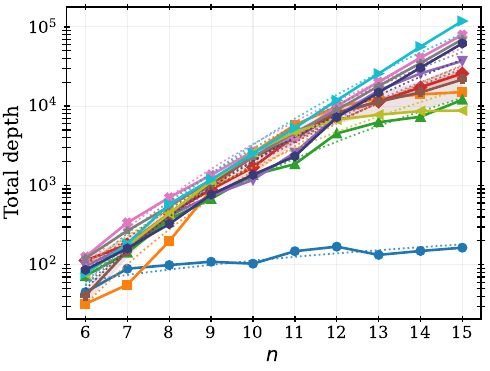}
&
\panelplot{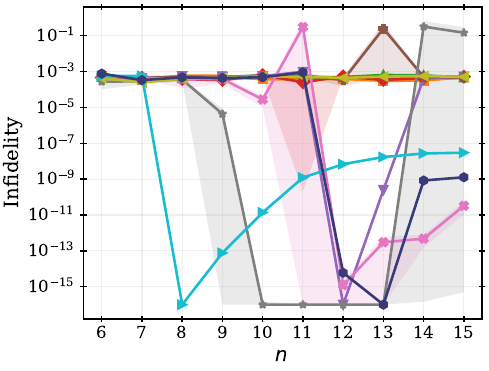}
\\[-0.5em]
{\small (g) Qiskit State Preparation}
&
{\small (h) Qiskit State Preparation}

\end{tabular}
\endgroup

\caption{
Circuit depth versus the number of qubits $n$ for BT-BE (a,b), BT-WHT (c,d), BT-Hybrid (e,f), and Qiskit state preparation (g,h) across the function family whose legends are shown in
\cref{tab:results-function-families}. Left panels show the total one- and two-qubit gate depth as well as scaling fits in dashed lines, whereas right panels show the state-preparation infidelity.
For stochastic families, curves plot the mean, with shaded regions indicating one quintile of deviation. While Qiskit scales as $\sim O(2^{cn})$ for some constant $c$ for most function families, our methods scale sub-linearly for all families besides Weierstrass states, plateauing for large $n$, thereby beating $\Theta(2^n/n)$, see text for details.
}
\label{fig:results-family-scaling-method-comparison}
\end{figure*}

\subsubsection{Construction of the target states}
For a one-dimensional function $f$, we evaluate the function on a grid
of $N=2^n$ points and construct the target quantum state 
\begin{equation}
\ket{v}
=
\frac{1}{\sqrt{
\sum_{k=0}^{2^n-1}|f(x_k)|^2
}}\sum_{j=0}^{2^n-1}
f(x_j)
\ket{j}.
\end{equation}
For a two-dimensional function $f(x,y)$, we use uniform grids containing
$2^{n_x}$ and $2^{n_y}$ points, for a total of $2^{n_x}2^{n_y}=2^n$ points where
\begin{equation}
n_x
=
\left\lceil\frac{n}{2}\right\rceil,
\qquad
n_y
=
\left\lfloor\frac{n}{2}\right\rfloor,
\end{equation}
and normalize the function over a flattened index. The benchmarks cover system sizes $n\in\{6,\ldots,20\}$. 
The fidelity of the prepared state $\ket{\psi_{\text{encoded}}}$ w.r.t. the target $\ket{\psi_{\text{target}}}$ is calculated for every generated circuit as
\begin{equation}
F=\left|\langle\psi_{\text{target}}|\psi_{\text{encoded}}\rangle\right|^2,
\end{equation}
from which we report the infidelity $\mathcal I=1-F$.
\subsubsection{Site Ordering}
The mapping which associates the index of a qubit in the register to the index of the layer in the binary tree representation (the layout) can be chosen arbitrarily.
It assigns one qubit to the root of the tree, one to layer 1, and so on.
This affects the distribution of significant Walsh spectral coefficients for \ac{BT-WHT} and the binary address layout for \ac{BT-BE} controls, 
with some layouts being trivial to encode, and others requiring more operations. 
Therefore the efficiency of our mutliplexor simplification routines depend strongly on the layout.

Thus a smart choice of the layout is important.
However, there are $n!$ possible layouts, so an exhaustive search for the optimal one is impossible. 
Instead, we randomly sample the space of layouts for every encoding, selecting the lowest depth configuration. 
The choice of a specific layout is absorbed into the inverse permutation applied to the instructions of the encoding circuit, so that the labeling of the qubits remains unchanged.
A more systematic algorithm to find approximately optimal layouts will be the subject of future studies. 
In some plots we fix the site ordering and vary other parameters to study convergence properties. 
We indicate when this is the case. 

\subsubsection{Algorithm Configurations}

The optimization of the quadrature basis of the \ac{BT-BE} algorithm is performed up to the 10-th level of the tree since it seems a good compromise between performance and execution time. 
The \ac{BT-BE} and BT-Hybrid require two ancillas for the final KG24 decomposition of the \ac{MCR} gates, while \ac{BT-WHT} only uses one ancilla directly in its decomposition. 
\subsection{Benchmark function families}
\label{sec:results-families}

The benchmark set is chosen to cover qualitatively different forms of
structure in the target amplitude vector, including smooth
localization, oscillations, multiscale behavior, support sparsity, and
one- and two-dimensional jump-diffusion distributions. The families
used in the analysis are summarized in
\cref{tab:results-function-families}.

All continuous functions are evaluated on a uniform
grid and the resulting discrete vector is normalized before state
preparation. For the random sparse states, we average over five instances, each with their own site ordering unless otherwise stated, and report the average depth, with the first quintile of deviation from the mean shown by shading. 

\subsubsection{Gaussians}
\begin{figure*}[!t]
\centering
\begingroup
\setlength{\tabcolsep}{0pt}
\renewcommand{\arraystretch}{1.0}

\newcommand{\panelplot}[1]{%
    \includegraphics[
        width=0.33\textwidth,
        keepaspectratio
    ]{#1}%
}

\begin{tabular}{@{}ccc@{}}

\panelplot{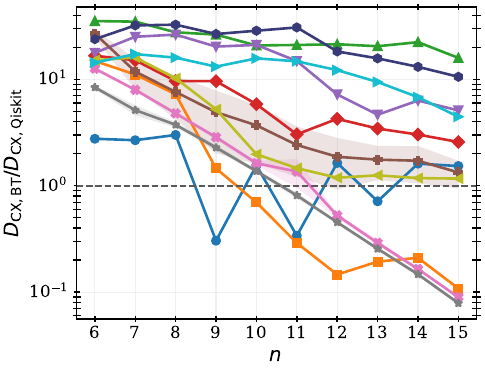} &
\panelplot{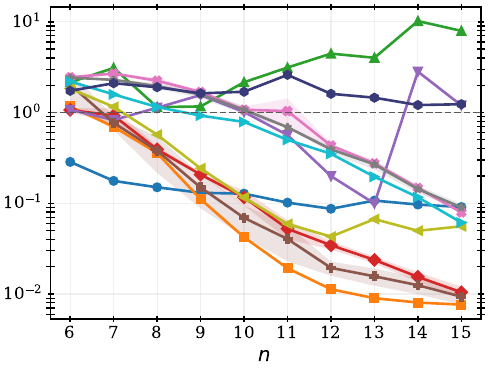} &
\panelplot{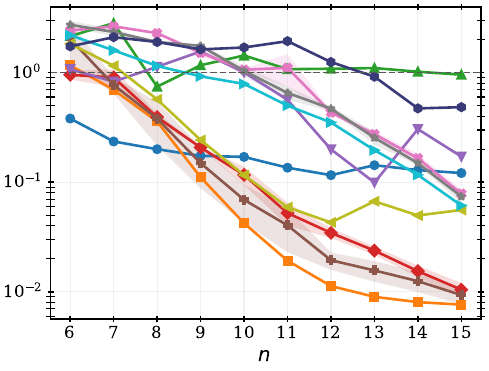}
\\[-0.5em]

{\small (a) BT-BE} &
{\small (b) BT-WHT} &
{\small (c) BT-Hybrid}

\end{tabular}

\endgroup

\caption{
Ratio of the CX-depth of each \ac{BT} method to that of Qiskit's state-preparation circuit as a function of the number of data qubits $n$, replotting the data from \cref{fig:results-family-scaling-method-comparison}. Panels (a), (b), and (c) correspond to the methods \ac{BT-BE}, \ac{BT-WHT}, BT-Hybrid, respectively. The function-family legend is provided in
\cref{tab:results-function-families}. We see that for most function families, BT-WHT beats qiskit by $n=15$, and when including BT-BE to construct a BT-Hybrid, it beats Qiskit for every function family.
}
\label{fig:results-family-scaling-cxonly-ratio-method-comparison}
\end{figure*}

The one-dimensional Gaussian is, up to normalization, 
\begin{equation}
\mathcal{N}(x;\mu,\sigma)
=
\exp
\left[
-\frac{(x-\mu)^2}{2\sigma^2}
\right],\ x\in [-1,1].
\end{equation}
Encodings of Gaussian functions into quantum states provide a basic test of smooth, localized structure. We set $\mu=0$ and $\sigma=1$, except for in \cref{fig:gaussian_heatmap}, where we vary both, and in \cref{fig:results-sum-og-gaussian-peaks-2d}, where we sum over a collection of Gaussians in 2D with different centers and widths.

\subsubsection{Lorentzians}
The Lorentzian is defined, up to normalization, by
\begin{equation}
f_{x_0,\gamma}(x)
=
\frac{\gamma^2}
{(x-x_0)^2+\gamma^2},\ x\in [-0.5,0.5]
\end{equation}
where $x_0$ is the center and $\gamma$ is the width parameter. Compared with a Gaussian, the Lorentzian decays more slowly away from its center and therefore retains substantially heavier tails.

\subsubsection{Sinusoids}
We explore the efficiency of our methods in capturing states dominated by a single oscillatory feature via sinusoids characterized by a single parameter $k$:
\begin{equation}
    f(x)=\sin(2\pi k x),\ x\in [0,1]
\end{equation}
We use $k=10$ in our benchmarks.

\subsubsection{Chebyshev polynomials}
The Chebyshev polynomials of the first kind, can be defined as
\begin{equation}
T_m(x)
=
\cos\!\left(m\arccos x\right),
\qquad
x\in[-1,1],
\end{equation}
where $m\in\{0,\ldots l\}$ is the integer polynomial degree. These functions display non-uniform oscillations which cluster near the boundaries, with a number of nodes equaling $m$.

\subsubsection{Harmonic-oscillator eigenstates}
The one-dimensional harmonic-oscillator eigenfunctions are, up to normalization
\begin{equation}
\begin{split}
\psi_{\ell,\alpha,x_0}(x)&=H_\ell\!\left(\sqrt{\alpha}(x-x_0)\right)\exp\left[-\frac{\alpha}{2}(x-x_0)^2\right]\\
H_{\ell}(x) &= (-1)^{\ell}e^{x^2}\frac{d^{\ell}}{dx^{\ell}}e^{-x^2},
\end{split}
\end{equation}
where $H_\ell$ is the physicists' Hermite polynomial of degree $\ell$, $\alpha$ controls the spatial width, and $x_0$ determines the center. The sampled vector is normalized again after discretization. The level $\ell$ controls the number of nodes, while $\alpha$ controls the degree of spatial localization. We take $x\in\left[-6,6\right]$.

\begin{figure*}
\centering

\begin{subfigure}[t]{0.32\textwidth}
    \centering
    \includegraphics[width=\linewidth]{
        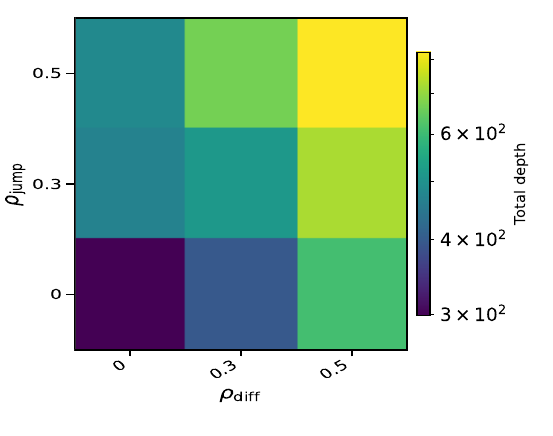
    }
    \caption{Correlated Merton}
    \label{fig:correlated_merton_heatmap}
\end{subfigure}
\hfill
\begin{subfigure}[t]{0.32\textwidth}
    \centering
    \includegraphics[width=\linewidth]{
        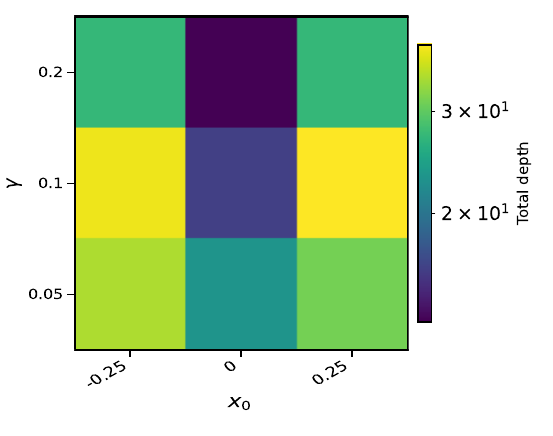
    }
    \caption{Lorentzian}
    \label{fig:lorentzian_heatmap}
\end{subfigure}
\hfill
\begin{subfigure}[t]{0.32\textwidth}
    \centering
    \includegraphics[width=\linewidth]{
        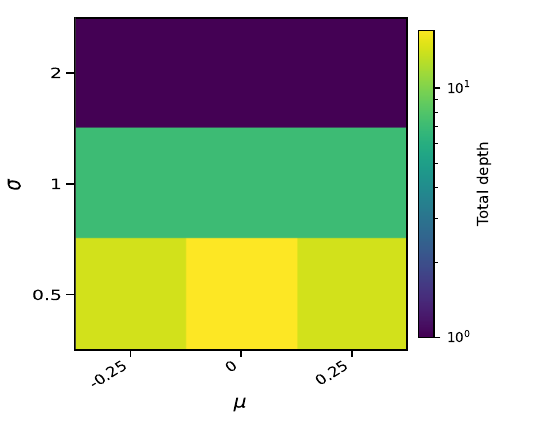
    }
    \caption{Gaussian}
    \label{fig:gaussian_heatmap}
\end{subfigure}

\caption{
BT-Hybrid parameter sweeps for
(a) Correlated Merton,
(b) Lorentzian, and
(c) Gaussian distributions.
All experiments use $n=15$ data qubits,
$L=10$, while the Walsh--Hadamard threshold
is adaptively chosen in order to keep the final infidelity at around $(5 \pm 1 )\cdot 10^{-3}$ .
Each heatmap shows the final CX+single-qubit
circuit depth on a logarithmic color scale.
For the Correlated Merton distribution,
$\lambda=0.5$ is held fixed.
}
\label{fig:bt_hybrid_parameter_sweeps}
\end{figure*}

\begin{figure}
    \centering
    \includegraphics[width=\linewidth]{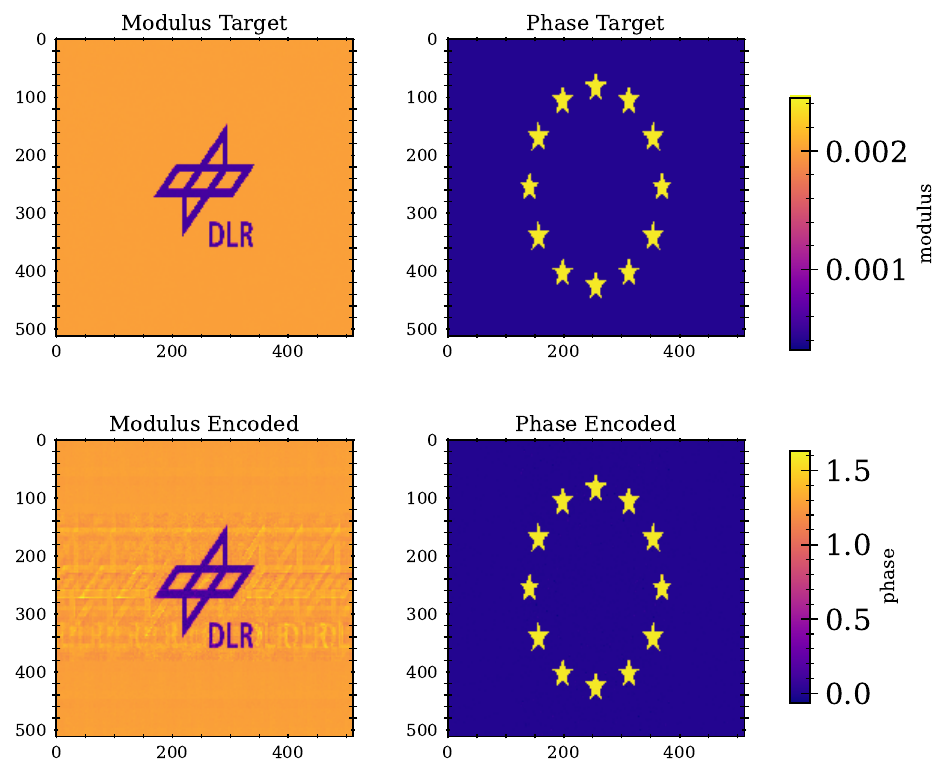}
    \label{dlr_eu_logo}
    \caption{Example of the encoding of arbitrary complex data into the state of $18$ qubits using the BT-Hybrid method. Here the DLR logo and the EU flag are respectively encoded into the modulus and phase of the amplitude vector with an infidelity of $8\cdot 10^{-4}$. See text in \cref{sec:arbitrary_complex} for details. }
\end{figure}

\subsubsection{Polynomial states}
The polynomial benchmark consists of one-dimensional
polynomials
\begin{equation}
    f(x) = \sum_{l=0}^m a_lx^m\null\quad x\in\left[-1,1\right],
\end{equation}
with the $m$ coefficients $a_l$ randomly sampled from the normal distribution $\mathcal{N}(\mu=0,\sigma=1)$. 

\subsubsection{Random sparse states}

We consider two distinct sparse-state families. In both cases a support
\begin{equation}
S
\subseteq
\{0,\ldots,2^n-1\}
\end{equation}
is uniformly randomly selected and amplitudes outside $S$ are set to zero. The Random Sparse Random-Amplitude (RSRA) test function picks random nonzero amplitudes $f_j$ for $j\in S$. The random values are pulled from a normal distribution with mean $0$ and standard deviation $1$. In contrast the Random Sparse Uniform-Amplitude (RSUA) test function sets all amplitudes within the support to be uniform, $f_j = \frac{1}{\sqrt{|\mathcal{S}|}}$, where $|\mathcal{S}|$ is the size of the support. 
We define the \textit{absolute sparsity} to be $|\mathcal{S}|$, while the \textit{relative sparsity} is defined as $|\mathcal{S}|/2^n$. We benchmark both RSRA and RSUA to separate the cost of the random support from the additional overhead to populate that support with random amplitudes. We fix $|\mathcal{S}|=32$ and average over five instantiations.

\subsubsection{Merton states}

The one-dimensional Merton states are obtained by discretizing the log-return distribution of the Merton jump-diffusion model \cite{Merton1976}. We turn the unconditional density
\begin{equation}
\label{Merton_definition}
p_{\mathrm{M}}(x)
=
e^{-\lambda T}\sum_{k=0}^{k_{\max}}
\frac{(\lambda T)^k}{k!}\,
\mathcal{N}\!\left(
x;\,
m_k,\,
v_k
\right),
\end{equation}
with
\begin{equation}
\begin{aligned}
m_k=&
\left(r-\lambda\kappa-\frac{\sigma^2}{2}\right)T
+k\mu_{\mathrm{jump}},\\
\text{and }v_k=&\sigma^2T+k\sigma_{\mathrm{jump}}^2
\end{aligned}
\end{equation}
into an amplitude by taking its square root, where 
\begin{equation}
    \kappa=\exp\left(\frac{\mu_\text{jump} + \sigma_\text{jump}^2 }{ 2}\right) - 1.
\end{equation}
In the analysis we truncate the sum in \cref{Merton_definition} to approximately 10 standard deviations above the Poisson mean, with additional protection when $\lambda T$ is small as in

\begin{equation}
k_{\max} =
\max\Big\{
10,\,
\left\lceil
\lambda T + 10\sqrt{\max(\lambda T,1)}
\right\rceil\Big\}.
\label{eq:merton-poisson-truncation}
\end{equation}
For grid points $x_i$, the corresponding normalized $n$-qubit state is
\begin{equation}
\ket{\psi_{\mathrm M}}
=
\frac{1}{\sqrt{\sum_i p_{\mathrm M}(x_i)}}
\sum_i \sqrt{p_{\mathrm M}(x_i)}\,\ket{i}.
\end{equation}
The parameter $T$ is kept fixed at $1$, $\sigma$ at $0.2$, $r$ at $0.02$. In the scaling study in $n$ (see \cref{fig:results-family-scaling-method-comparison},\cref{fig:results-family-scaling-cxonly-ratio-method-comparison}), the parameters $\lambda$ and $\sigma_{\mathrm{jump}}$ and $\mu_{\mathrm{jump}}$ are fixed according to the values in \cref{tab:results-function-families}.

\subsubsection{Correlated Merton states}

We consider a bivariate extension of the Merton
jump-diffusion model, in which two correlated processes
share a common Poisson jump count
\cite{johannes2010mcmc,makarov2023multiasset}.
The resulting joint probability density is a Poisson
mixture of bivariate Gaussian distributions,
\begin{equation}
\label{eq:correlated-merton-density}
p_{\mathrm{CM}}(x,y)
=
e^{-\lambda T}
\sum_{k=0}^{\infty}
\frac{(\lambda T)^k}{k!}\,
\mathcal N\!\left(
\begin{bmatrix}x\\y\end{bmatrix};
\boldsymbol{m}_k,\Sigma_k
\right),
\end{equation}
where the conditional mean vector is
\begin{equation}
\boldsymbol{m}_k =
\begin{bmatrix}
\mu_1 T + k\mu_{\mathrm{jump},1}\\
\mu_2 T + k\mu_{\mathrm{jump},2}
\end{bmatrix},
\end{equation}
and the covariance matrix is
\begin{align}
\Sigma_k &=
T\begin{bmatrix}
\sigma_1^2 & \rho_{\mathrm d}\sigma_1\sigma_2\\
\rho_{\mathrm d}\sigma_1\sigma_2 & \sigma_2^2
\end{bmatrix}
\nonumber\\
&\quad+
k\begin{bmatrix}
\sigma_{\mathrm{jump},1}^2 &
\rho_{\mathrm j}\sigma_{\mathrm{jump},1}
\sigma_{\mathrm{jump},2}\\
\rho_{\mathrm j}\sigma_{\mathrm{jump},1}
\sigma_{\mathrm{jump},2} &
\sigma_{\mathrm{jump},2}^2
\end{bmatrix}.
\end{align}
The parameters $\rho_{\mathrm d}$ and $\rho_{\mathrm j}$
independently specify the correlations of the diffusion
increments and jump sizes, respectively.

The infinite Poisson mixture is truncated at the smallest
nonnegative integer $k_{\max}$ for which the discarded
probability mass does not exceed
$\varepsilon=10^{-12}$, namely,
\begin{equation}
\label{eq:correlated-merton-truncation}
k_{\max}
=
\min\left\{
K\in\mathbb{N}_0:
e^{-\lambda T}
\sum_{k=K+1}^{\infty}
\frac{(\lambda T)^k}{k!}
\leq \varepsilon
\right\}.
\end{equation}
We denote the corresponding truncated density by
$p_{\mathrm{CM}}^{(k_{\max})}(x,y)$.

Although the full continuous density is normalized over
$\mathbb{R}^2$, its values sampled on a finite grid
do not generally sum to unity. Therefore, the sampled
probabilities must be normalized before being encoded
as quantum amplitudes.

For grid points $(x_i,y_j)$, the normalized
$(n_x+n_y)$-qubit state is
\begin{equation}
\label{eq:correlated-merton-state}
\ket{\psi_{\mathrm{CM}}}
=
\frac{1}{\sqrt{\mathcal Z}}
\sum_{i,j}
\sqrt{
p_{\mathrm{CM}}^{(k_{\max})}(x_i,y_j)
}
\ket{i}\ket{j},
\end{equation}
where
\begin{equation}
\mathcal Z
=
\sum_{i,j}
p_{\mathrm{CM}}^{(k_{\max})}(x_i,y_j),
\end{equation}
and the two registers contain $n_x$ and $n_y$
qubits, respectively. Unlike a tensor product of two independently prepared
Merton states, this construction captures the
statistical dependence between the two variables
arising from correlated diffusion increments,
correlated jump sizes, and their shared jump count.

\subsubsection{Weierstrass-type states}
The one-dimensional Weierstrass function is
\begin{equation}
W_{M,a,b}(x)
=
\sum_{k=0}^{M}
a^k\cos\!\left(b^k\pi x\right),\ x \in [-1,1]
\end{equation}
where $M$ controls the number of frequency components, $a$ controls
the decay of their amplitudes, and $b$ controls the growth of their
frequencies. The value of these parameters we used in our scaling study in $n$ is defined in \cref{tab:results-function-families}. This family of functions combines oscillations over several length scales and therefore engender a large degree of entanglement between qubits when the function is encoded in a quantics representation as we do here.

\begin{figure}
\centering
\begin{subfigure}[t]{0.49\textwidth}
    \centering
    \includegraphics[width=\linewidth]{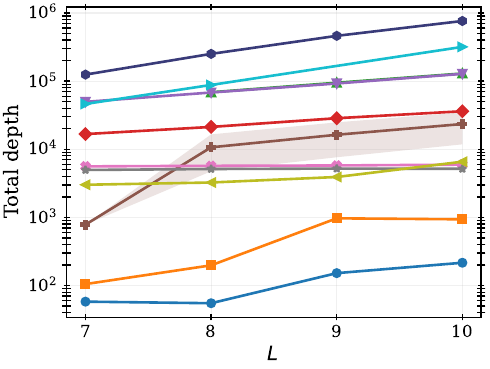}
    \caption{Dependence of the $\text{CX}$ +single qubit rotations depth on the binary precision $L$ at fixed $n=15$ for
    BT-BE.}
    \label{fig:results-L-sweep_depth}
\end{subfigure}
\hfill
\begin{subfigure}[t]{0.49\textwidth}
    \centering
    \includegraphics[width=\linewidth]{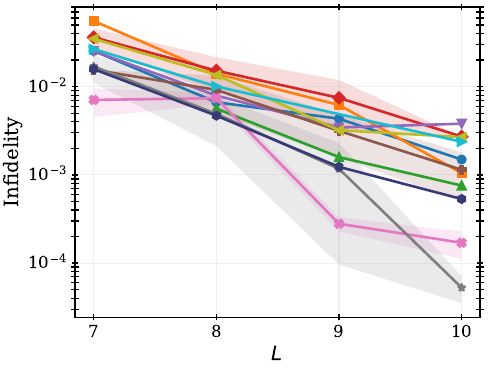}
    \caption{Dependence of the infidelity on the binary precision $L$ at fixed $n=15$ for
    BT-BE.
    }
    \label{fig:results-L-sweep_infidelity}
\end{subfigure}
\caption{Dependence of circuit $\text{CX}$ +single qubit rotations depth and infidelity on the binary precision $L$ at fixed $n=15$.}
\label{fig:results-L-sweep}
\end{figure}

\begin{figure}
    \centering
    \includegraphics{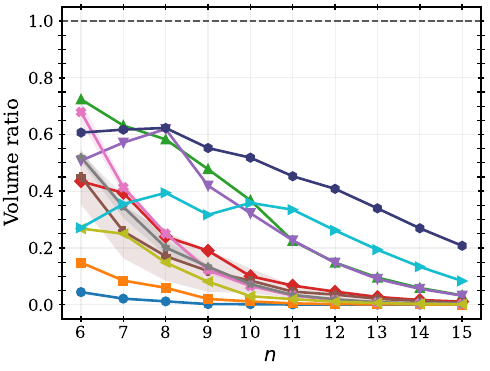}
    \caption{Degree of Multiplexor simplification, as measured by the control-volume ratio, \cref{eq:volume_ratio}, as a function of the number of data qubits n for BT-BE with L=10. The horizontal dashed line marks a volume ratio of one, indicating no simplification.}
    \label{fig:Bt-BE_volume_ratio}
\end{figure}

\subsection{Encoding Results}
\label{sec:encoding_results}

In \cref{fig:results-family-scaling-method-comparison} we compare the efficiency of the three BT methods and of Qiskit's state preparation by considering the total depth of the encoding circuits as a function of the system size $n$. Here the ordering is optimized at each value of $n$. For BT-BE the binary precision is kept fixed at $L=10$ and the optimal basis is recomputed at each point using \cref{alg:optimal-generalized-basis}. For BT-WHT and BT-Hybrid the WHT thresholds are chosen in order to keep the infidelity close to $5\cdot 10^{-3}$, with $L=9$ for Hybrid. The Qiskit implementation of Iten et. al's encoding method \cite{Iten_2016} is, in contrast to our own methods, exact. Thus to provide a fair comparison, after obtaining the abstract state preparation unitary, we tune the approximation level when transpiling into the one- and two-qubit gate set. This has the effect of selectively removing gates in the unitary composition in order to achieve the same value of infidelity as our methods. This is not always possible to do in a smooth manner, as different circuits can be sensitive to small changes, hence the sharp discontinuities for some examples. Despite this tuning, the resulting Qiskit circuit depths still scale as $O(2^{cn})$ for some constants $c$ for each family besides Gaussian, Lorentzian, and Merton. By contrast, besides the highly entangled Weierstrass function, all of our BT methods display an asymptotic scaling plateau at the roughly the same level of infidelity, thus, for these particular relatively structured inputs, beating the theoretical tight bounded scaling of $\Theta(2^n/n)$ for exact generic data encoding \cite{Sun2023}. 

In \cref{fig:results-family-scaling-cxonly-ratio-method-comparison} we show the ratio of the CX-circuit depth of each BT method to that of Qiskit's state-preparation circuit as a function of the number of data qubits $n$. BT-BE does not provide any advantage over Qiskit' s state preparation method, with the exception of Lorentzian, Random Support Random Amplitudes (RSRA) and  Random Support Uniform Amplitudes (RSUA), where it achieves shorter circuit depths for $n> 10$. For the Gaussian, the advantage is only present for certain odd values of $n$ and not for the even ones. We conjecture that this is due to with the alignment and anti-alignment of the angle-decompositions in the Boolean representation, as DSOP is sensitive to the layout of the non-zero entries in powers of two.

Conversely, BT-WHT and especially BT-Hybrid clearly outperform Qiskit for most of the functions in our analysis and for large enough $n$. The latter in particular, which relies on BT-BE as a subroutine, provides circuits whose CX-only depths are from one to two orders of magnitude smaller than Qiskit's at $n=15$. The most significant improvements are achieved for large values of $n$ for the Lorentzian, the polynomial of degree $2$ and the Chebyshev polynomial with $m=2$, with CX-only depths that are roughly $100$ times smaller than Qiskit's at $n=15$. While BT-WHT alone does not reduce the depth over Qiskit for the Hermite polynomial and the Weierstrass function, the hybrid approach gives shallower circuits in both cases. Finally, in the case of a single sine wave with frequency $k=10$, both BT-WHT and BT-BE give significantly deeper circuits than Qiskit's method, while BT-Hybrid shows comparable depths, asymptotically beating it. This example in particular can be prepared much more efficiently using other methods, such as QFT encoding \cite{moosa2024fourier}.
For completeness, in \cref{fig:bt_hybrid_parameter_sweeps} we show the results of a parameter sweep as heatmaps. We analyse the encoding depth associated to the correlated Merton \cref{fig:correlated_merton_heatmap}, the Lorentzian \cref{fig:lorentzian_heatmap} and the Gaussian \cref{fig:gaussian_heatmap} functions using BT-Hybrid at fixed infidelity $(5\pm 1)\cdot 10^{-3} $. In
the first heatmap we observe how both $\rho_{\text{jump}}$ and $\rho_{\text{diff}}$ contribute to the correlation, with the circuit depth increasing monotonically with both parameters. The heatmap associated to the Lorentzian shows how shifts of the peak to the left or to the right of $x=0$ of identical length produce identical depths. While for $x_0=0$ the parameter $\gamma$ seems to be inversely proportional to the depth, for $|x_0|=0.25$ the depth reaches a peak for the intermediate value in our sweep $\gamma=0.1$, Finally, for the Gaussian profile, the depth is almost invariant under shifts of the peak and it rapidly converges to $4$ for $\sigma=2$. 

\begin{figure}
\centering
\begin{subfigure}[t]{0.49\textwidth}
    \centering
    \includegraphics[width=\linewidth]{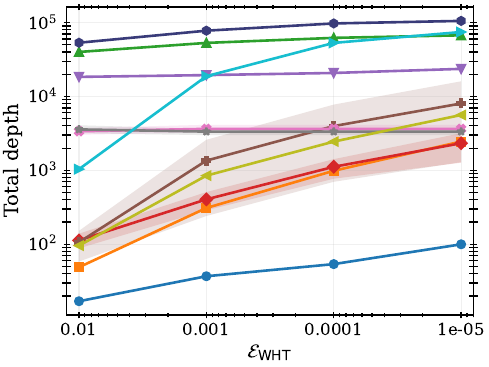}
    \caption{Total depth versus $\mathcal{E}_\mathrm{WHT}$ for BT-WHT at n=15. 
}
    \label{fig:results-eps-sweep_depth}
\end{subfigure}
\hfill
\begin{subfigure}[t]{0.49\textwidth}
    \centering
    \includegraphics[width=\linewidth]{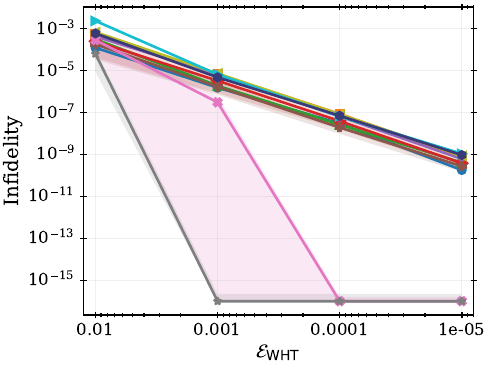}
    \caption{Infidelity versus $\mathcal{E}_\mathrm{WHT}$ for BT-WHT at n=15.
}
    \label{fig:results-eps-sweep_infidelity}
\end{subfigure}
\caption{Total depth and infidelity versus $\mathcal{E}_\mathrm{WHT}$ for BT-WHT at n=15. The WHT Threshold is defined in \cref{eq:relative_error_def}. The total depth growth in panel(a) depends on the structure of the data, while the infidelity in panel (b) scales proportionally to the WHT threshold squared, except for the sparse functions where they are eventually exactly captured.}
\label{fig:results-eps-sweep}
\end{figure}

\subsection{Arbitrary Complex Input}
\label{sec:arbitrary_complex}

In order to illustrate the performance of the BT-Hybrid method on a given complex input, in \cref{dlr_eu_logo} we rescaled the image of the DLR logo and of the European Union flag over $2^9 \times 2^9$ pixels and we then encoded the first into the modulus and the second into the phase part of the amplitude vector of a state of $18$ qubits. The final state is described by the function 
\begin{equation}
    f(x,y)= \mathcal M(x,y)e^{i\frac{\pi}{2} u(x,y)} ,
\end{equation}
where the first $9$ qubits encode the $x$ positions, the last $9$ encode the $y$ position and $M(x,y),u(x,y)$ are the values of the rescaled DLR logo and European flag respectively at position $(x,y)\in [2^9]\times [2^9]$. The modulus function is normalized s.t. $\sum_{x,y}M(x,y)^2=1$ while the phase function is binary with $u(x,y)\in \{0,1\}$.
The circuit's one- and two-qubit gate depth is $851,318$ while the infidelity is $8 \cdot 10^{-4}$. The visual effect of patchy infidelity is due to truncating the angle-vectors within the \ac{WHT} basis, which is composed of a series of step functions. Therefore elimination of some step functions leads to discontinuities in angles and subsequently the reconstructed data.
Encoding the same function with Qiskit's state preparation for the same size would take a prohibitively large amount of time, we therefore compared the total depth of the two approaches only for $n=14$ at fixed infidelity $7\cdot 10^{-4}$. At this size, Qiskit's depth $59,802$ is almost identical to the one given by our method $58,422$.
\subsection{BT-BE control volume reduction}
In this section, we analyse the level of simplification BT-BE achieves on different families of functions by reducing both the amount of controls and the number of \ac{MCR} gates. We do this by studying the control volume ratio, defined as the ratio between simplified control volume $\mathcal{V}_{\text{BT-BE}}(L)$ and the unsimplified or trivial one $\mathcal{V}_{\text{trivial}}$. The first is the sum of the amount of control qubits over MCR gates computed after the \ac{BT-BE} simplification but before the decomposition into two and one qubit gates. The second is the same quantity calculated over the full unsimplified multiplexer. Both volumes take into consideration both the modulus and the phase multiplexer ladders in \cref{fig:multiplexer_ladder}.
For the modulus, the formulas for the two volumes are

\begin{align}
&&\mathcal{V}_{\text{BT-BE},\mathcal{M}}(L)=&\sum_{l=1}^{n-1}\sum_{i=1}^L\sum_{\vc c \in \mathcal{G}^{(l)}_i}m(\vc c)\\
\text{and}&&\mathcal{V}_{\text{trivial},\mathcal{M}}=&\sum_{l=1}^{n-1} l|S_i^{(l)}|,
\end{align}
where $\mathcal{S}_i^{(l)}$ is the support and  $\mathcal{G}_i^{(l)}$ the DSOP decomposition of the $i$-th column of the binary matrix $\vc{B}_\mathcal{M}^{(l)}$ associated to the $l$-th level  introduced in \cref{level_binary_matrix_BT_BE}, while $m(\vc c)$ is the amount of controls of the generalized control string $\vc c$ as in \cref{eq:amount_of_controls}.
The definitions of $\mathcal{V}_{\text{BT-BE},\zeta}(L)$ and $\mathcal{V}_{\text{trivial},\zeta}$ are analogous to the ones for the modulus, with the binary matrices now  associated to the phase angle level $\vc{B}_\zeta^{(l)}$. Finally,  the total volume ratio is
\begin{equation}\label{eq:volume_ratio}
    \mathcal{R}_{\mathcal{V}}(L)=\frac{ \mathcal{V}_{\text{BT-BE},\mathcal{M}}(L)+ \mathcal{V}_{\text{BT-BE},\zeta}(L)}{\mathcal{V}_{\text{trivial},\mathcal{M}}+\mathcal{V}_{\text{trivial},\zeta}}.
\end{equation}
In \cref{fig:Bt-BE_volume_ratio} we show the relation between the volume ratio and the system size for the different families of functions listed in \cref{tab:results-function-families}.  
During the decomposition we replace the multiplexer with $L$ multiplexers with fixed angles. If the subsequent simplification does not significantly reduce the number of \ac{MCR} gates and simplify their controls, the final circuit volume can be larger than the original.
We see that this is not the case for any of the analyzed functions, which all show a reduction of more than $70 \%$ already at $n=15$ with respect to the unsimplified multiplexer circuit.
\subsection{Dependence of depth and infidelity on $L$}
In \cref{fig:results-L-sweep} we show how the binary precision $L$ impacts the depth of the final circuit after decomposition into $\text{CX}$ and single qubit gates and its infidelity with respect to the target. Here the ordering of the qubits is optimized once for $L=10$ and kept constant for every other value. We notice how for most of the function we analyzed, the growth of their depths as functions of $L$ is slower than the reduction in their infidelities, which decrease monotonically with $L$ except that for the Hermite polynomial. We conjecture this is due to the basis optimization that is done independently for every value of $L$ and that can lead to a better or worse accuracy.

\begin{figure*}
    \centering
\includegraphics[width=\textwidth]{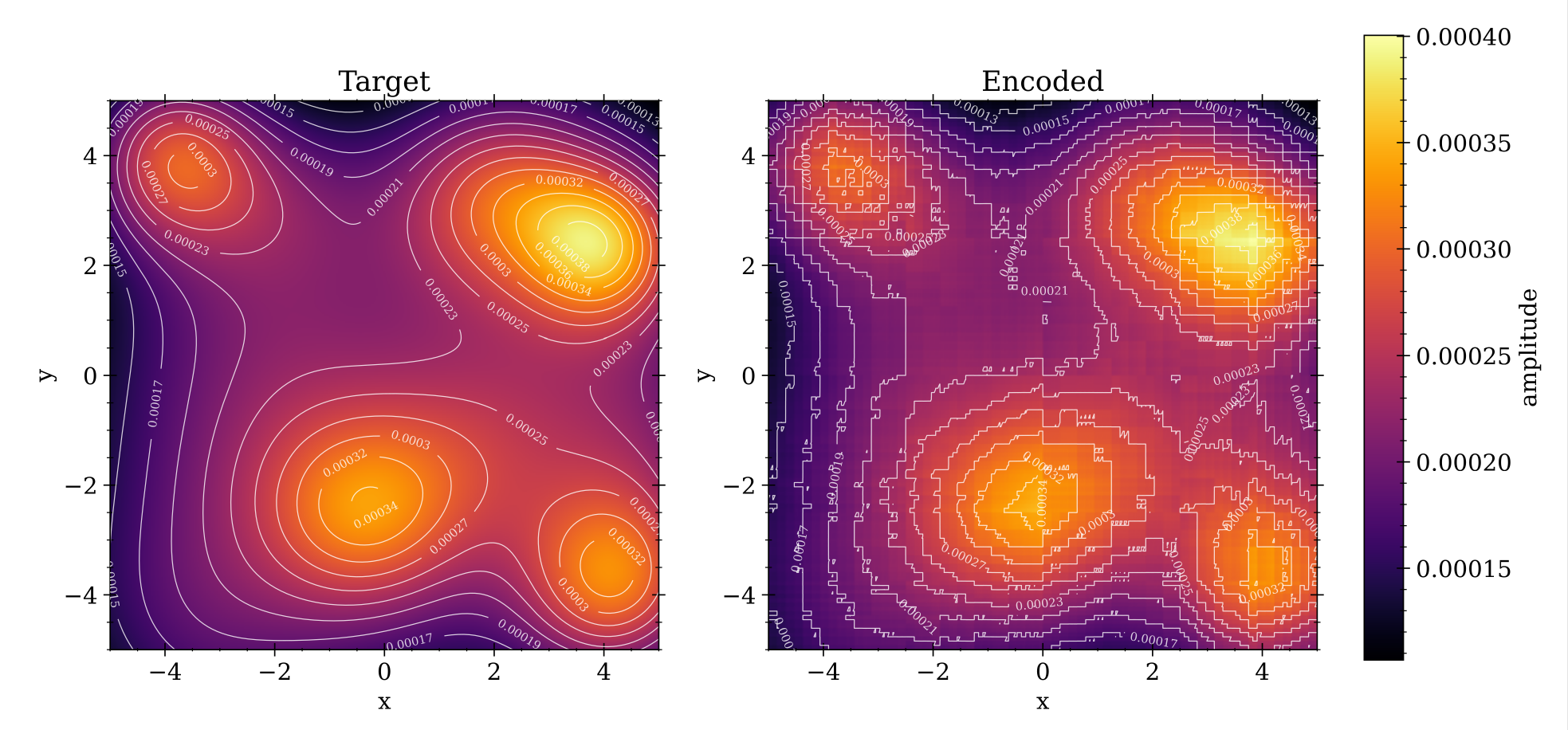}
    \caption{Comparison between the target (left) and encoded (right) states for a weighted sum of fifty two-dimensional Gaussian peaks sampled on a grid of size $2^{12} \times 2^{12}= 16,777,216$ points discretizing the interval $[-5,5] \times [-5,5]$. The infidelity is $2.7\cdot 10^{-4}$ while the one- and two-qubit gate circuit depth is $1087$. We furthermore extend to $2^{18}\times 2^{18}$ with depth $1128$ at infidelity $6.5\cdot 10^{-4}$. For details see \cref{sec:sum_Gaus}.}
    \label{fig:results-sum-og-gaussian-peaks-2d}
\end{figure*}
\begin{figure}
    \centering
    \includegraphics[width=\linewidth]{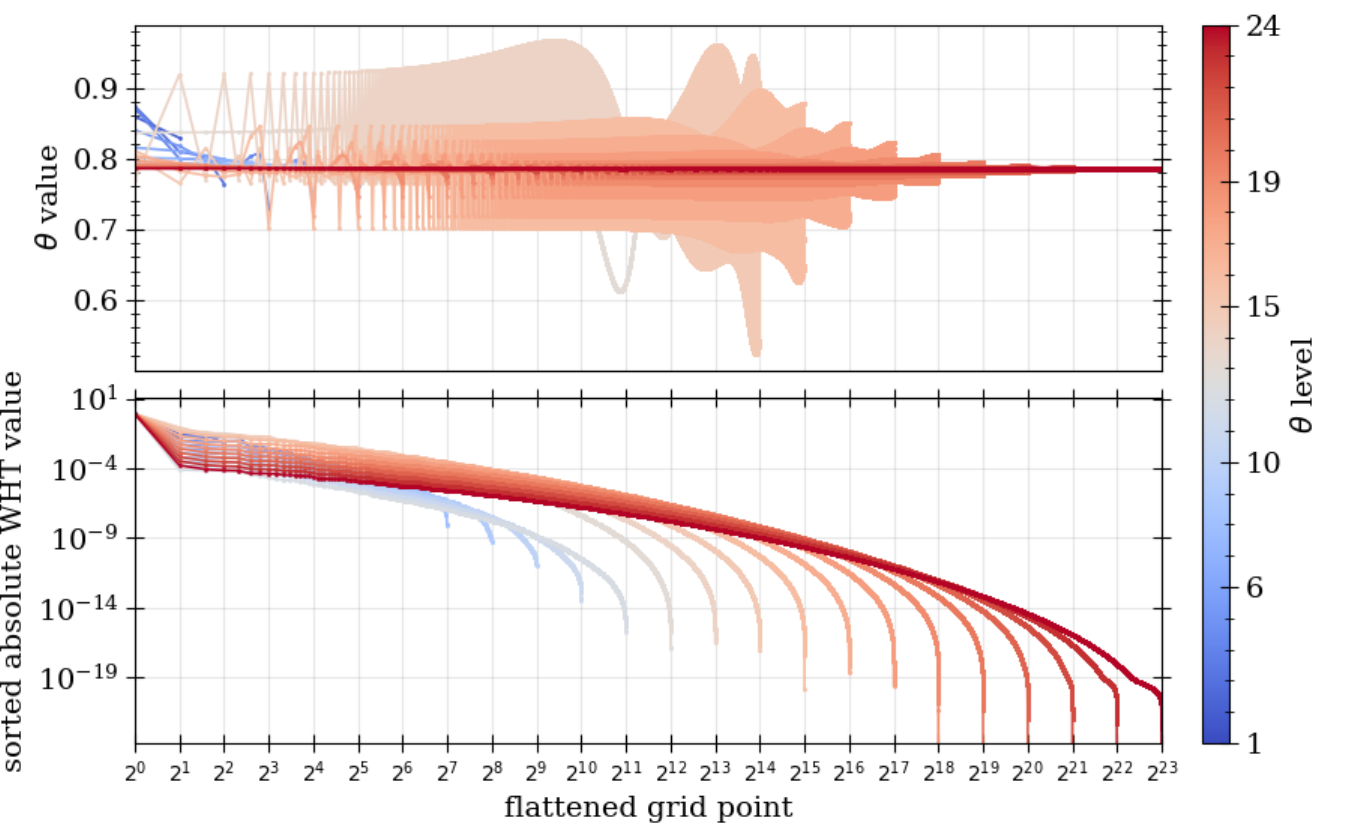} \caption{The top panel shows the angle layers $\vc{\theta}$ from the tree root to layer 23, indicated via coloration going from blue to red, calculated from the sum over ten two-dimensional Gaussian peaks shown in \cref{fig:results-sum-og-gaussian-peaks-2d}, with the deeper layers clearly becoming smoother and flatter. The bottom panel shows the corresponding Walsh spectrum. For legibility in presentation, we plot the absolute values of the components, which are sorted according in descending order. The dominant peak at deeper layers is subsequently the constant Walsh function.}
    \label{fig:levels_recosntructed_from_TCI}
\end{figure}

\subsection{BT-WHT dependence on $\mathcal{E}_{\mathrm{WHT}}$}
As mentioned in \cref{sec:BT-WHT}, the infidelity of the approximated encoding provided by the BT-WHT and consequently BT-Hybrid method depend on the parameter $\mathcal{E}_{\mathrm{WHT}}$, defined in \cref{eq:relative_error_def}. In particular, an efficiently computable upper bound is given in \cref{eq:relative_error_upperbound}. In this section we show how both the encoding circuit depth and the infidelity vary with the threshold $\mathcal{E}_{\mathrm{WHT}}$. In \cref{fig:results-eps-sweep_depth} we see how the total circuit depth at $n=15$ associated to different functions in our analysis changes with the relative error threshold. Based on the rapidity of the growth we can divide the function into two groups: the group with slow growing depth and the one with fast growing depth with $\mathcal{E}_{\mathrm{WHT}}$, The first group includes the Weierstrass function, the sine with $k=10$, the Hermite polynomial, the Gaussian and both types of random sparse states. The second group is composed of the Lorentzian, the polynomial of degree $2$, the Merton, the correlated Merton and the Chebyshev polynomial. Notably, with the exception of the sparse states, the improvement in infidelity decreases uniformly proportional to $\mathcal{E}_{\mathrm{WHT}}^{2}$, as shown in \cref{fig:results-eps-sweep_infidelity}. This behavior is in agreement with the formula in \cref{eq:relative_error_upperbound}. For the sparse function families, after a certain threshold they are exactly captured.

\subsection{Sum over Gaussians}
\label{sec:sum_Gaus}
A superposition of Gaussians is a common state used in many applications, for example in quantum chemistry \cite{Huggins2025}. We study the encoding of a sum of $M=50$ two-dimensional Gaussian peaks with random center and random standard deviation $\sigma$. The centers are sampled uniformly randomly over the range $[-5, 5] \times [-5, 5]$ while the standard deviation is sampled uniformly over $[1, 5]$ and the coefficients $\alpha_j$ are pulled from random uniform distributions and later normalized. The two-dimensional interval is then discretized using a uniform quantics grid with $2^{12}\times 2^{12}$ points. Here we do not optimize over site ordering, the first $12$ qubit encode x and the last $12$ qubits encode y. The function calculated on the grid is 
\begin{equation}
\label{eq:sum_of_gaussians}
    \psi(x,y)=\sum_{j=1}^{M=50} \alpha_j e^{-\frac{(x-x_j)^2+(y-y_j)^2}{2\sigma_j}},
\end{equation}
before normalization. In \cref{fig:results-sum-og-gaussian-peaks-2d} we show the result of the encoding for the BT-Hybrid method with $\mathcal{E}_\mathrm{WHT}=8\times 10^{-3}$ and $L=10$. The total depth after decomposition into CX + single qubit rotations is $1087$ with an infidelity of $2.7\times 10^{-4}$. 

\cref{fig:levels_recosntructed_from_TCI} shows the angle levels associated to the modulus of the function from $l=1$ to $24$ in the top panel as well as the sorted absolute magnitude of the Walsh spectrum in the bottom panel. For smooth functions such as these, deeper angle layers become progressively flatter while their \ac{TT} representation become lower-rank (from blue to dark red). We reconstruct their dense representation here for visualization. This gives both greater compressibility and a sparser Walsh spectrum.
The MPS of the state has a max bond dimension of $24$ with the original ordering of the qubits and a cutoff of $10^{-10}$ for the SVD spectrum. It is constructed by applying TCI to \cref{eq:sum_of_gaussians}. We tested the encoding up to $n=36$, where the circuit depth is $1128$ and infidelity is $6.5\cdot 10^{-4}$. Note that storing this vector as a complex vector would require over 1 TB of memory.  
A comparison with Qiskit state preparation method could not be done because the time and memory requirements are prohibitively expensive. We therefore used Qiskit's method to encode the same function but discretized on a much smaller grid of size $2^6 \times 2^6$. Despite being $2^{12}$ times smaller in size, the resulting circuit is over $10.8$ times larger, with a depth of $11740$ and infidelity $4 \cdot 10^{-4}$. 

\subsection{Experiment on IQM Emerald}
\label{sec:experiment}

\begin{figure}
    \centering
    \includegraphics[width=\linewidth]{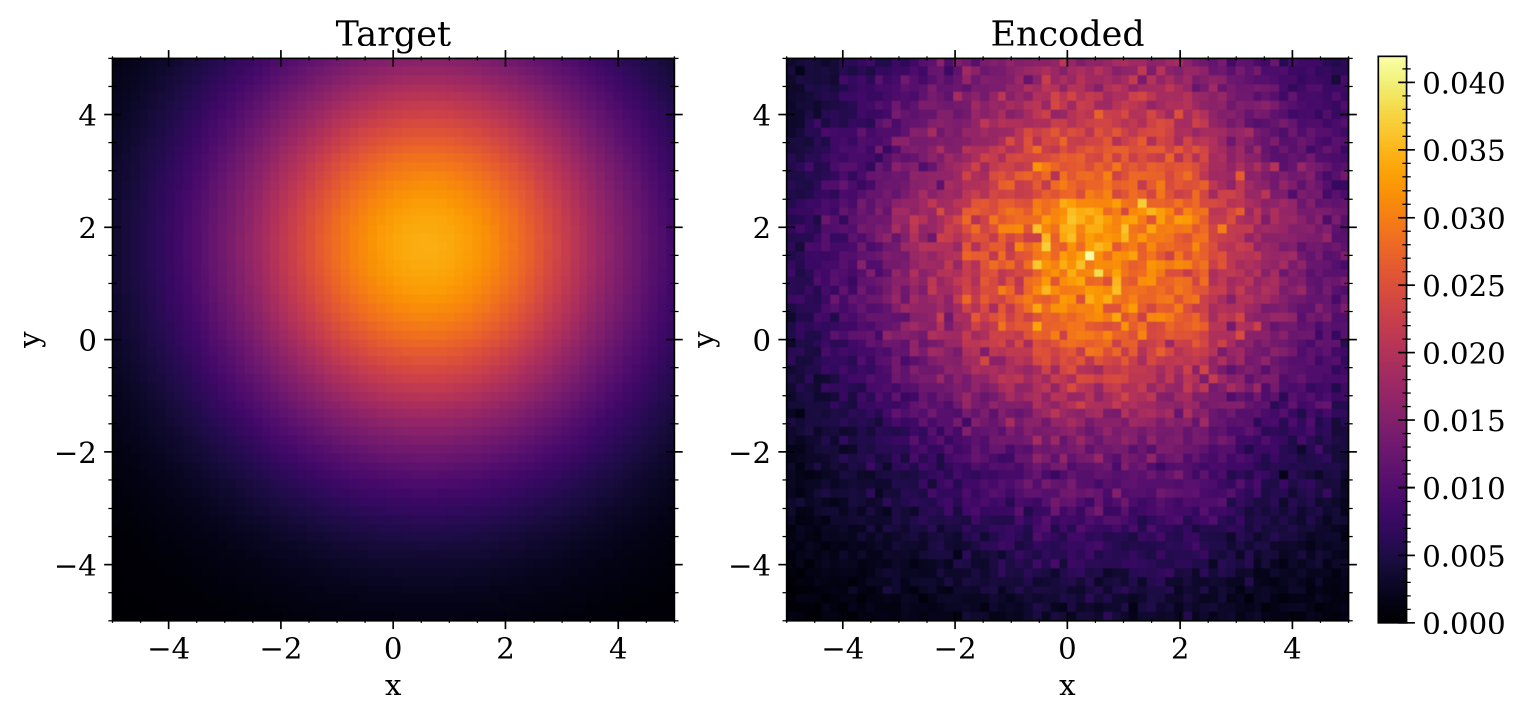}
    \caption{Target quantum state's amplitude squared, left, encoding a Gaussian function, and experimentally measured sample distribution, right, on IQM device `Emerald'. The state is encoded over $20$ qubits in a $2^{10}\times 2^{10}$ grid. Therefore, the measured amplitudes are binned into a total of $64\times 64$ bins for visibility. For ease of comparison, the target state is also plotted over the same set of bins. See \cref{sec:experiment} for details.
    }
    \label{fig:gaussian_encoding_experiment}
\end{figure}

In \cref{fig:gaussian_encoding_experiment} we show the results of experiments on the IQM Emerald quantum processing unit \cite{IQMEmerald}. This is a superconducting qubit device with square lattice connectivity. Its native gate set is two-qubit controlled Z (CZ) and single qubit phased-X rotation (PRX) gates. At the time the circuits were executed on the 12$^{\text{th}}$ of October 2026, these displayed a median PRX gate fidelity of 99.97\% and median CZ gate fidelity of 99.72\%, while the qubits had a median T1 time of 48.96 $\mu$s and median T2 time of 14.78 $\mu$s as determined by Ramsey characterization. To reduce the effect of coherent noise channels we use 16 Pauli Twirled instantiations of the transpiled circuit\cite{Wallman2016,Ware2021}, and utilize dynamical decoupling with a minimal sequence of two X gates during idle times \cite{Ezzell2023}. We take $16,000$ shots per circuit, totaling $256,000$ shots. \\
We encode a single off-set Gaussian into a statevector over $n=20$ qubits via an $(x,\ y)$ grid with $2^{10}\times 2^{10}$ grid points using BT-Hybrid with $\mathcal{E}_{\mathrm{WHT}}=5\cdot 10^{-3}$ and $L=9$. Upon transpilation to the native gate set and routing to the square lattice connectivity, the transpiled circuits had a total depth of 56 and a CZ-depth of 26. Since full state tomography would be infeasible, we simply measure in the computational basis, sort the amplitude into the appropriate $x$ and $y$ index position and then sum together $2^4\times 2^4$ patches of shot counts in order to construct a total of $64\times 64$ bins. Binning the target state over the same coarse grid, and given that the (binned) target function $\psi$ and (binned) encoded quantum state $\psi^*$ are composed entirely of real probability amplitudes, we can estimate the infidelity as
\begin{equation}
    \mathcal{I} = 1 - \left(\sum_i |\psi(\vc{r}_i)||\psi^*(\vc{r}_i)|\right)^2,
\end{equation}
where the sum occurs over the bins $\vc{r}_i$. For this configuration of bin sizes we find an infidelity of $\mathcal{I}=0.011\pm 0.002$, where the uncertainty comes from the shot noise $1/\sqrt{N_s}$ for number of shots $N_s$.

\section{Conclusion and Outlook}\label{sec:conclusions}
We have developed a practical and scalable set of methods for deterministic amplitude-encoding large complex data with tunable trade-offs between state-fidelity and circuit depth. We demonstrated that for several function families our approaches produce one- and two-qubit gate circuits which are up to two orders of magnitude smaller in depth than Iten, et. al's encoding method as implemented in Qiskit \cite{Iten_2016,qiskit}, while scaling asymptotically better than the theoretical tight bound of $\Theta(2^n/n)$ expected for exact encoding, at a modest cost in infidelity. We furthermore demonstrated that this reduction in circuit depth allows us to prepare states on IQM hardware with high accuracy. We are able to do so by 1) systematically eliminating the multiplexer angles via truncation in their binary-tree representation, 2) formally mapping the problem of multiplexer simplification to Boolean logic problems which admit highly-optimized heuristic solutions, 3) decomposing remaining angles via a cycle encoding of their \ac{WHT} spectrum with a single ancilla, and 4) formulating the angle calculation and encoding in terms of tensor-network representations, allowing us to go beyond classical memory constraints.
As highlighted by our analysis, in particular the study on image encoding and states formed by sums of Gaussians up to 36 qubits, our method can prepare an arbitrary quantum state of moderate size and can scale to larger numbers of qubits when the underlying data exhibit exploitable structure. Compared with standard generic-data state-preparation methods, such as those implemented in Qiskit, our approaches achieve varying degrees of circuit-depth reduction across all the functions analyzed, with the Hybrid method beating Qiskit across all function classes at large numbers of qubits. 

We expect our work to be useful for many quantum algorithms which require loading of initial data or reference states such as quantum machine learning \cite{Morales2026,Ji2026}, quantum many-body simulation \cite{Fauseweh2024}, quantum chemistry \cite{Fomichev2024,Huggins2025} and simulating partial differential equations \cite{Childs2021,Sato2025} to name a few. Since our method accepts a tensor-network as a starting point, this provides a natural interface to the hybrid quantum/classical tensor quantum programming paradigm \cite{Termanova2024} where classical tensor-network based algorithms are propagated to the point of saturating classical resources before being transferred to a quantum computer. Furthermore, since our algorithm is not restricted to nearest-neighbor interactions as many \ac{TT} encoders are, it engenders itself to hardware with long range interactions like ion traps and neutral atoms, decreasing the total depth.
Finally, we expect our multiplexer simplification routines to be useful in circuit simplification beyond quantum state preparation, such as in the quantum Shannon or block ZXZ decompositions \cite{Shende_2006,Krol2024}.

Overall, this work makes significant strides in addressing a critical bottleneck in realizing near-term quantum utility, while paving the way for larger scale applications as hardware continues to advance towards fault-tolerance.

\section{Acknowledgments}
VP acknowledges funding by the DLR Quantum Fellowship Program. PS and ME acknowledge funding by the project Algorithms for quantum computer development in hardware-software codesign (ALQU) \cite{ALQU}. Both programs were made possible by the DLR Quantum Computing Initiative (QCI) and the German Federal Ministry of Research, Technology and Space (BMFTR). KL acknowledges the Q-Neko project, which has received funding from the European Union’s Horizon Europe research and innovation programme under Grant Agreement No. 101241875. We acknowledge IQM for providing access to their quantum computing resources.

\textbf{AI use} AI was utilized for the optimization and organization of some of the code base used in the paper. 

\textbf{Author Contributions}
Conceptualization: VP, KL; Formal Analysis: VP, ME, PS, KL; Funding Acquisition: ME, MF; Software: VP, KL; Supervision: ME, MF, KL; Writing - original draft: VP; Writing - review and editing: VP, PS, MF, ME, KL

The original contributions presented in this study, including quantum circuits and raw data for every plot, are available upon reasonable request to the authors.

\bibliography{main.bib}

\appendix
\renewcommand{\thesection}{\Roman{section}}
\renewcommand{\thesubsection}{\Alph{subsection}}
\renewcommand{\thesubsubsection}{\arabic{subsubsection}}
\crefalias{section}{appendix}
\crefalias{subsection}{appendix}
\crefalias{subsubsection}{appendix}
\section{Separable states in the binary-tree representation}
\label{app:separable-state-trees}

A normalized $n$-qubit product state can be written, up to a global phase, in terms of $2n$ real parameters, $\alpha_q,\ \beta_q$, as
\begin{equation}
\label{eq:separable-state}
\ket{\Psi}
=
\bigotimes_{q=0}^{n-1}
\left[
e^{-\frac{i}{2}\beta_q}\cos(\alpha_q)\ket{0}_q
+
e^{\frac{i}{2}\beta_q}\sin(\alpha_q)\ket{1}_q
\right].
\end{equation}
Its amplitude associated with the bit string ${\vc{j}=(j_0,\ldots,j_{n-1})}$ \begin{equation}
    \ket{\Psi}=\sum_{\vc{j} \in \{0,1\}^n} v_{\vc{j}} \ket{\vc{j}}
\end{equation} is given by 
\begin{equation}
\label{eq:product-state-amplitudes}
v_{\vc{j}}
=
\prod_{q=0}^{n-1}
\left[e^{-\frac{i}{2}\beta_q}\cos(\alpha_q)\right]^{1-j_q}
\left[e^{\frac{i}{2}\beta_q}\sin(\alpha_q)\right]^{j_q}.
\end{equation}
For example, the first and last amplitudes are
\begin{align}
v_{\vc{0}}
&=
e^{-\frac{i}{2}\sum_{q=0}^{n-1}\beta_q}\prod_{q=0}^{n-1}\cos(\alpha_q),\\
v_{\vc{1}}
&=
e^{\frac{i}{2}\sum_{q=0}^{n-1}\beta_q}
\prod_{q=0}^{n-1}\sin(\alpha_q),
\end{align}
where $\vc{0}=(0,\ldots,0)$ and $\vc{1}=(1,\ldots,1)$.

Because the amplitude in \cref{eq:product-state-amplitudes} factorizes over the qubits, the relative weight and phase associated with the bit $j_l$ are independent of the preceding prefix $\vc{j}[l]$. Consequently, the binary-tree angles are constant across each level
\begin{equation}
\label{eq:separable-layerwise-angles}
\theta_{\vc{p}}^{(l)}=\alpha_l,
\qquad
\phi_{\vc{p}}^{(l)}=\frac{\beta_l}{2},
\qquad
\forall\vc{p}\in\{0,1\}^l,
\end{equation}
for every node whose associated subtree has nonzero norm. 
Angles associated with zero-norm subtrees are physically irrelevant and can be assigned according to a given convention, which we exploit in the \ac{BT-WHT} algorithm to reduce gate-count.

Consistent with the definition of a product state, \Cref{eq:separable-layerwise-angles} shows that the $2(2^n-1)$ labels of the full binary tree reduce to only $2n$ independent parameters corresponding to independent single-qubit modulus and phase rotations. Assuming an input $\ket{0}^{\otimes n}$ state, with the standard convention
\begin{equation}
R_y(\vartheta)\ket{0}
=
\cos(\vartheta/2)\ket{0}
+
\sin(\vartheta/2)\ket{1},
\end{equation}
the required modulus rotation on qubit $q$ is $R_y(2\alpha_q)$, followed by
\begin{equation}
R_z(\beta_q)
=
\begin{pmatrix}
e^{-\frac{i}{2}\beta_q} & 0\\
0 & e^{\frac{i}{2}\beta_q}
\end{pmatrix},
\end{equation}
thus trivially preparing \cref{eq:separable-state}.

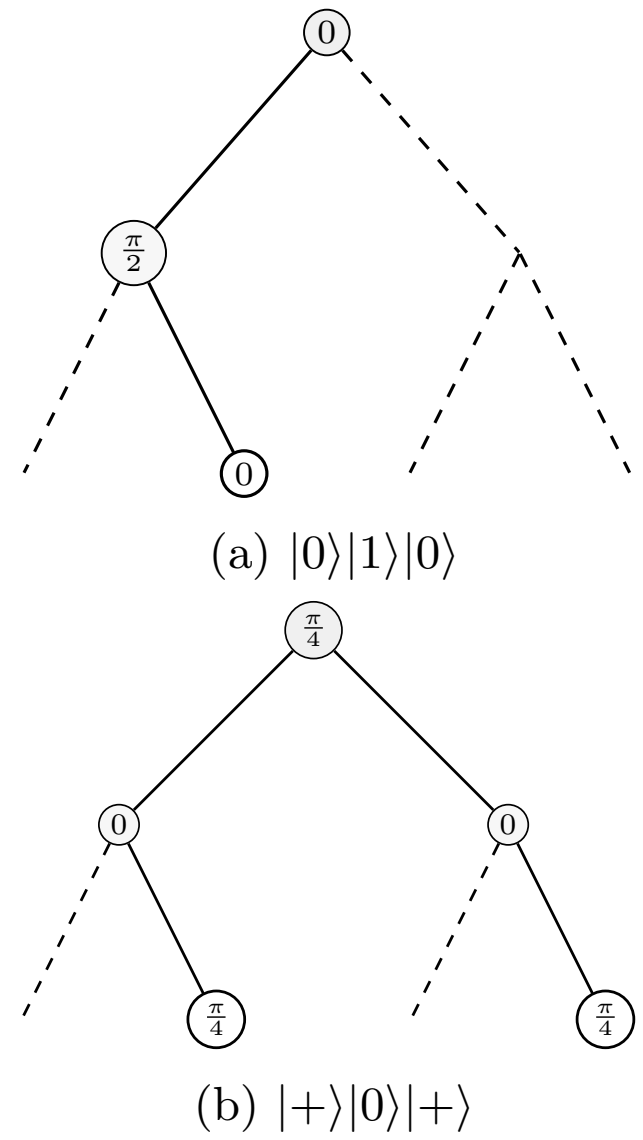
\begin{figure}[!t]
\centering
\begin{subfigure}[b]{0.5\columnwidth}
\centering
\resizebox{\linewidth}{!}{
\begin{tikzpicture}[
    grow=down,
    level distance=16mm,
    level 1/.style={sibling distance=28mm},
    level 2/.style={sibling distance=16mm},
    every node/.style={
        draw,
        circle,
        align=center,
        inner sep=1.2pt,
        font=\scriptsize
    },
    root/.style={
        fill=gray!12,
        font=\scriptsize\bfseries
    },
    internal/.style={
        fill=gray!6
    },
    leaf/.style={
        fill=white
    },
    missingnode/.style={
        draw=none,
        fill=none,
        inner sep=0pt,
        minimum size=0pt
    },
    edge from parent/.style={
        draw,
        line width=0.65pt
    }
]

\node[root] {$0$}
    child {
        node[internal] {$\frac{\pi}{2}$}
        child {
            node[missingnode] {}
            edge from parent[dashed]
        }
        child {
            node[leaf] {$0$}
        }
    }
    child {
        node[missingnode] (missingright) {}
        child {
            node[missingnode] {}
            edge from parent[dashed]
        }
        child {
            node[missingnode] {}
            edge from parent[dashed]
        }
        edge from parent[dashed]
    };

\end{tikzpicture}
}
\caption{$\ket{0}\ket{1}\ket{0}$}
\label{fig:separable_state_tree_001}
\end{subfigure}
\hfill
\begin{subfigure}[b]{0.5\columnwidth}
\centering
\resizebox{\linewidth}{!}{
\begin{tikzpicture}[
    grow=down,
    level distance=16mm,
    level 1/.style={sibling distance=32mm},
    level 2/.style={sibling distance=16mm},
    every node/.style={
        draw,
        circle,
        align=center,
        inner sep=1.2pt,
        font=\scriptsize
    },
    root/.style={
        fill=gray!12,
        font=\scriptsize\bfseries
    },
    internal/.style={
        fill=gray!6
    },
    leaf/.style={
        fill=white
    },
    missingnode/.style={
        draw=none,
        fill=none,
        inner sep=0pt,
        minimum size=0pt
    },
    edge from parent/.style={
        draw,
        line width=0.65pt
    }
]

\node[root] {$\frac{\pi}{4}$}
    child {
        node[internal] {$0$}
        child {
            node[missingnode] {}
            edge from parent[dashed]
        }
        child {
            node[leaf] {$\frac{\pi}{4}$}
        }
    }
    child {
        node[internal] {$0$}
        child {
            node[missingnode] {}
            edge from parent[dashed]
        }
        child {
            node[leaf] {$\frac{\pi}{4}$}
        }
    };

\end{tikzpicture}
}
\caption{$\ket{+}\ket{0}\ket{+}$}
\label{fig:separable_state_tree_I0I}
\end{subfigure}
\caption{Binary-tree representations of two product states. The tree topology reflects the support of each state, while the angle labels are consistent with the layerwise structure in \cref{eq:separable-layerwise-angles}.  }
\label{fig:state_comparison}
\end{figure}

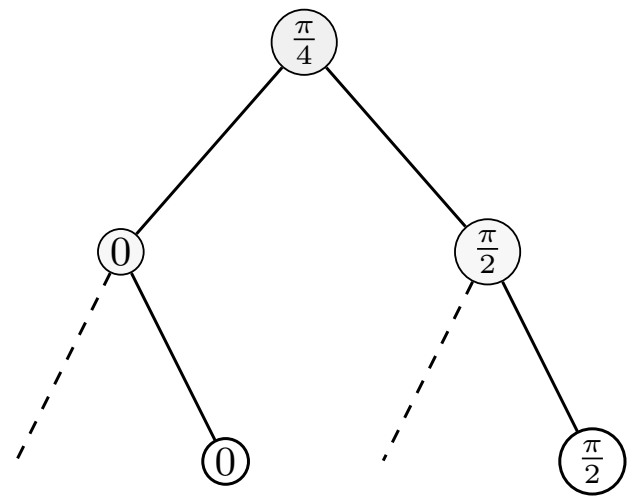
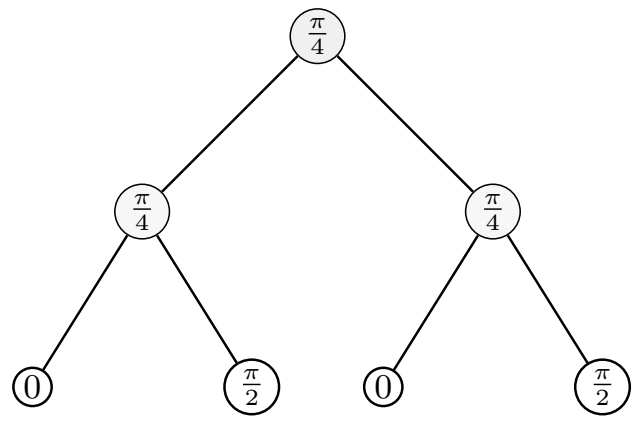
\begin{figure}[!t]
\centering
\begin{subfigure}[b]{0.5\columnwidth}
\centering
\resizebox{\linewidth}{!}{
\begin{tikzpicture}[
    grow=down,
    level distance=16mm,
    level 1/.style={sibling distance=28mm},
    level 2/.style={sibling distance=16mm},
    every node/.style={
        draw,
        circle,
        align=center,
        inner sep=0.9pt,
        font=\small
    },
    root/.style={
        fill=gray!12,
        font=\small\bfseries
    },
    internal/.style={
        fill=gray!6
    },
    leaf/.style={
        fill=white
    },
    missingnode/.style={
        draw=none,
        fill=none,
        inner sep=0pt,
        minimum size=0pt
    },
    edge from parent/.style={
        draw,
        line width=0.65pt
    }
]

\node[root] {$\frac{\pi}{4}$}
    child {node[internal] {$0$}
        child {
            node[missingnode] {}
            edge from parent[dashed]
        }
        child {node[leaf] {$0$}}
    }
    child {node[internal] {$\frac{\pi}{2}$}
        child {
            node[missingnode] {}
            edge from parent[dashed]
        }
        child {node[leaf] {$\frac{\pi}{2}$}}
    };

\end{tikzpicture}
}
\caption{$\left(\ket{000}+\ket{111}\right)/\sqrt{2}$}
\label{fig:entangled_ghz}
\end{subfigure}
\hfill
\begin{subfigure}[b]{0.5\columnwidth}
\centering
\resizebox{\linewidth}{!}{
\begin{tikzpicture}[
    grow=down,
    level distance=16mm,
    level 1/.style={sibling distance=32mm},
    level 2/.style={sibling distance=20mm},
    every node/.style={
        draw,
        circle,
        align=center,
        inner sep=0.9pt,
        font=\small
    },
    root/.style={
        fill=gray!12,
        font=\small\bfseries
    },
    internal/.style={
        fill=gray!6
    },
    leaf/.style={
        fill=white
    },
    edge from parent/.style={
        draw,
        line width=0.65pt
    }
]

\node[root] {$\frac{\pi}{4}$}
    child {node[internal] {$\frac{\pi}{4}$}
        child {node[leaf] {$0$}}
        child {node[leaf] {$\frac{\pi}{2}$}}
    }
    child {node[internal] {$\frac{\pi}{4}$}
        child {node[leaf] {$0$}}
        child {node[leaf] {$\frac{\pi}{2}$}}
    };

\end{tikzpicture}
}
\caption{$\ket{+}\otimes
\left(\ket{00}+\ket{11}\right)/\sqrt{2}$}
\label{fig:entangled_partial_ghz}
\end{subfigure}
\caption{Binary-tree representations of two states containing entanglement. Correlations are reflected by variations of the angle labels within at least one level; they should not be inferred solely from whether the tree topology is balanced.}
\label{fig:state_comparison_entangled}
\end{figure}

Conversely, variation of the angles within a level signals correlations between the qubits. 
In \cref{fig:state_comparison} we show two examples of trees associated to the separable states $\ket{001} $ and $\ket{+0+}$, whereas \cref{fig:state_comparison_entangled} shows two examples of entangled states.

\section{Simplification of a uniform full-weight multiplexer}
\label{sec:properties_of_the_multiplexer}

In \cref{eq:modulus_multiplexer_ladder_as_product} we show how to write the encoding circuit for the moduli (and analogously for the phases) as a product of multiplexers of increasing size $l \in \{0,1,\dots n-1\}$. 
In the following sections of the appendix we will focus on how to reduce the depth of these multiplexers by simplifying their controls. 
Without loss of generality we focus on those associated with the moduli, since the same reasoning applies to phases when one replaces the $C^lR_y$ gates with $C^lR_z$.

The most trivial decomposition for a multiplexer $\mathcal{M}_l$, acting on a total of $l+1$ qubits, with $l$ control qubits and depending on an angle vector $\vc{\theta}^{(l)}$ (reshaped into a tensor with $l$ two-dimensional indices) is a product over $2^l$ \acp{MCR}
\begin{align}
\label{eq:multiplexer_as_product}
\mathcal{M}_l(\vc{\theta})=&\prod_{\vc{j}[l]=(0,\ldots, 0)}^{(1,\ldots, 1)}  C^lR_y(\vc{j}[l],\theta_{\vc{j}[l]}),
\end{align}
where the \ac{MCR} for $R_y$ is defined as
\begin{equation}
\label{eq:multi_controlled_rotation}
C^lR_y(\vc{j}[l],\theta_{\vc{j}[l]}):= (\mathbf{P}_{\vc{j}[l]}\otimes R_y (\theta_{\vc{j}[l]})+(\I-\mathbf{P}_{\vc{j}[l]}) \otimes I)
\end{equation}
and $\mathbf{P}_{\vc{j}[l]}=\ket{\vc{j}[l]}\bra{\vc{j}[l]}$. To avoid notational clutter for explaining simplifications at a given level $l$, we drop let $\vc{j}[l]=\vc{j}$ for the rest of this section.  We refer to $\vc{j}$ as the \textit{control string}.
  
This formula defines the multiplexer as a product of $2^l$ fully controlled $R_y$ rotations with arguments given by the elements of the angle vector $\vc{\theta}^{(l)}$, as can be seen for $l=2$ in \cref{fig:multiplexer}. However, when multiplying such \acp{MCR} together, we can exploit the structure of the angle vector $\vc{\theta} ^{(l)}$ to discover and eliminate redundant controls of these rotations, thus reducing their number. In particular for \acp{MCR} sharing the same axis, and thus commuting, they can be pairwise eliminated as sketched in \cref{fig:boolean_simplification_colored}. The core problem which our algorithms must then tackle is how to determine of which groups of \acp{MCR} can be eliminated simultaneously in order to produce the smallest number of gate applications. However, crucially, the order of pairwise elimination can change the final gate count. Given that there are generally $2^l$ possible \acp{MCR} and thus
\begin{align}
\sim \binom{2^l}{2}=2^{l-1}\left(2^l-1\right)    
\end{align}
choices to be made in the early stages the problem becomes quickly challenging and it is in fact proven to be NP-Hard (see \cite{Bernasconi2013,ESOP_article} ).
In this work we introduce two different heuristic approaches to achieve this simplification: the \ac{BT-BE} and the \ac{BT-WHT} algorithms. They are in fact complementary and can be combined in what we call a hybrid algorithm. In this section of the appendix we introduce some basic properties of the multi-controlled rotations that will be later used to prove other results related to the two algorithms.

In the rest of the appendix we will make use of the concept of discrete support of a real vector
$v\in \mathbb{R}^n$defined as 
\begin{equation}
\label{eq:def_support}
    \mathcal{S}(v)=\{j \in [n] | v_j\neq 0\}.
\end{equation}

\subsection{Connection between control strings, projectors, index sets and Boolean functions}

\subsubsection{Decompositions of a uniform, fixed angle multiplexer}
\label{decomposition_uniform_equal_angle_mux}

Given a fixed angle multiplexer $ {\mathcal{M}}_l(\vc{\theta})$ with $ \vc{\theta} \in [0,2\pi]^{2^l} $ s.t.
\begin{equation}
\label{theta_vec_definition}
    \theta_j =\begin{cases}
    \theta \in [0,2\pi] & j \in \mathcal{S}\\
   0 & \text{else}
    \end{cases},
\end{equation}
its most direct decomposition is obtained by assigning one fully controlled \ac{MCR} to each element of its support  $\mathcal{S}(\vc{\theta})$ \cref{eq:def_support}.
Since all of these gates commute, the order is irrelevant and we can therefore collect them in a set, which we call the trivial decomposition and is defined as
\begin{equation}
    \mathcal{G}_{\mathrm{trivial}}(\vc{\theta})
    =
    \bigl\{\vc{j}
        \,\big|\,
        \theta_j \neq 0.
    \bigr\}
\end{equation}
More in general, a decomposition $\mathcal{G}\subseteq \{0,1,\I\}^l$ is defined as a set of generalized control strings of \ac{MCR} gates which decompose the multiplexer $ {\mathcal{M}}_l$ as 
\begin{equation}
    {\mathcal{M}}_l= \prod_{\vc{c} \in \mathcal{G}}C^lR(\vc{c},\theta). 
\end{equation}
For convenience we refer to it just as decomposition.
Each generalized control string $\vc{c} \in \{0,1,\I\}^l$ is associated to a set of binary-representation integers which correspond to the computational basis elements which the simplified multiplexer will affect. We denote this set of indices as $C_{\vc{c}}\subseteq\{0,1\}^l$. It is obtained by substituting every $\I$ symbol in the string with two entries in the set $C_{\vc{c}}$ containing $0$ and $1$ in its place. For example the index set associated to the control string $\vc{c}=(\I,0,\I)$ is $C_{\vc{c}}=\{000,001,100,101\}$. It follows naturally that the size of the set is exponential in the amount of $\I$ symbols in $\vc{c}$, 
\begin{equation}
    |C_{\vc{c}}|=2^{\#_{\I}(\vc{c})},
\end{equation}
where $\#_{\I}()$ counts the amount of $\I$ symbols in a given string. Therefore for each decomposition $\mathcal{G}\subseteq \{0,1,\I\}^l$ we can define a class $\mathcal{C}_{\mathcal{G}}\subseteq \mathcal{P}(\{0,1\}^l)$ containing all of the index sets corresponding to its generalized control strings, which we call a decomposition class (in this context a set of sets)
\begin{equation}
    \mathcal{C}_{\mathcal{G}}:=\{C_{\vc{c}} , \vc{c} \in \mathcal{G}\},
\end{equation}
where $\mathcal{P}(\{0,1\}^l)$ indicates the power set of $ \{0,1\}^l$.
It follows that every decomposition class $\mathcal{C}_\mathcal{G}$ represents a cover of the support set
\begin{equation}
  \mathcal{S}(\theta) =\bigcup_{\vc{c} \in \mathcal{G}} C_{\vc{c}}=\bigcup_{C \in \mathcal{C}_{\mathcal G}} C.
\end{equation}
The trivial decomposition $\mathcal{G}_\text{trivial}$ and the associated  trivial decomposition class $\mathcal{C}_\text{trivial}:=\mathcal{C}_{\mathcal{G}_\text{trivial}}$ represent a special case, with the generalized strings being fully controlled (and therefore Boolean) and the size of each index set being exactly one. 
Since any simplification merges or removes terms, but cannot introduce
more terms than this singleton decomposition, the trivial decomposition
is the largest possible decomposition. Moreover,
$\mathcal{C}_{\mathrm{trivial}}(\vc{\theta})$ is disjoint, hence it represents a partition of $\mathcal{S}(\theta)$.
\subsubsection{Map between generalized control strings and fully separable Boolean vectors}
\label{sec:map_control_string_boolean}
For the sake of later efficient numerical manipulation, we can map every generalized control string $\vc{c} \in \{0,1,\I\}^l$ to a binary vector  living in $\mathbb{F}_2^{2^l}\coloneqq\{0,1\}^{2^l}$ using the map 
\begin{equation}\label{c_string_to_binary}
\begin{split}
    \mathfrak{b} :&\{0,1,\mathbf{I}\}^l \to \mathbb{F}_2^{2^l},\\
    &\mathfrak{b}(\vc{c}) = \bigotimes_{i=0}^{l-1} \vc{T}(c_i),
\end{split}
\end{equation}
where 
\begin{equation}
\label{binary_2_vectors}
\vc{T}(c_i) =
    \begin{cases}
        (1,0) &\qquad \text{if } c_i=0\\
        (0,1) &\qquad \text{if } c_i=1\\
        (1,1) &\qquad \text{if } c_i=\mathbf{I},
    \end{cases}
\end{equation}
and $\otimes$ is the standard Kronecker outer product. This mapping has the benefit that the entries of the index set $C_{\vc{c}}$ are precisely the non-zero indices, that is to say, the support of the vector $\mathfrak{b}(\vc{c})$, written in binary. So, continuing with the previous example,
\begin{equation}
\mathfrak{b}(\I0\I) =(1,1)\otimes (1,0) \otimes(1,1)= (1,1,0,0,0,1,1,0).
\end{equation}
Because of their definition in terms of a single Kronecker outer product, we call such Boolean vectors \textit{fully separable}, in analogy to quantum states. By construction, such fully separable Boolean vectors correspond to single \ac{MCR} gates. Note that not every Boolean vector living in the $2^l$ space that $\mathfrak{b}$ maps onto is fully separable. Below, we will define how the multiplication of \ac{MCR} gates sharing the same angles correspond to compositions of these Boolean vectors using element-wise logical operations. Thus our problem of how to decompose multiplexers in terms of a minimal number of \ac{MCR} gates shifts onto a problem of constructing binary vectors from a minimal number of separable component vectors.
\subsubsection{The indicator vector associated to the multiplexer}
The support $\mathcal{S}(\theta)$ of a multiplexer with constant angle can be encoded by a Boolean vector
$\vc{b}(\vc{\theta})\in\{0,1\}^{2^l}$. This vector records which control
strings occur in the trivial circuit decomposition: its $j$th component
is equal to one precisely when the \ac{MCR} associated with control string
$j$ has a nonzero rotation angle. Thus,
\begin{equation}
\label{eq:boolean_indicator}
    b_j(\vc{\theta})
    =
    \begin{cases}
        1, & j \in \mathcal{S}(\vc{\theta}),\\
        0, & j \notin \mathcal{S}(\vc{\theta}).
    \end{cases}
\end{equation}
In other words, $\vc{b}(\vc{\theta})$ is the indicator vector of
$\mathcal{S}(\vc{\theta})$. It discards the numerical values of the rotation
angles and retains only the combinatorial structure of the circuit,
i.e. which fully controlled rotations are present.
In terms of the Boolean vectors, we can find a \ac{SOP} decomposition of $\vc{b}(\vc{\theta})$ \cite{gimpel1965reduction}
\begin{equation}
\vc{b}(\vc{\theta}) =\bigvee_{\vc{c} \in \mathcal{G}} \mathfrak{b}(\vc{c}),
\end{equation}
where $\vee$ is the element wise logical OR operator acting on the Boolean vector $\mathfrak{b}(c)$, which is a fully separable binary vector that admits a decomposition of the form in \cref{c_string_to_binary} for some $\vc{c} \in \{0,1,\I\}^l$
. This problem is known to be NP-hard, but a good (even if often sub optimal) solution can be efficiently found using the well known heuristic based algorithm Espresso \cite{brayton1984espresso}. 

As we will see in the next section, the simplification of the multiplexer involves the solution of two problems which are different but closely related to \ac{SOP}, namely the \ac{DSOP} and the \ac{ESOP}. Both of them are known to be even harder than \ac{SOP} due to their additional constraints \cite{Bernasconi2013,ESOP_article}.

The geometric interpretation of all of these maps and of the related simplification process is explored in more detail in \cref{sec:appendix geometric interp}.

\subsubsection{Alternative representations for the projectors}
\label{sec:control_strings_to_index_sets}
  In contrast to equation \cref{eq:multi_controlled_rotation}, using the generalized control strings one can represent any partially controlled $R_y$, which is written as
\begin{equation}
    \begin{split}
        C^lR_y(\vc{c}, \theta) &= (\mathbf{P}_{\vc{c}}\otimes R_y (\theta)+(\I-\mathbf{P}_{\vc{c}}) \otimes I).
    \end{split}
\end{equation}
Here the projector $\mathbf{P}_{\vc{c}}$ has the identity anywhere a $I$ character appears in $\vc{c}$, for example
\begin{equation}
\label{projector_def}
    \mathbf{P}_{0\I1}= \mathbf{P}_0 \otimes \mathbf{I} \otimes \mathbf{P}_1 =\mathbf{P}_{001}+\mathbf{P}_{011}.
\end{equation}

Following the example from above we have the three equivalent representations
\begin{equation}
\begin{aligned}
    P_{0\I1}=&P_{\{001,011\}}=P_{(0,1,0,1,0,0,0,0)}\\
    =&\ket{001}\bra{001}+\ket{011}\bra{011}.
\end{aligned}
\end{equation}
We hence have three different and equivalent ways to indicate a given projector
\begin{itemize}
    \item in terms of the generalized control string $\vc{c} \in \{0,1,\I\}$ as $P_{\vc{c} }$,
     \item using the associated index set $C_{ \vc{c}} \in \{0,1\}^l$ as $P_{C }$ (see \cref{decomposition_uniform_equal_angle_mux}),
      \item in terms of the indicator Boolean vector $\mathfrak{b}_{ \vc{c}} \in \{0,1\}^{2^l}$ as $P_{\mathfrak{b} }$ (see \cref{sec:map_control_string_boolean}).
    
\end{itemize}
The first representation has the advantage of being directly connected to the \acp{MCR} gates, the second one allows us to describe the simplification of the multiplexer in formalism of set theory and finally the third one is particularly suitable for a fast implementation on the computer in terms of basic binary operations.

\subsection{Properties of a product of controlled rotations} 
\subsubsection{Product rule}
The product rule for uniform \acp{MCR} with binary control strings $\vc{b}',\vc{b} \in \{0,1\}^l$ of length $l$ and $\vc{b}\neq\vc{b}'$ is
\begin{align}
\label{eq:product_rule_MCR}
C^l& R_y(\vc{b},\theta)C^l R_y(\vc{b}',\theta') \nonumber \\
 =& \mathbf{P}_{\vc{b}}\otimes R_y (\theta)+\mathbf{P}_{\vc{b}'}\otimes R_y (\theta')+ (\I- (\mathbf{P}_{\vc{b}}+\mathbf{P}_{\vc{b}'}))\otimes \I.
\end{align}
For $\theta=\theta'$ we have 
\begin{align}
\label{eq:product_rule_MCR_constant_arguments}
&C^l R_y({\vc{b}},\theta)C^l R_y({\vc{b}'},\theta) \nonumber \\
=& (\mathbf{P}_{\vc{b}}+\mathbf{P}_{\vc{b}'})\otimes R_y (\theta)+ (\I- (\mathbf{P}_{\vc{b}}+\mathbf{P}_{\vc{b}'}))\otimes \I.
\end{align}
In other words, whenever the angles agree, the projector of the product is the sum of the projectors. Thus, whenever two control strings  differ in exactly one position and agree everywhere else, this sum reduces to the identity operator that position. In the formalism introduced in \cite{Pagni2025}, this is indicated by an $\I$ symbol in a \textit{generalized} control string, which now belongs to the set $\vc{c}\in \{0,1,\I\}^l$. In terms of circuit diagrams this is illustrated in \cref{fig:boolean_simplification_colored}. 

In the following subsections, we will consider a special kind of uniform multiplexer, where the angle of the rotations is fixed to $\theta \in [0,2\pi]$.

\subsubsection{Generalized product rule of uniform controlled gates}
An important consequence of the introduction of the $\I$ symbol is that it makes projectors like the one in the example \cref{projector_def} not orthogonal anymore, leading to the following  multiplication rule across the three equivalent representations 
\begin{align}
\label{eq:generalized_product_rule_control_strings}
P_{\vc{c}} P_{\vc{c}'}
&=\bigotimes_{m \in [l]}P_{c_m}P_{c'_m}\\
\label{eq:generalized_product_rule_control_strings2}
P_{C}P_{ C'}&=P_{C\cap C'}\\
\label{eq:generalized_product_rule_control_strings3}
P_{\vc{b}} P_{\vc{b}'}&=P_{\vc{b} \land \vc{b}'}.
\end{align}
For the representation using index sets, the product corresponds to a set intersection operation while in terms of the binary vectors $\mathfrak{b}({\vc{c}})$, $\mathfrak{b}({\vc{c}}') \in [0,1]^{2^l}$ associated to the control strings $\vc{c},\vc{c}'\in \{0,1,\I\}^l$ by \cref{c_string_to_binary}, the intersection corresponds to the AND operation. We further find
\begin{equation}
\begin{aligned}
P_a P_b=&0 \quad &a\neq b ,a,b \in \{0,1\},\\
\text{and }P_a P_{\I}=&  P_{\I}P_a=P_a, &\forall a \in \{0,1\} 
\label{eq:product_rule_single_qubit_projectors}
\end{aligned}
\end{equation}
for the product of single qubit projectors on generalized control strings.
The generalized product rule for two multi-controlled rotations around the same axis is therefore 
\begin{equation}
\begin{aligned}
&C^lR(\vc{c},\theta)  C^lR(\vc{c}',\theta')\\
=&\hphantom{{}+{}}(P_{\vc{c}}\otimes R(\theta)+(\I - P_{\vc{c}})\otimes \I) (P_{\vc{c}'}\otimes R(\theta') \\
&+(\I - P_{\vc{c}'})\otimes \I) \\
=&\hphantom{{}+{}}P_{\vc{c}}P_{\vc{c}'}\left[R(\theta+\theta')-R(\theta)-R(\theta')+\I\right]\\
&+P_{\vc{c}}\otimes R(\theta)+P_{\vc{c}'}\otimes R(\theta')\\
&+(\I-P_{\vc{c}}-P_{\vc{c}'})\otimes \I,
\end{aligned}
\end{equation}
which when $\theta = \theta'$ simplifies to 
\begin{equation}
\begin{aligned}
\label{eq:generalized_multi_controlled_gate_product_rule_same_angle}
C^lR(\vc{c},\theta)  C^lR(\vc{c}',\theta) 
=&P_{\vc{c}}P_{\vc{c}'}\left[R(2\theta)-2R(\theta)+\I\right]\\
&+\left(P_{\vc{c}}+P_{\vc{c}'}\right)\otimes R(\theta)\\
&+\left[\I-(P_{\vc{c}}+P_{\vc{c}'})\right]\otimes \I.
\end{aligned}
\end{equation}
\Cref{eq:generalized_multi_controlled_gate_product_rule_same_angle} reduces to \cref{eq:product_rule_MCR_constant_arguments} whenever the index sets associated to the control strings $\vc{c}$ and $\vc{c}'$ do not overlap.

We can write the rule for the sum of projectors expressed in terms of the index sets by splitting it into the disjoint and the overlapping part 
\begin{align}
    P_{C}+P_{C'}&=P_{(C\cup C') \setminus (C\cap C') }+2P_{C\cap C' }\nonumber\\
    &=P_{C\triangle C' }+2P_{C\cap C' },
    \label{sum_of_projectors_sets}
\end{align}
where we introduced the symmetric difference operator $C\triangle C' =(C\cup C') \setminus( C\cap C') $.
This, in terms of the Boolean indicators can be written as 
\begin{equation}
    P_{\vc{b}}+P_{\vc{b}'}=P_{\vc{b} \oplus \vc{b}'}+2P_{\vc{b}\land \vc{b}'},
\end{equation}
where $\oplus$ denotes entrywise addition modulo 2. Using \cref{sum_of_projectors_sets} we can now write 
\begin{align}
C^lR(\vc{c},\theta)  C^lR(\vc{c}',\theta)
=&\hphantom{{}+{}}P_{C}P_{C'}\left[R^2(\theta)-2R(\theta)+\I\right]\nonumber \\
&+\left(P_{C}+P_{C'}\right)\otimes R(\theta) \nonumber\\
&+\left[\I-(P_{C}+P_{C'})\right]\otimes \I\nonumber\\
=&\hphantom{{}+{}}P_{C \cap C'}\left[R^2(\theta)-2R(\theta)+\I\right]\nonumber \\
&+\left(P_{C\triangle C' }+ 2P_{C\cap C'}\right)\otimes R(\theta)\nonumber\\
&+\left[\I-(P_{C\triangle C' }+ 2P_{C\cap C'})\right]\otimes \I\nonumber\\
=&\hphantom{{}+{}}\left(P_{C\triangle C' }\right)R(\theta)\nonumber\\
&+\left[\I-(P_{C\triangle C'  })\right]\otimes \I\nonumber\\
&+\left(P_{C\cap C'}\right)\otimes\left( R^2(\theta)-\I\right). \label{eq:split_generalized_multi_controlled_gate_product_rule_same_angle}
\end{align}
In terms of the associated Boolean vectors $\vc{b}=\vc{b}_{\vc{c}},\vc{b}'=\vc{b}_{\vc{c}'},$ the rule becomes
\begin{equation}
\begin{aligned}
C^lR(\vc{c},\theta)  C^lR(\vc{c}',\theta)=\I &+P_{\vc{b}\oplus \vc{b}' }\otimes\left( R-\I\right)\\
&+P_{\vc{b}\land \vc{b}' }\otimes\left( R^2-\I\right)\label{eq:split_generalized_multi_controlled_gate_product_rule_same_angle_boolean_vectors},
\end{aligned}
\end{equation}
where $R=R(\theta)$.
Applying it repeatedly over a product with $T\geq|\mathcal{G}(\vc{\theta})|\geq 2$ terms with the same angle we have the following general formula.

\begin{lemma}[Product rule for equal-angle multi-controlled rotations]
\label{lmm:binary_manyterms_MCR_product_rule_same_angle}
Let
$\mathcal{G}(\vc{\theta})=\{\vc{c}_1(\vc{\theta}),\dots,\vc{c}_T(\vc{\theta})\}$, where $\vc{\theta} $ is of the kind in \cref{theta_vec_definition},
be a collection of $T=|\mathcal{G}(\vc{\theta})|$ control strings, and consider
\begin{equation}
\label{eq:def_UT}
    U_T
    :=
    \prod_{\vc{c}\in\mathcal{G}(\theta)}
    C^lR(\vc{c},\theta).
\end{equation}
Let $\vc{b}_j$ denote the Boolean indicator vector associated with the
control string $\vc{c}_j$. Define
\begin{equation}
    \vc{w}_r
    :=
    \bigoplus_{\substack{
        A\subseteq\{1,\dots,T\}\\
        |A|=2^r
    }}
    \bigwedge_{j\in A}\vc{b}_j,
    \qquad
    r=0,\dots,\lceil\log_2 T\rceil -1.
\end{equation}

where the AND operation $\land$ is elementwise. In particular,
\begin{equation}
    \vc{w}_0
    =
    \bigoplus_{j=1}^T \vc{b}_j.
\end{equation}
Then
\begin{align}
    U_T
    &=
    \left(
        \I + P_{\vc{w}_0}\otimes\bigl(R(\theta)-\I\bigr)
    \right)\times
    \nonumber\\
    &\quad\times
    \prod_{r=1}^{\lceil\log_2 T\rceil}
    \left(
        \I
        +
        \sum_{l=0}^{2^{r-1}-1}
        P_{\vc{w}_r}
        \otimes
        R^{2l}(\theta)
        \bigl(R^2(\theta)-\I\bigr)
    \right).
    \label{eq:binary_manyterms_MCR_product_rule_same_angle_final}
\end{align}

Here $P_{\vc{w}_r}$ denotes the projector associated with the Boolean
indicator $\vc{w}_r$ (see \cref{sec:control_strings_to_index_sets}).

\end{lemma}

\begin{proof}
Given a decomposition $\mathcal{G}=\{\vc{c}_1,\dots,\vc{c}_{T}\} $ of the multiplexer $ {M^{(l)}}$ of size $|\mathcal{G}|=T$ and associated Boolean indicators $B_\mathcal{G}=\{\vc{b}_1=\vc{b}_{\vc{c}_1}, \dots,\vc{b}_T:=\vc{b}_{\vc{c}_T}\} $ we define
\begin{equation}
    w_\mathcal{G}(k)=\sum_{j=1}^T b_{j,k},\quad k \in [2^l],
\end{equation} 
which measures the amount of control strings in $\mathcal{G}$ that cover $k$ (their associated indicators have a one at position $k$). This is equivalent to the time the fully controlled string $(k_1,\dots,k_l)$ appears in the version of $\vc{c}_j $ where we expand every $\I$ symbol.

We observe that 
\begin{equation}
    C^lR(\vc{c}_j,\theta ) \ket{k}\otimes \ket{\psi} = \ket{k}\otimes R^{b_{j,k}} \ket{\psi},
    \label{U_as_power_of_rotations}
\end{equation}
where $b_{j,k}$ is the $k$-th entry 
Applying $U_T$ to a state and substituting \cref{U_as_power_of_rotations} into its definition \cref{eq:def_UT} we get
\begin{align}
    &U_T  \ket{k}\otimes \ket{\psi} =\prod_{j=1}^T C^lR(\vc{c}_j,\theta ) \ket{k}\otimes \ket{\psi}\nonumber \\
    &= \ket{k}\otimes \prod_{j=1}^T R^{b_{j,k}} \ket{\psi}=\ket{k}\otimes  R^{\sum_{j=1}^Tb_{j,k}} \ket{\psi}\nonumber \\
    &=\ket{k}\otimes  R^{w_\mathcal{G}(k)} \ket{\psi}.
    \label{eq:first_equation_for_U_T}
\end{align}
If we consider the projector 
\begin{equation}
    \Pi_s :=\sum_{\substack{k\in\{0,1\}^l \\ w_\mathcal{G}(k)=s}}\ket{k}\bra{k}
\end{equation}
we use the completeness of a basis of projectors and the previous result to write 
\begin{equation}
    U_T=\sum_{s=0}^T \Pi_s \otimes R^s.
\end{equation}
Let $w_r(k) $ be the $r$-th binary digit of $w_{\mathcal{G}}(k)$.
In the following we will show that it can be written as
\begin{equation}
   w_r(k)=\bigoplus_{\substack{A \subseteq\{1,\dots T\}\\ |A|=2^r}} \bigwedge_{j \in A}  b_{j,k}
   \label{digit_of_w_as_boolean_predicate}
\end{equation}
in terms of the Boolean indicators associated to the generalized control strings in $\mathcal{G}$.
This can be done by considering the matrix $b_{j,k}$ with $j\in [T],k \in [2^l]$, whose rows are the Boolean indicator vectors, and defining the support $\mathcal{S}_k$ of each column $k$ as in \cref{eq:def_support}.
This immediately gives the relation 
\begin{equation}
\label{abs_Sk}
    |\mathcal{S}_k|= w(k).
\end{equation}
Using the fact that
\begin{equation*}
\bigwedge_{j \in A\subseteq \mathcal{S}_k}  b_{j,k} =1\text{ and }\bigwedge_{j \in A\nsubseteq \mathcal{S}_k}  b_{j,k} =0,    
\end{equation*}
we can rewrite the RHS of \cref{digit_of_w_as_boolean_predicate} using \cref{abs_Sk} as
\begin{equation}
\begin{split}
\bigoplus_{\substack{A \subseteq\{1,\dots T\}\\ |A|=2^r}} &\bigwedge_{j \in A}  b_{j,k} = \bigoplus_{\substack{A \subseteq \mathcal{S}_k\\ |A|=2^r}} 1=\binom{w(k)}{2^r} \mod 2.
   \end{split}
\end{equation}
Using Lucas' theorem \cite{Lucas1878} we can write for the binary digits 
\begin{equation}
   w_r(k)=\binom{w(k)}{2^r}\bmod 2 
\end{equation} 
and conclude that \cref{digit_of_w_as_boolean_predicate} is correct. We associate a projector
\begin{equation}
\label{P_omega_r}
P_{\vc{w}_r}
:=
\mathrm{diag}(\vc{w}_r). 
\end{equation}
with the Boolean vector $\vc{w}_r \in\{0,1\}^{2^l} $ with components $w_r(k)$ for $k=0,1,...,2^l-1$.

The next step it to rewrite $U_T$ as a product. In order to do so we first define the projector controlled rotation
\begin{equation}
    C R(P_{\vc{w}_r},2^r \theta)
    =\I + P_{\vc{w}_r}\otimes (R^{2^r}(\theta)-\I),\quad r \in \{0,\dots,L\},
\end{equation}
where we used $L:=\lceil \log_2(T)\rceil-1$ and \cref{P_omega_r}.

This operator acts on states as 
\begin{equation}
     C R(P_{\vc{w}_r},2^r \theta) \ket{k}\otimes \ket{\psi}= \ket{k}\otimes R^{2^r w_r(k)}\ket{\psi},
\end{equation}
therefore taking a product over the $r$ values we obtain 
\begin{align}
   & \prod_{r=0}^{L} C R(P_{\vc{w}_r},2^r \theta) \ket{k}\otimes \ket{\psi}= \ket{k}\otimes\prod_{r=0}^{L} R^{2^r w_r(k)}\ket{\psi}\nonumber \\
   &=\ket{k}\otimes R^{\sum_{r=0}^{L}2^r w_r(k)}\ket{\psi}=\ket{k}\otimes R^{w_\mathcal{G}(k)}\ket{\psi},
\end{align}
where again we used the properties of uniform rotations.

By comparing this equation with \cref{eq:first_equation_for_U_T} we can write \begin{align}
&U_T=\prod_{r=0}^L  C R(P_{\vc{w}_r},2^r \theta)\\
&= \prod_{r=0}^{\lceil \log_2(T)-1\rceil} \left[\I + P_{\vc{w}_r}\otimes (R^{2^r}-\I)\right]\nonumber \\
    &=\left[\I + P_{\vc{w}_0}\otimes (R-\I)\right]\prod_{r=1}^{\lceil \log_2(T)\rceil -1} \left[\I + P_{\vc{w}_r}\otimes (R^{2^r}-\I)\right].
\end{align}
We finally use the series identity 
\begin{equation}
    (R^{2^r}-\I)=( R^2-\I)\sum_{m=0}^{2^{r-1}-1}R^{2m}
\end{equation} 
to conclude the proof and obtain 
\begin{align}
U_T=&\prod_{r=0}^L C R(P_{\vc{w}_r},2^r \theta)\\
=& \prod_{r=0}^{\lceil \log_2(T)\rceil -1} \left[\I + P_{\vc{w}_r}\otimes (R^{2^r}-\I)\right]\nonumber \\
    =&\left(\I + P_{\vc{w}_0}\otimes (R-\I)\right)\times\nonumber \\
    &\times\prod_{r=1}^{\lceil \log_2(T)\rceil -1} \left(\I + \sum_{m=0}^{2^{r-1}-1} P_{\vc{w}_r}\otimes R^{2m}(R^2-\I)\right).
\end{align}
\end{proof}

\subsection{Simplification of the controls as a Boolean reduction problem}
\begin{figure*}[!t]
    \centering

     \begin{subfigure}[t]{0.48\textwidth}
        \centering
        \includegraphics[width=\linewidth]{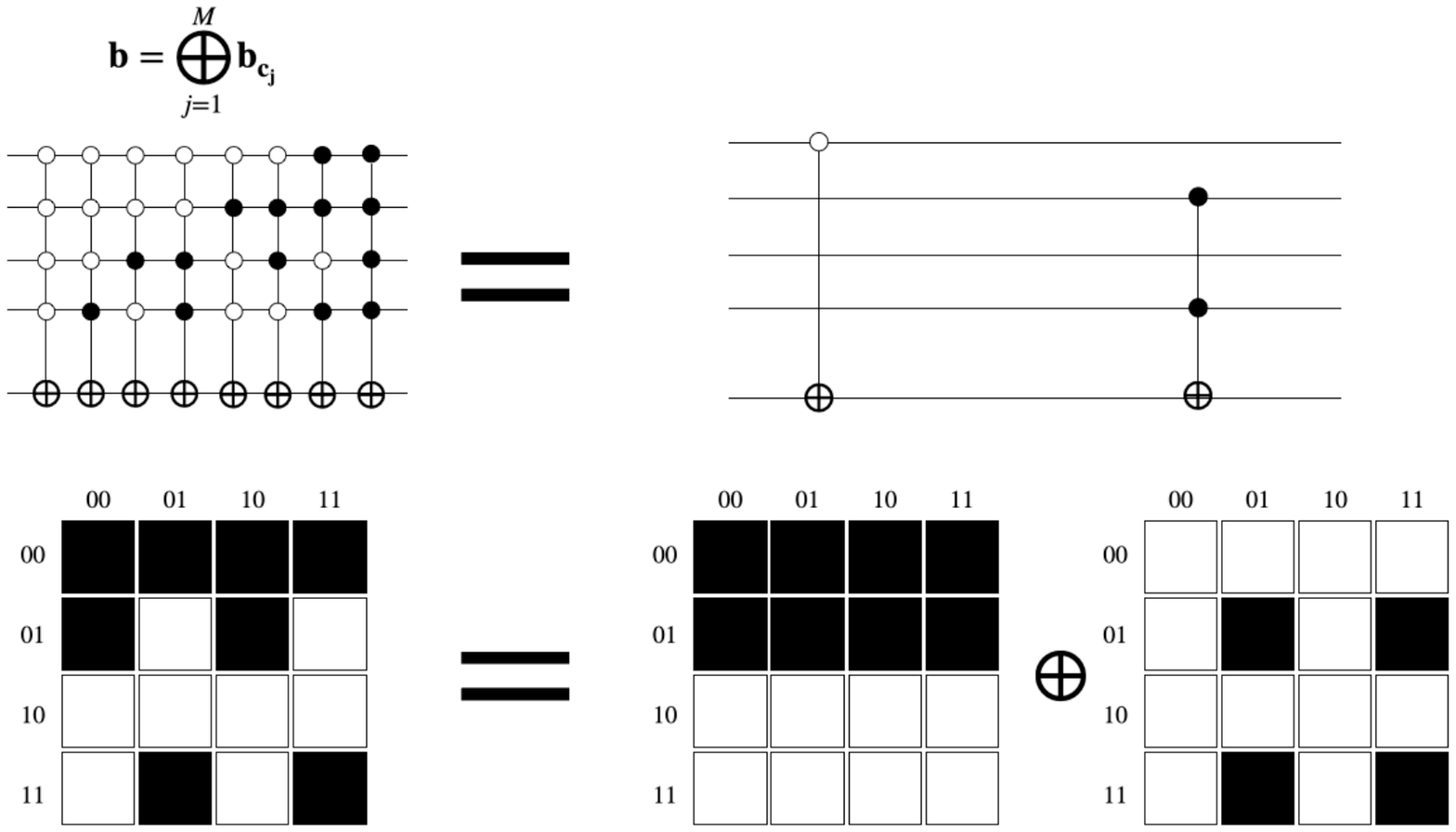}
        \caption{ESOP decomposition of the vector $\vc{b}$ for multi-controlled operations that are self-inverse, such as MCX. While the initial vector is associated to the trivial circuit on the left, the decomposition provides an alternative simplified circuit (on the right) that is equivalent to the the trivial one (on the left).}
        \label{fig:esop_example}
    \end{subfigure}
    \hfill
    \begin{subfigure}[t]{0.48\textwidth}
        \centering
        \includegraphics[width=\linewidth]{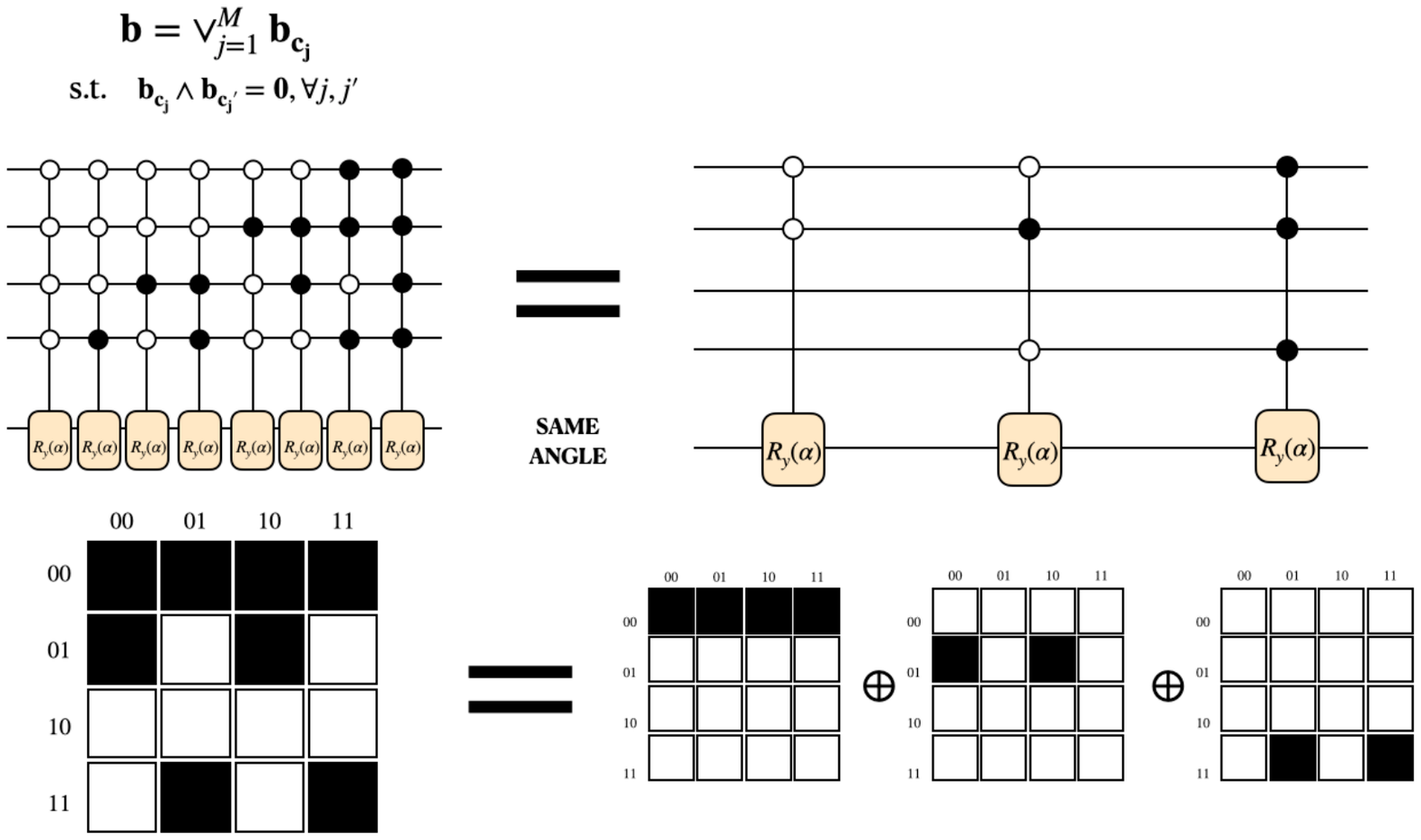}
        \caption{DSOP decomposition of the vector $\vc{b}$ for multi-controlled operations that are not self-inverse. While the initial vector is associated to the trivial circuit on the left, the decomposition provides an alternative simplified circuit (on the right) that is equivalent to the trivial one (on the left).}
        \label{fig:dsop_example}
    \end{subfigure}
    \caption{ Example of a DSOP and an ESOP decompositions of the unsimplied circuit on the left. The decomposition, ESOP for self-inverse gates and DSOP for the others, defines the equivalent simplified circuit on the right. }
    \label{fig:DSOP_ESOP}
\end{figure*}

By applying the product rule for constant arguments \cref{eq:binary_manyterms_MCR_product_rule_same_angle_final}, we can decompose a uniform multiplexer whose non zero angles are constant. There are two cases when the formula simplifies and we can reduce the problem of finding a small decomposition $\mathcal{G}(\vc{\theta})$ to a well known Boolean simplification problem that we can efficiently perform on a computer: 
\begin{itemize}
    \item When the operator is self-inverse $R^2=\I$.
    \item When we restrict our search for the best decomposition to those whose associated sets are disjoint.
\end{itemize}
These result is summarized in the following theorem:
\begin{theorem}
Given a uniform fixed-angle multiplexer $U_T$, defined as in
\cref{lmm:binary_manyterms_MCR_product_rule_same_angle}, and a decomposition
$
    \mathcal{G}(\vc\theta)
    =
    \{\vc{c}_1(\vc\theta),\dots,\vc{c}_T(\vc\theta)\}
$
of control strings, suppose that either
\begin{enumerate}
    \item[\textup{(a)}] $R^2(\theta)=R(2\theta)=\I$, which corresponds to
    $
        \theta = 4k\pi,
        \qquad k\in\mathbb{N},
    $
    as, for example, for multi-controlled Pauli-$X$ gates; or

    \item[\textup{(b)}] the Boolean indicator vectors $\vc{b}_j$ associated with the
    control strings $\vc{c}_j$ are supported on pairwise disjoint index sets.
\end{enumerate}
Then the multiplexer can be written as
\begin{equation}
    U_L
    =
    \I
    +
    P_{\vc{w}_0}\otimes\bigl(R(\theta)-\I\bigr).
\end{equation}
In case~\textup{(a)}, we have
\begin{equation}
    \vc{w}_0
    =
    \bigoplus_{j=1}^{T}\vc{b}_j,
\end{equation}
and therefore $\{\vc{b}_j\}_{j=1}^{T}$ constitutes a valid ESOP of
$\vc{w}_0$. In case~\textup{(b)}, we have
\begin{equation}
    \vc{w}_0
    =
    \bigvee_{j=1}^{T}\vc{b}_j,
\end{equation}
and $\{\vc{b}_j\}_{j=1}^{T}$ constitutes a valid DSOP of $\vc{w}_0$.
\end{theorem}
\begin{proof}
Considering \cref{lmm:binary_manyterms_MCR_product_rule_same_angle} and in particular \cref{eq:binary_manyterms_MCR_product_rule_same_angle_final}, the second factor in the product associated to $r\geq1$
\begin{equation}
     \prod_{r=1}^{\lceil\log_2 T\rceil}
    \left(
        \I
        +
        \sum_{l=0}^{2^{r-1}-1}
        P_{\vc{w}_r}
        \otimes
        R^{2l}(\theta)
        \bigl(R^2(\theta)-\I\bigr)
    \right)
\end{equation} 
becomes identity for two distinct reasons:
\begin{itemize}
    \item[(a)] in the self-inverse case we have $R^2(\theta)-I=0$
    \item[(b)] in the case of disjoint $\vc{b}_j $ boolean indicators we have that $\bigwedge_{j\in A}\vc{b}_j$ vanishes for every $A$ of size $|A|=2^r\geq 1$ and hence $r\geq 1$. This makes the projector $P_{\vc{w}_r}$ vanish as well. 
\end{itemize}
Hence, in both cases we have \begin{equation}
    U_L
    =
    \I
    +
    P_{\vc{w}_0}\otimes\bigl(R(\theta)-\I\bigr).
\end{equation}
In case $\textup{(a)}$ by definition we have 
\begin{equation}
    \vc{w}_0
    =
    \bigoplus_{j=1}^{T}\vc{b}_j,
\end{equation}
while for  $\textup{(b)}$, being the boolean vectors disjoint, the XOR reduces to OR giving
\begin{equation}
    \vc{w}_0
    =
    \bigvee_{j=1}^{T}\vc{b}_j,
\end{equation}
\end{proof}
\subsubsection{Simplification of a multiplexer of \ac{MCX} gates} 
As explained above and in more detail in numerous works, \cite{Fazel2007,Giulia2019,Matthew2019,Zheng2025,Pagni2025}, whenever the multi-controlled rotations are self-inverse, such as for example when dealing with multi-controlled X gates $C^l X$, the problem of finding a small decomposition $\mathcal{G}( \theta)$ in terms of control strings and amount of gates reduces to obtaining a small \ac{ESOP} decomposition of the Boolean vector $\vc{b}(\vc{\theta})$. The latter consists of the smallest set of fully separable Boolean vectors $\vc{b}_{\vc{c}}$, corresponding to each control string in $\mathcal{G}(\vc{\theta})$ via the map \cref{c_string_to_binary}, whose XOR gives 
\begin{equation}
    \vc{b} = \bigoplus_{\vc{c}\in \mathcal{G}} \mathfrak{b}(\vc{c}).
\end{equation}
This can be solved using the algorithm provided together with the paper or, alternatively, using another heuristic method presented in \cite{ESOP_article}.
\subsubsection{Simplification of uniform a multiplexer with constant arbitrary angle}
\label{app:multiplexer_simplification_DSOP}
Whenever the non zero arguments of the rotations are constant but the gates are not self-inverse, we can restrict ourselves to decompositions $\mathcal{G}(\vc{\theta})$ that represent partitions (their elements are disjoint) of $S(\vc{\theta})$. This way the complicated term in \cref{eq:binary_manyterms_MCR_product_rule_same_angle_final} vanishes because every AND operation gives zero. In this case the formula relates the problem of finding a good decomposition to finding a \ac{DSOP} decomposition of the indicator Boolean vector $\vc{b}( \theta)$. Given that all the sets are disjoint, the XOR and OR operation coincide
\begin{equation}
    \vc{b} = \bigoplus_{\vc{c}\in \mathcal{G}_\textbf{Partition}} \vc{b}_{\vc{c}}=\bigvee_{\vc{c}\in \mathcal{G}_\textbf{Partition}} \vc{b}_{\vc{c}}.
\end{equation}
The problem can also be restated in term of  sets as the following: given a Boolean function 
\begin{equation}
    b:\{0,1\}^n \to \{0,1\}
\end{equation}
with support set
\begin{equation}
\mathcal{S}=\{j \in \{0,1\}^n | b_j=1\}
\end{equation}
we want to 
find the decomposition $\mathcal{G}(\theta) \subseteq \{0,1,\I\}^l$ with the smallest amount $M=|\mathcal{G}|$ of generalized control strings $\vc{c}_0, \vc{c}_1,\dots ,\vc{c}_{M-1}$  such that the class containing their associated index sets $\mathcal{C}_\mathcal{G} =\{C_{\vc{c}} , \vc{c} \in \mathcal{G}\}$ provides a partition $\mathcal{G}$ of the support $S$ i.e.

\begin{equation}
\label{eq:partition_of_the_support}
    \bigsqcup_{\vc{c} \in \mathcal{G}}^M C_{\vc{c}}=  S,
\end{equation}
where $\bigsqcup$ stands for disjoint set union. 
In this work, both in the self-inverse and in the disjoint cases, the Boolean function we start from is always the one associated to the trivial decomposition of the sum of projectors, where every term is fully controlled. We can then perform \ac{ESOP} or \ac{DSOP} in order to optimize its size and consequently simplify the multiplexer.
Both the \ac{DSOP} and the \ac{ESOP} problems are known to be \ac{NP}-hard and they are considered an even harder version of the Boolean cube decomposition or \ac{SOP} \cite{gimpel1965reduction},
for which a good solution can be efficiently found using the well known heuristic based algorithm Espresso \cite{brayton1984espresso}.

A good (sub-optimal) \ac{DSOP} decomposition can be found using the algorithm in \cite{Pagni2025} for \ac{ESOP} by adding some constraints to its greedy search, that is considering only full cubes \cite{Pagni2025}, where a cube is the term we use to indicate the geometric counterpart of a generalized control string, as explained in detail in \cref{sec:appendix geometric interp}. On the benchmark instances reported in \cite{Bernasconi2013}, this approach produces decompositions that are comparable to, and in some cases slightly shorter than, those obtained using the \ac{DSOP}-4 algorithm. However, the method proposed in \cite{Balasubramanian} consistently produces shorter decompositions while requiring less computation time. For this reason, we adopt it as our preferred \ac{DSOP} synthesis method. 
A summary of the different scenarios is given in \cref{fig:DSOP_ESOP}.

\section{The Binary-Tree Walsh--Hadamard Transform Algorithm}
\label{app:BT-WHT}

In this appendix, we explain in detail how to exploit the
Walsh--Hadamard transform to find a possibly shorter decomposition of a given
multiplexer circuit into a sequence of one- and two-qubit gates.

\subsection{Comparison with previous methods}

A well-known method for decomposing a uniform multiplexer was given in
Refs.~\cite{Mottonen,Shende_2006}. The authors present a procedure based
on a Gray-code ordering and an averaging-and-differencing recursion
that is algebraically equivalent to a \ac{WHT}. For an
$l$-controlled multiplexer, the method gives an ancilla-free
decomposition containing $2^l$ single-qubit $R_y$ rotations and $2^l$
CX gates. Additional cancellations may be possible when neighboring
multiplexers are compiled jointly.

Here, we introduce a different approach that is explicitly based on
the \ac{WHT} and uses one additional clean qubit. This method permits a controlled truncation of
small spectral components, thereby trading circuit depth for accuracy.
In \cite{Welch_2014} a very similar approach is introduced and extended in \cite{Zhang2024}, however, there are some differences both in the implementation and in the optimization of the method, such as the presence of one ancilla. 
\subsection{Connection between the WHT of the angles and the decomposition}

In the following, we focus on the multiplexers associated
with the moduli, $ {\mathcal M}_l$ for
$l\in\{0,\ldots,n-1\}$. The same reasoning applies to the phases after
replacing the $C^lR_y$ gates with $C^lR_z$ gates.

In our formalism, the angles of the multiplexer ladder in
\cref{fig:multiplexer_ladder} are computed as nodes of a binary-tree
graph, whose levels correspond to the different multiplexers. For
example, the angles of the multiplexer acting on three qubits are
represented by one of the levels of the tree in
\cref{fig:angles_tree}.

Given a multiplexer acting on $l$ control qubits and one target qubit,
whose angles are given by the $l$-th layer of the binary tree, the
first step consists of completing any structurally undefined nodes. 
Such missing values occur only on inactive branches carrying zero
probability mass and therefore do not affect the state prepared by the
multiplexer. They can consequently be chosen so as to favor a
compressible Walsh representation, in such a way that the sparsity of the Walsh spectrum does not change. We stress that this procedure is only applied to the dense version of the algorithm, as in the TT formulation for large levels we directly apply the Hadamard MPO operator to the MPS of the layer, which we learn using TCI.

Whenever a layer contains such missing values,
\cref{alg:walsh-interpolation} fills them using a dyadic-replication
heuristic designed to preserve Walsh sparsity. This intuitively consists of copying existing values of the angle vector into the missing positions, This is done according to the structure of the slices we obtain when we reshape it as a $(2,2,\dots,2)$ tensor and fix one or more indices. For example, the vector

\begin{equation}
\vc\theta_{\mathrm{init}}
=
(a,\texttt{None},b,\texttt{None})
\end{equation}
is completed as

\begin{equation}
\vc\theta_{\mathrm{interp}}
=
(a,a,b,b).
\end{equation}

\begin{figure}[tbph]
\begin{algorithm}[H]
\caption{Recursive Walsh interpolation}
\label{alg:walsh-interpolation}
\begin{algorithmic}[1]

\REQUIRE
$\mathbf v\in(\mathbb R\cup\{\textit{None}\})^N$,
where $N=2^l$ and at least one entry is known

\ENSURE
A vector containing no missing entries

\STATE \textbf{Main routine}

\WHILE{$\mathbf v$ contains \textit{None}}
    \STATE
    $\mathbf v\gets\textsc{RecursiveWalsh}(\mathbf v)$
\ENDWHILE

\RETURN $\mathbf v$

\STATE \textbf{Recursive routine}
$\textsc{RecursiveWalsh}(\mathbf v)$

\STATE $N\gets\operatorname{length}(\mathbf v)$

\IF{$N=1$}
    \RETURN $\mathbf v$
\ENDIF

\STATE Split
$\mathbf v\gets(\mathbf v_0,\mathbf v_1)$,
where
$\mathbf v_0=\mathbf v_{[1:N/2]}$
and
$\mathbf v_1=\mathbf v_{[N/2+1:N]}$

\STATE
$m_i\gets
\#\{x\in\mathbf v_i:x=\textit{None}\}$
for $i\in\{0,1\}$

\IF{$m_0=0$}
    \IF{$m_1=N/2$}
        \RETURN $(\mathbf v_0,\mathbf v_0)$
    \ELSE
        \STATE
        $\mathbf v_1\gets
        \textsc{RecursiveWalsh}(\mathbf v_1)$
        \RETURN $(\mathbf v_0,\mathbf v_1)$
    \ENDIF

\ELSIF{$m_1=0$}
    \IF{$m_0=N/2$}
        \RETURN $(\mathbf v_1,\mathbf v_1)$
    \ELSE
        \STATE
        $\mathbf v_0\gets
        \textsc{RecursiveWalsh}(\mathbf v_0)$
        \RETURN $(\mathbf v_0,\mathbf v_1)$
    \ENDIF

\ELSIF{$m_0=N/2$}
    \STATE
    $\mathbf v_1\gets
    \textsc{RecursiveWalsh}(\mathbf v_1)$
    \RETURN $(\mathbf v_1,\mathbf v_1)$

\ELSIF{$m_1=N/2$}
    \STATE
    $\mathbf v_0\gets
    \textsc{RecursiveWalsh}(\mathbf v_0)$
    \RETURN $(\mathbf v_0,\mathbf v_0)$

\ELSE
    \STATE
    $\mathbf v_0\gets
    \textsc{RecursiveWalsh}(\mathbf v_0)$
    \STATE
    $\mathbf v_1\gets
    \textsc{RecursiveWalsh}(\mathbf v_1)$
    \RETURN $(\mathbf v_0,\mathbf v_1)$
\ENDIF

\end{algorithmic}
\end{algorithm}
\end{figure}
Once we possess a completed angle vector containing $2^l$ values, we
compute its unnormalized Walsh--Hadamard transform. In case of larger system sizes this operation can be done entirely within the TT representation, as long as the ranks of the layers of angles are small enough. We define

\begin{equation}
\mathcal H^{(l)}
=
\frac{1}{\sqrt{2^l}}\begin{bmatrix}
1&1\\
1&-1
\end{bmatrix}^{\otimes l},
\end{equation}
or, equivalently,

\begin{equation}
\label{eq:app-definition-of-H-element}
\mathcal H_{\vc{j},\vc{k}}^{(l)}
=
\frac{(-1)^{\langle\vc{j},\vc{k}\rangle_2}}{\sqrt{2^l}},
\qquad
\vc{j},\vc{k}\in\{0,1\}^l,
\end{equation}
where

\begin{equation}
\langle\vc{j},\vc{k}\rangle_2
=
\bigoplus_{m=0}^{l-1}j_mk_m
\end{equation}
is the binary inner product. The Walsh spectrum is

\begin{equation}
\label{eq:app-direct-WHT-formula}
\vc{w}^{(l)}
=
\mathcal H^{(l)}\vc\theta^{(l)}.
\end{equation}
Since

\begin{equation}
\mathcal H^{(l)}\mathcal H^{(l)}
=\I,
\end{equation}
the inverse transform is

\begin{equation}
\theta_{\vc{j}}^{(l)}
=
\frac{1}{\sqrt{2^l}}
\sum_{\vc{k}\in\{0,1\}^l}
(-1)^{\langle\vc{j},\vc{k}\rangle_2}
w_{\vc{k}}^{(l)}.
\end{equation}
To simplify the notation, we define

\begin{equation}
\label{eq:app-normalized-Walsh-coefficient}
W_{\vc{k}}^{(l)}
=
\frac{w_{\vc{k}}^{(l)}}{\sqrt{2^l}}.
\end{equation}
The inverse transform can then be written as

\begin{equation}
\label{eq:app-inverse-WHT}
\theta_{\vc{j}}^{(l)}
=
\sum_{\vc{k}\in\{0,1\}^l}
(-1)^{\langle\vc{j},\vc{k}\rangle_2}
W_{\vc{k}}^{(l)}.
\end{equation}
Thus, $w_{\vc{k}}^{(l)}$ denotes an entry of the unnormalized Walsh
spectrum, whereas $W_{\vc{k}}^{(l)}$ denotes the corresponding
coefficient entering the circuit decomposition as a rotation angle.
\subsection{Association of each spectral element with its decomposition}\label{sec:Walsh-spectral-circuit-proof}

Each component of the Walsh spectrum is associated with a sequence of
one- and two-qubit gates according to the binary representation of its
index. The decomposition proceeds conceptually in two steps. First,
the properties of the Walsh functions are used to rewrite the
multiplexer in terms of one- and two-qubit gates. Second, some of these
gates are merged in order to reduce the total depth.

\subsubsection{Direct decomposition of the multiplexer based on the WHT}
\label{app:btwht-direct-decomposition}

Using the inverse transform in \cref{eq:app-inverse-WHT}, we substitute
$\theta_{\vc{j}}^{(l)}$ into the multiplexer formula
\begin{align}
 {\mathcal M}_l
&=
\prod_{\vc{j}\in\{0,1\}^l}
C^lR_y\left(
\vc{j},
\theta_{\vc{j}}^{(l)}
\right)
\nonumber\\
&=
\prod_{\vc{j}\in\{0,1\}^l}
C^lR_y\left(
\vc{j},
\sum_{\vc{k}\in\{0,1\}^l}
(-1)^{\langle\vc{j},\vc{k}\rangle_2}
W_{\vc{k}}^{(l)}
\right).
\end{align}
Rotations about the same axis commute and their angles add. We can
therefore turn the sum over $\vc{k}$ into a product and reorder the
resulting rotations as

\begin{align}
 {\mathcal M}_l
&=
\prod_{\vc{j}\in\{0,1\}^l}
\prod_{\vc{k}\in\{0,1\}^l}
C^lR_y\left(
\vc{j},
(-1)^{\langle\vc{j},\vc{k}\rangle_2}
W_{\vc{k}}^{(l)}
\right)
\nonumber\\
&=
\prod_{\vc{k}\in\{0,1\}^l}
\prod_{\vc{j}\in\{0,1\}^l}
C^lR_y\left(
\vc{j},
(-1)^{\langle\vc{j},\vc{k}\rangle_2}
W_{\vc{k}}^{(l)}
\right).
\end{align}
We then divide the product over $\vc{j}$ according to the value of the
binary inner product

\begin{align}
\mathcal M_l
&=
\prod_{\vc{k}\in\{0,1\}^l}
\left[
\prod_{\substack{
\vc{j}\in\{0,1\}^l\\
\langle\vc{j},\vc{k}\rangle_2=0
}}
C^lR_y\left(
\vc{j},
W_{\vc{k}}^{(l)}
\right)
\right]\times
\nonumber\\
&\hspace{1.0cm}\times
\left[
\prod_{\substack{
\vc{j}\in\{0,1\}^l\\
\langle\vc{j},\vc{k}\rangle_2=1
}}
C^lR_y\left(
\vc{j},
-W_{\vc{k}}^{(l)}
\right)
\right]
\nonumber\\
&=
\prod_{\vc{k}\in\{0,1\}^l}
\mathcal W_{\vc{k}}^{(l)}(0)
\mathcal W_{\vc{k}}^{(l)}(1),
\end{align}
where
\begin{equation}
\label{eq:app-Walsh-component}
\mathcal W_{\vc{k}}^{(l)}(a)
=
\prod_{\substack{
\vc{j}\in\{0,1\}^l\\
\langle\vc{j},\vc{k}\rangle_2=a
}}
C^lR_y\left(
\vc{j},
(-1)^aW_{\vc{k}}^{(l)}
\right),
\qquad
a\in\{0,1\}.
\end{equation}
These two operators represent the positive and negative piecewise
constant parts of the Walsh function associated with the index
$\vc{k}$.
Given a binary string $\vc{j}\in\{0,1\}^l$, we define the set of
positions at which it has value $a\in\{0,1\}$ as

\begin{equation}
\label{eq:app-definition-of-O-sets}
O_{\vc{j}}(a)
=
\{
m\in\{0,\ldots,l-1\}:j_m=a
\}.
\end{equation}
We also define the set of addresses of Hamming weight $p$ and length $l$ as

\begin{equation}
\label{eq:app-Hamming-weight-sets}
\mathbf H_p^{(l)}=\{\vc{k}\in\{0,1\}^l:|\vc{k}|_{\mathrm H}=p\}.
\end{equation}
We will now consider the three cases $\mathbf H_0^{(l)},\mathbf H_1^{(l)}$ and $\mathbf H_{\geq 2}^{(l)}$ separately as they require a different treatment in the decomposition.
\subsubsection{Walsh components of Hamming weight zero}
For $\vc{k}=\vc{0}$, the condition

\begin{equation}
\langle\vc{j},\vc{0}\rangle_2=0
\end{equation}
is satisfied for every $\vc{j}\in\{0,1\}^l$. Therefore,

\begin{equation}
\mathcal W_{\vc{0}}^{(l)}(0)
=
\prod_{\vc{j}\in\{0,1\}^l}
C^lR_y(
\vc{j},
W_{\vc{0}}^{(l)}
),
\end{equation}
whereas

\begin{equation}
\mathcal W_{\vc{0}}^{(l)}(1)
=
\I.
\end{equation}
The product in $ {\mathcal W}_{\vc{0}}^{(l)}(0)$ contains all
possible control strings, and all its rotations have the same angle.
We can therefore apply
\cref{eq:binary_manyterms_MCR_product_rule_same_angle_final} and sum the
corresponding disjoint projectors. This reduces the product to the
single-qubit rotation

\begin{equation}
\mathcal W_{\vc{0}}^{(l)}(0)
=
R_y\left(
W_{\vc{0}}^{(l)}
\right).
\end{equation}
In the generalized-control-string formalism, this corresponds to the
decomposition

\begin{equation}
\mathcal G_{\vc{0}}
=
\{
\I\I...\I
\}.
\end{equation}
This operation is the first term in the direct decomposition shown in
\cref{fig:walsh_decomposition_k2_full}.

\subsubsection{Walsh components of Hamming weight one}

We next consider the indices $\vc{k}\in\mathbf H_1^{(l)}$. For $l=2$, for example,

\begin{equation}
\mathbf H_1^{(2)}
=
\{
(0,1),(1,0)
\}.
\end{equation}
We denote by $\mathbf H_{1,m}^{(l)}$ the singleton containing the
one-hot vector whose nonzero entry is at position $m$. For example 
\begin{equation}
\mathbf H_{1,1}^{(2)}
=
\{
(0,1)
\}.
\end{equation}
For
$\vc{k}\in\mathbf H_{1,m}^{(l)}$, the binary inner product satisfies

\begin{equation}
\langle\vc{j},\vc{k}\rangle_2
=
j_m.
\end{equation}
The set of binary strings for which this inner product has value
$a\in\{0,1\}$ is therefore the slice set
\begin{equation}
\begin{aligned}
C_{l,m}(a)
=&
\{
\vc{j}\in\{0,1\}^l,k\in\mathbf H_{1,m}^{(l)} :
\langle\vc{j},\vc{k}\rangle_2=a
\}\\
=&
\{
\vc{j}\in\{0,1\}^l:
j_m=a
\}.
\end{aligned}
\end{equation}
For example, for $l=3$ and $m=0$,

\begin{equation}
C_{3,0}(a)
=
\{
(a,0,0),
(a,0,1),
(a,1,0),
(a,1,1)
\}.
\end{equation}
For generic $l$ and $m$, this set can be written as

\begin{equation}
C_{l,m}(a)
=
\{0,1\}^m
\times
\{a\}
\times
\{0,1\}^{l-m-1}.
\end{equation}
Considered as the only term in the union forming the support set in
\cref{eq:partition_of_the_support}, it admits the trivial \ac{DSOP}
decomposition
\begin{equation}
\mathcal G_{l,m}(a)
=
\{
\I^m a\I^{l-m-1}
\}.
\end{equation}
It follows that the two products
$ {\mathcal W}_{\vc{k}}^{(l)}(0)$ and $ {\mathcal W}_{\vc{k}}^{(l)}(1)$ reduce to two singly
controlled rotations. The first is zero-controlled on the $m$-th
qubit 

\begin{align}
 {\mathcal W}_{\vc{k}}^{(l)}(0)
&=
\prod_{\vc{c}\in\mathcal G_{1,m}(0)}
C^lR_y(
\vc{c},
W_{\vc{k}}^{(l)}
)
\nonumber\\
&=
C^lR_y(
\I^m0\I^{l-m-1},
W_{\vc{k}}^{(l)}
)
\nonumber\\
&=
C^{0_m}R_y(
W_{\vc{k}}^{(l)}
),
\end{align}
while the second is one-controlled on the same qubit and has the opposite
angle

\begin{align}
 {\mathcal W}_{\vc{k}}^{(l)}(1)
&=
\prod_{\vc{c}\in\mathcal G_{1,m}(1)}
C^lR_y(
\vc{c},
-W_{\vc{k}}^{(l)}
)
\nonumber\\
&=
C^lR_y(
\I^m1\I^{l-m-1},
-W_{\vc{k}}^{(l)}
)
\nonumber\\
&=
C^{1_m}R_y(
-W_{\vc{k}}^{(l)}
).
\end{align}
An example of these two terms for $\vc{k}=(0,1)$ and $\vc{k}=(1,0)$ is shown in
\cref{fig:walsh_decomposition_k2_full}.

\subsubsection{Walsh components of Hamming weight greater than one}

Finally, consider $\vc{k}\in\mathbf H_p^{(l)}$ with
$p\in\{2,\ldots,l\}$. An example is shown in the
right-hand part of \cref{fig:walsh_decomposition_k2_full}. In this case, we use one clean ancilla qubit to
compute the binary inner product
$\langle\vc{j},\vc{k}\rangle_2$. The relevant control positions are
those in the support $S_{\vc{k}}$, whose cardinality is

\begin{equation}
|S_{\vc{k}}|
=
p.
\end{equation}
Let anc denote the ancilla and let $t$ denote the multiplexer target.
For every $\vc{k}\in\mathbf H_p^{(l)}$, we define the CX ladder

\begin{equation}
\label{eq:app-parity-ladder}
\mathcal L_{\vc{k}}
=
\prod_{m\in S_{\vc{k}}}
\mathrm{CX}(q_m,\text{anc}).
\end{equation}
On a computational-basis state, this ladder acts as

\begin{equation}
\mathcal L_{\vc{k}}
\ket{\vc{j}}\ket{0}_{\text{anc}}
=
\ket{\vc{j}}
\ket{\langle\vc{j},\vc{k}\rangle_2}_\text{anc}.
\end{equation}
We define the ancilla-assisted operator

\begin{equation}
U_{\vc{k}}^{(l)}
=
\mathcal L_{\vc{k}}
C_\text{anc}^0R_y(
t,
W_{\vc{k}}^{(l)}
)
C_\text{anc}^1R_y(
t,
-W_{\vc{k}}^{(l)}
)
\mathcal L_{\vc{k}}.
\end{equation}
For every state $\ket{\psi}$ of the control and target registers,

\begin{equation}
U_{\vc{k}}^{(l)}
\ket{\psi}\ket{0}_\text{anc}
=
\mathcal W_{\vc{k}}^{(l)}(0)
\mathcal W_{\vc{k}}^{(l)}(1)
\ket{\psi}\ket{0}_\text{anc}.
\end{equation}
Thus, the first ladder computes the required parity on the ancilla, the
two controlled rotations select the appropriate sign, and the second
ladder returns the ancilla to $\ket{0}_\text{anc}$. 

\subsection{From the direct to the simplified decomposition}
\label{app:btwht-simplification}

The direct decomposition described above, and illustrated for $l=2$ in
\cref{fig:walsh_decomposition_k2_full}, can be simplified in three
steps.

First, all zero-controlled rotations can be removed and their angles
added to the initial single-qubit rotation. This follows from
\begin{equation}
\label{eq:two_controlled_to_one_controlled_rot}
C^0R_y(W)
C^1R_y(-W)
=
R_y(W)
C^1R_y(-2W),
\end{equation}
for a given angle $W\in[0,2\pi].$
The angle of the initial single-qubit rotation is therefore

\begin{equation}
\label{eq:app-alpha-zero}
\alpha_{\vc{0}}^{(l)}
=
\sum_{\vc{k}\in\{0,1\}^l}
W_{\vc{k}}^{(l)}.
\end{equation}
Second, the angles of the remaining one-controlled rotations are

\begin{equation}
\label{eq:app-alpha-nonzero}
\alpha_{\vc{k}}^{(l)}
=
-2W_{\vc{k}}^{(l)},
\qquad
\vc{k}\neq\vc{0}
\end{equation}
again as a direct consequence of \cref{eq:two_controlled_to_one_controlled_rot}, where $W_{k}^{(l)}$ is the k-th element of the normalized Walsh spectrum of level $l$.
Third, instead of computing and uncomputing
$\langle\vc{j},\vc{k}\rangle_2$ independently for every Walsh address of
Hamming weight at least two, we order the corresponding commuting
operations so as to minimize the number of CX gates. Let
\begin{equation}
\mathbf H_{\geq2}^{(l)}
\coloneqq
\bigcup_{p=2}^{l}\mathbf H_p^{(l)}
\end{equation}
and let
\begin{equation}
M_{\geq2}
=
\left|\mathbf H_{\geq2}^{(l)}\right|.
\end{equation}
For an ordering $\sigma$ of these addresses, we consider the cycle

\begin{equation}
\mathcal P_\sigma
=
(
\vc{0},
\vc{k}_{\sigma_1},
\ldots,
\vc{k}_{\sigma_{M_{\geq2}}},
\vc{0}
).
\end{equation}
The ancilla initially stores the parity associated with $\vc{0}$. A
transition from $\vc{k}$ to $\vc{k}'$ requires CX gates at the positions
where the two addresses differ and therefore costs

\begin{equation}
|\vc{k}\oplus\vc{k}'|_{\mathrm H}
\end{equation}
CX gates. The total CX cost of the cycle is

\begin{align}
\label{eq:app-Hamming-cycle-cost}
D_{\mathrm{CX},\geq2}(\sigma)
&=
|\vc{k}_{\sigma_1}|_{\mathrm H}
+
\sum_{r=1}^{M_{\geq2}-1}
|\vc{k}_{\sigma_r}\oplus
\vc{k}_{\sigma_{r+1}}|_{\mathrm H}
\nonumber\\
&\quad+
|\vc{k}_{\sigma_{M_{\geq2}}}|_{\mathrm H}.
\end{align}
The first term computes the parity associated with the first address,
while the final term returns the ancilla to $\ket{0}$. Since all the
CX gates in this part of the circuit share the ancilla, the CX count
also gives the CX depth of this subcircuit.

The ordering that minimizes the total Hamming distance also minimizes
this contribution to the circuit depth. The result of these three
simplifications is illustrated for $l=2$ in
\cref{fig:walsh_decomposition_k2_simplified} and for $l=3$ in
\cref{fig:walsh_decomposition_k3}.

\subsubsection{Minimization of the total depth}
\label{app:minimization_of_BTWHT_error}

The final depth depends on the truncation applied to the Walsh
spectrum. Given a threshold $\epsilon>0$, we define the retained
support as

\begin{equation}
\mathcal K_\epsilon^{(l)}
=
\Big\{
\vc{k}\in\{0,1\}^l:
\left|W_{\vc{k}}^{(l)}\right|
\geq\epsilon
\Big\}.
\end{equation}
We divide the retained addresses according to their Hamming weight as

\begin{equation}
\mathbf H_{\epsilon,p}^{(l)}
=
\big\{
\vc{k}\in\mathcal K_\epsilon^{(l)}:
|\vc{k}|_{\mathrm H}=p
\big\}.
\end{equation}
The retained higher-weight support is

\begin{equation}
\mathbf H_{\epsilon,\geq2}^{(l)}
=
\bigcup_{p=2}^{l}
\mathbf H_{\epsilon,p}^{(l)},
\end{equation}
and its size is

\begin{equation}
M_{\epsilon,\geq2}
=
\left|\mathbf H_{\epsilon,\geq2}^{(l)}\right|.
\end{equation}
Be

\begin{equation}
\delta_0
=
\begin{cases}
1, & \alpha_{\vc{0}}^{(l)}\neq0,\\
0, & \alpha_{\vc{0}}^{(l)}=0.
\end{cases}
\end{equation}
If a single-qubit or singly controlled rotation is counted as one
native gate layer, the depth is

\begin{equation}
\label{eq:app-BTWHT-native-depth}
D_{\mathrm{native}}(\epsilon,\sigma)
=
\delta_0
+
\left|\mathbf H_{\epsilon,1}^{(l)}\right|
+
\left|\mathbf H_{\epsilon,\geq2}^{(l)}\right|
+
D_{\mathrm{CX},\epsilon,\geq2}(\sigma).
\end{equation}
This is a depth in the native gate set containing singly controlled
rotations. A depth in the $\{\mathrm{CX},R_x,R_z,R_y\}$ basis requires
the controlled rotations to be decomposed further.

For $M_{\epsilon,\geq2}\geq1$, the first and last transitions in the
cycle each have Hamming weight at least two, while every transition
between distinct retained addresses has Hamming weight at least one.
Furthermore, the length of a closed walk on a hypercube has even
parity. We therefore obtain the lower bound 

\begin{equation}
\label{eq:app-BTWHT-transition-lower-bound}
D_{\mathrm{CX},\epsilon,\geq2}(\sigma)
\geq
2
\left\lceil
\frac{M_{\epsilon,\geq2}+3}{2}
\right\rceil.
\end{equation}

For a fixed Walsh spectrum, the cost function is invariant under a
permutation of qubit labeling, because the Hamming weight remains invariant. However, the binary-tree angle
vectors, and hence their Walsh spectra, depend on the ordering of the
qubits in the original amplitude tensor. The final depth can therefore
be optimized over qubit permutations, for example using Monte Carlo
sampling \cite{metropolis1949monte}.

Finally, the truncation error can be expressed directly in terms of
the discarded Walsh coefficients. Let
$\widetilde{\vc\theta}^{(l)}$ be the angle vector reconstructed from
the retained support. Since

\begin{equation}
\mathcal H^{(l)}
\mathcal H^{(l)}
=\I,
\end{equation}
we obtain

\begin{align}
\label{eq:app-BTWHT-truncation-error}
\left\|
\vc\theta^{(l)}
-
\widetilde{\vc\theta}^{(l)}
\right\|_2^2
&=
2^l
\sum_{\vc{k}\notin\mathcal K_\epsilon^{(l)}}
\left|W_{\vc{k}}^{(l)}\right|^2
\nonumber\\
&=
\sum_{\vc{k}\notin\mathcal K_\epsilon^{(l)}}
\left|w_{\vc{k}}^{(l)}\right|^2,
\end{align}
where we used the property 
\begin{equation}
  \sum_{\vc{b} \in \{0,1\}^l}(-1)^{\langle\vc{b},\vc{c}\rangle}=\delta_{\vc{b},\vc{0}}.
\end{equation}
This relation gives the explicit trade-off between the Walsh
truncation and the error in the multiplexer angle vector.
This error can be related to the error on the single entry of encoded state with the upper bound 
\begin{equation}
\label{eq:upperbound_pointwise_amplitude_abs_error}
    |\tilde v_{j}-v_{ j}| = \frac{M_j(\theta)}{2}n\mathcal{E}_Z + \sqrt{\frac{n}{e}}\mathcal{E}_M+ O(\mathcal{E}_M^2 + \mathcal{E}_Z^2 ),\forall j \in [2^n]
\end{equation}
where we assume an error across the levels of the tree for both the modulus
\begin{equation}
\mathcal{E}_{M}=\left\|\vc\theta^{(l)}-\widetilde{\vc\theta}^{(l)}\right\|_2^2,\quad\forall l \in [n]    
\end{equation}
and the phase 
\begin{equation}
\mathcal{E}_{Z}=\left\|\vc\phi^{(l)}-\widetilde{\vc\phi}^{(l)}\right\|_2^2,\quad\forall l \in [n]     
\end{equation}
and $M_{\vc{j}}$ is the modulus of the $\vc{j}$-th entry. 
\begin{proof}[Proof of \cref{eq:upperbound_pointwise_amplitude_abs_error}]
\label{proof:absolute_error_upperbound}
We start by considering a tiny variation of the trees $\tilde{\vc{\theta}}=\vc{\theta} + \vc{\epsilon} $ and $\tilde{\vc{\phi}}=\vc{\phi} + \vc{\delta}$ with 
\begin{align}
    \epsilon:=& \max_j |\epsilon_j|\\
    \delta:=& \max_j |\delta_j|
\end{align}
such that
\begin{equation}
\label{eq:def_epsilon}
    |\tilde \theta_j - \theta_j|:= |\epsilon_j|\leq \epsilon
\end{equation}
and 
\begin{equation}
    |\tilde \phi_j - \phi_j|:= |\delta_j|\leq \delta.
\end{equation}
Every amplitude can be decomposed as
\begin{equation}
    \tilde v_{j}:=v_{j}(\tilde \theta,\tilde \phi)= M_{j}(\tilde \theta) Z_{j}(\tilde \phi)
\end{equation}
where $Z_{j}(\tilde \phi):= e^{i \zeta_j(\tilde \phi)}$.
The effect of the small variation on the modulus is
\begin{equation}
\label{eq:error_on_the_modulus_first_order}
\begin{split}
    &M(\tilde \theta)_{j}=M_{j}(\theta+\epsilon)=\prod_{k=0}^{n-1}G_{j_k}(\theta_{j_{[k]}}+\epsilon_{j_{[k]}})\\
    &=\prod_{k=0}^{n-1}\left[G_{j_k}(\theta_{j_{[k]}})+(-1)^{j_k+1}\epsilon_{j_{[k]}} G_{j_k+1}(\theta_{j_{[k]}}) + O\left(\epsilon_{j_{[k]}}^2\right)\right ]\\
    &=\prod_{k=0}^{n-1}G_{j_k}(\theta_{j_{[k]}})  \\
    &+ \sum_{k=0}^{n-1} (-1)^{j_k+1}\epsilon_{j_{[k]}} G_{j_k+1}(\theta_{j_{[k]}}) \prod_{\substack{l=0\\ l\neq k}}G_{j_l}(\theta_{j_{[l]}})+O\left(\epsilon_{j_{[k]}}^2\right)\\
   &=M_{j}(\theta) +  \Delta^{[1]}_{M,j}+O\left(\epsilon_{j_{[k]}}^2\right),
\end{split}
\end{equation}
where we used the definition of the modulus in \cref{eq:modulus_tensor_angles} and we introduced
\begin{equation}
    \Delta^{[1]}_{M,j}:= \sum_{k=0}^{n-1} (-1)^{j_k+1}\epsilon_{j_{[k]}}G_{j_k+1}(\theta_{j_{[k]}}) \prod_{\substack{l=0\\ l\neq k}}G_{j_l}(\theta_{j_{[l]}})
\end{equation}
We now focus on this first order term, which we can upper bound with the sum of the moduli, and we use the fact that the pointwise variation of the tree has absolute value smaller than $\epsilon$:
\begin{equation}
\begin{split}
    \sum_{k=0}^{n-1}& (-1)^{j_k+1}\epsilon_{j_{[k]}}G_{j_k+1}(\theta_{j_{[k]}}) \prod_{\substack{l=0\\ l\neq k}}G_{j_l}(\theta_{j_{[l]}})\\
    &\leq \sum_{k=0}^{n-1} |G_{j_k+1}(\theta_{j_{[k]}})||\epsilon_{j_{[k]}}| \prod_{\substack{l=0\\ l\neq k}}|G_{j_l}(\theta_{j_{[l]}})|\\
     &\leq \epsilon \sum_{k=0}^{n-1} |G_{j_k+1}(\theta_{j_{[k]}})| \prod_{\substack{l=0\\ l\neq k}}|G_{j_l}(\theta_{j_{[l]}})|.
    \end{split}
\end{equation}
The term 
\begin{equation}
    f_{\vc j}(\theta)=\sum_{k=0}^{n-1}f_{\vc j,k}(\theta)
\end{equation}
where
\begin{equation*}
f_{\vc j,k}(\theta)=|G_{j_k+1}(\theta_{j_{[k]}}) \prod_{\substack{l=0\\ l\neq k}}G_{j_l}(\theta_{j_{[l]}})| 
\end{equation*}
becomes a sum of $n$ terms that are products of $n-1$ $\sin$ factors and one $\cos$ when $j=0$. Any other combination of $\sin$ or $\cos$ terms can be reduced to these, since we are considering the maximum over all the possible angles in the range $[0,\frac{\pi}{2}]$.

For example, for $n=2$ and $j=(01)$ the associated variation of the modulus is 
\begin{equation}
  f_{01}(\theta_r,\theta_0)= |\sin(\theta_r)\sin(\theta_0)|+|\cos(\theta_r)\cos(\theta_0)|.
\end{equation}
By changing the value of $\theta_r$ we can turn it to
\begin{equation}
\begin{split}
&  f_{01}\left(\theta_r,\theta_0\right)\left(\theta_r \pm\frac{\pi}{2},\theta_0\right)\\
&=\left|\sin\left(\theta_r\pm\frac{\pi}{2}\right)\sin\left(\theta_0\right)\right|+\left|\cos\left(\theta_r\pm\frac{\pi}{2}\right)\cos\left(\theta_0\right)\right|\\
&=|\cos(\theta_r)\sin(\theta_0)|+|\sin(\theta_r)\cos(\theta_0)|\\
&= f_{11}(\theta_r,\theta_0).
\end{split}
\end{equation}
Since for any $j$, $f_{\vc j}$ can be seen as a shifted version of $f_{\vc 1}$, we can find an upper bound for $f_{\vc j}$ that does not depend on $j$ but only on $n$ by considering the configuration associated to $f_{\vc 1}$ in 
\begin{equation}
\label{eq:upper_bound_for_sin_cos_prod}
   f_{\vc 1}= f_n(x_1,\dots x_n) := \sum_{k=1}^n \sqrt{1-x_k^2}\prod_{j\neq k} x_j\leq \sqrt{\frac{(n-1)^{n-1}}{n^{n-2}}}
\end{equation}
 in both cases, where $|\sin(\theta_k)|=x_k \in [0,1] \forall k \in \{1\dots n\},n\in \mathbb{N},n\geq 3$.
 \begin{proof}[Proof of \cref{eq:upper_bound_for_sin_cos_prod}]
 We define 
 \begin{equation}
    \begin{split}
        f_n(x_1,\dots x_n) := \sum_{k=1}^n \sqrt{1-x_k^2}\prod_{j\neq k} x_j\\
        = \sum_{k=1}^n \frac{\sqrt{1-x_k^2}}{x_k}\prod_{j} x_j \\
      =\prod_{j} x_j\sum_{k=1}^n \frac{\sqrt{1-x_k^2}}{x_k}\\
      =  \prod_{j} \frac{1}{\sqrt{1+t_j^2}}\sum_{k=1}^n t_k
        \end{split}
 \end{equation}
 where $t_k:=\frac{\sqrt{1-x_k^2}}{x_k}$. The case where $x_k=0$ is excluded as it gives the value $1$. We will compare this value wiht the maximum we find later in the proof to show it is not the maximum in this interval.
 This function is everywhere positive in the interval $x_k \in [0,1] \forall k$, therefore we can take its $\log$
  \begin{equation}
    \begin{split}
         \log f_n(x_1,\dots x_n)=-\frac{1}{2}\sum_j \log\left(1+t^2_j\right) +\log\left(\sum_k t_k\right)
        \end{split}
 \end{equation}
  and maximize it.
  The gradient of the $\log $ function is 
  \begin{equation}
      \begin{split}
       \partial_{t_m} \log f_n =- \frac{t_m}{1+t_m^2} + \frac{1}{\sum_k t_k}
      \end{split}
  \end{equation}
  and it vanishes for 
    \begin{equation}
     t_m^2-\left(\sum_k t_k\right)t_m+1=(t_m-t_0)(t_m-t_1)=0
  \end{equation}
  The two possible solutions are the same for every $m \in \{1\dots n\}$ and they must fulfill 
  \begin{equation}
      \begin{split}
          t_0+t_1=\sum_k t_k\\
          t_0t_1=1.
      \end{split}
      \end{equation}
For $n\geq 3$ the first equation cannot be fulfilled unless all the coordinates assume the same value 
  \begin{equation}
        t_m=t_0 =\frac{1}{\sqrt{n-1}},\forall m \in \{1,\dots n\}
\end{equation}
which implies 
  \begin{equation}
        x_m=\frac{1}{1+t^2_0} =\sqrt \frac{n-1}{n},\forall m \in \{1,\dots n\}
\end{equation}
and substituting back into $f_n$ we get 
  \begin{equation}
        f_n(x_m=t_0,\forall m)=\sqrt{\frac{(n-1)^{n-1}}{n^{n-2}}},n \in \mathbb{N},n\geq 3.
\end{equation}\\
What remains to check are the eigenvalues of the Hessian matrix at $\vc x=x_0 \vc 1$.
The Hessian matrix of the logarithm of f is 
\begin{equation}
    K_{m,l}:=\partial_m\partial_lf_n(t)= \delta_{m,l}c_m + a\vc 1^t \vc1. 
\end{equation}
where 
\begin{align}
  c_m&=\frac{t_m^2-1}{1+t_m^2}\\
  a&=-\frac{1}{(\sum_k t_k)^2}
\end{align}
Using the Lemma of the determinant 
\begin{equation}
    \det(B+a\vc 1^t \vc1) = \det(B) ( 1+ a \vc 1^t B^{-1}\vc 1),
\end{equation}
we have 
\begin{equation}
\begin{split}
   \det(K-\lambda I) &= \det(D -\lambda I + a\vc 1^t \vc1)\\
   &=\det(D-\lambda I) ( 1+ a \vc 1^t (D-\lambda I)^{-1}\vc 1)\\
   &=\prod_m(c_m-\lambda)\left(1+a\sum_k\frac{1}{c_m-\lambda}\right),
  \end{split}
\end{equation}
 When we consider $\vc t=t_0 \vc 1$ we get that 
 \begin{equation}
     c_m=c_0=\frac{(2-n)(n-1)}{n^2},\forall m \in \{1,\dots n\}
 \end{equation}
 and the characteristic equation becomes 
 \begin{equation}
 \begin{split}
     (c_0-\lambda)^n \left(1+\frac{an}{(c_0-\lambda)}\right)\\
     =(c_0-\lambda)^{n-1}(c_0+An -\lambda).
  \end{split}
 \end{equation}
 The eigenvalues are 
 \begin{equation}
 \begin{split}
     \lambda_0 = c_0 \quad \hspace{.9cm}&\text{multiplicity}=n-1\\
     \lambda_1 = c_0+An \quad&\text{multiplicity}= 1.
      \end{split}
 \end{equation}
 The condition for $\vc x =x_0 \vc 1$ to be a maximum is that 
 \begin{equation}
 \begin{split}
      c_0&=\frac{(2-n)(n-1)}{n^2}< 0\iff n>2 \lor n<1 \\
      c_0&<An \iff \frac{(2-n)(n-1)}{n^2}< \frac{n-1}{n} \iff n>1.
  \end{split}
 \end{equation}
 This conditions are both met for $n\geq 3$, which concludes the proof.
 \end{proof}
 In the large $n$ limit \cref{eq:upper_bound_for_sin_cos_prod} tends to $\sqrt{\frac{n}{e}}$, so that we can write 
\begin{equation}
  \sum_{k=0}^{n-1} \left|G_{j_k+1}(\theta_{j_{[k]}})\right|\left|\epsilon_{j_{[k]}}\right| \prod_{\substack{l=0\\ l\neq k}}\left|G_{j_l}(\theta_{j_{[l]}})\right|
    \leq \epsilon\sqrt{\frac{n}{e}}.
  \end{equation}
and for the entire modulus term we have 
\begin{equation}
\left|\Delta^{[1]}_{M,j}\right|:=\left|M_j-\tilde{M}_j\right|\leq \epsilon \sqrt{\frac{n}{e}}+O(\epsilon^2).
\end{equation}
For the phase term we have 
\begin{equation}
\label{eq:error_phase_first_order}
    \begin{split}
        Z(\phi + \delta)&=  \prod_{k=0}^{n-1}e^{-\frac{i}{2}(-1)^{j_k}\left(\phi_{j_{[k]}}+ \delta_{j_{[k]}}\right)}\\
        &= \prod_{k=0}^{n-1}e^{-\frac{i}{2}(-1)^{j_k}\phi_{j_{[k]}}} e^{-\frac{i}{2}(-1)^{j_k}\delta_{j_{[k]}}}\\
        &=Z(\phi) e^{-\frac{i}{2} \sum_{k=0}^{n-1}(-1)^{j_k}\delta_{j_{[k]}}}\\
        &=Z(\phi) \left(1-\frac{i}{2} \sum_{k=0}^{n-1}(-1)^{j_k}\delta_{j_{[k]}} + O(\delta^2)\right).
    \end{split}
\end{equation}
For the variation of the phase we get 
\begin{equation}
\Delta^{[1]}_{Z,j}:=Z_j-\tilde{Z}_j= -\frac{i}{2}Z_j(\phi)\sum_{k=0}^{n-1}(-1)^{j_k}\delta_{j_{[k]}} + O(\delta^2),
\end{equation}
whose modulus is upper bounded by 
\begin{equation}
\begin{split}
\left|\Delta^{[1]}_{Z,j}\right|&= \frac{1}{2}\left|Z_j(\phi)\sum_{k=0}^{n-1}(-1)^{j_k}\delta_{j_{[k]}}\right| + O(\delta^2)\\
&\leq \frac{1}{2}|Z_j(\phi)|\sum_{k=0}^{n-1}\left|(-1)^{j_k}\delta_{j_{[k]}}\right| + O(\delta^2)\\
&\leq \frac{1}{2}n\delta + O(\delta^2).
\end{split}
\end{equation}
Putting everything together, the total variation of the entry is 
\begin{equation}
    \begin{split}
        v_j& - \tilde{v}_j\\ &=(M_j(\theta)+\Delta^{[1]}_{M,j}(\theta))(Z_j(\phi)+\Delta_{Z,j}(\phi))
        -M_j(\theta)Z_j(\phi)\\
        &=M_j(\theta)\Delta_{Z,j}(\phi)+Z_j(\phi)\Delta^{[1]}_{M,j}(\theta)
        +O(\delta \epsilon).
    \end{split}
\end{equation}
We can upper bound its absolute value with the sum of the absolute values of the two terms 
\begin{equation}
\label{intermediate_abs_diff_amplitude_upperbound}
    \begin{split}
        | v_j - \tilde{v}_j|&=\left|M_j(\theta)\Delta_{Z,j}(\phi)+Z_j(\phi)\Delta^{[1]}_{M,j}(\theta)\right|+O(\delta \epsilon)\\
        &\leq \left|M_j(\theta)\Delta_{Z,j}(\phi)\right|+\left|Z_j(\phi)\Delta^{[1]}_{M,j}(\theta)\right|+O(\delta \epsilon)\\
        &\leq M_j(\theta)\left|\Delta_{Z,j}(\phi)\right|+\left|\Delta^{[1]}_{M,j}(\theta)\right|+O(\delta \epsilon)\\
        &\leq \frac{M_j(\theta)}{2}n\delta + \sqrt{\frac{n}{e}} \epsilon+ O(\epsilon^2 + \delta^2 ).
        \end{split}
\end{equation}
Finally, since the max of the absolute value of the pointwise $\theta$ variation is smaller than $\epsilon$ (or $\delta$ in the
case of $\phi$) 
we have 
\begin{equation}
    \epsilon \leq \sqrt{\sum_{j} (\tilde {\theta}_j-\theta_j)^2 }=\mathcal{E}_M
\end{equation}
and 
\begin{equation}
    \delta \leq \sqrt{\sum_{j} (\tilde {\phi}_j-\phi_j)^2 }=\mathcal{E}_\mathcal{A},
\end{equation}
which allows us to upper bound \cref{intermediate_abs_diff_amplitude_upperbound} and write \cref{eq:upperbound_pointwise_amplitude_abs_error}.
\end{proof}
When, as in \cref{sec:results}, we keep instead the relative error fixed for every layer of the tree following \cref{eq:relative_error_def}, we get for the real valued functions the following upper bound for the infidelity
\begin{equation}
\label{eq:relative_error_upperbound}
    \mathcal{I}\leq \mathcal{E}_{\mathrm{WHT}}^2\sum_{k=0}^{n-1}\left\|\vc \theta^{(k)}\right\|^2_2 (2k+1)+o\left(\mathcal{E}^2_{\mathrm{WHT}}\right),
\end{equation}
where small $o\left(\mathcal{E}^2_{\mathrm{WHT}}\right)$, indicates a strict upper bound.

\begin{proof}[Proof of \cref{eq:relative_error_upperbound}]
In order to upper bound the infidelity, we need to extend the formula for the error on the modulus \cref{eq:error_on_the_modulus_first_order} to second order, which gives
\begin{equation}
\label{proof:Taylor}
   \tilde M_j= M_{j}(\theta+\epsilon)=M_{j}(\theta) +  \Delta^{[1]}_{M,j}+\Delta^{[2]}_{M,j}+\Delta^{[1,1]}_{M,j}+O\left(\epsilon_\text{max}^3\right),
\end{equation}
where
\begin{equation}
    \Delta^{[2]}_{M,j}=\sum_{k=0}^{n-1}(-1)^{j_k}\frac{\epsilon^2_{j_{[k]}}}{2}G_{j_k}\prod_{\substack{l=0\\l\neq k}}^{n-1}G_{j_l},
\end{equation}
\begin{equation}
    \Delta^{[1,1]}_{M,j}=\sum_{k,k'=0}^{n-1}(-1)^{j_k+j_{k'}}\epsilon_{j_{[k]}}\epsilon_{j_{[k']}}G_{j_k}G_{j_{k'}}\prod_{\substack{l=0\\l\notin \{k,k'\}}}^{n-1}G_{j_l},
\end{equation}
\begin{equation}
    \epsilon_\text{max}:=\max_k \max_{\vc j} |\epsilon_{j_{[k]}}|
\end{equation}
and all the other quantities are defined as in \cref{proof:absolute_error_upperbound}.
Imposing the normalization of the approximated state $\ket{\tilde v}$ gives
\begin{equation}
\begin{split}
  1&= \bra{\tilde v}\tilde v\rangle =\sum_{j} \tilde v^*_j \tilde v_j=\sum_{j} \tilde{M}_j \tilde Z^*_j \tilde{M}_j \tilde Z_j=\sum_{j} \tilde{M}_j^2.
\end{split}
\end{equation}
Using \cref{proof:Taylor} we find
\begin{equation}
\begin{split}
  1&=\sum_j\left[M_{j}+  \Delta^{[1]}_{M,j}+\Delta^{[2]}_{M,j}+\Delta^{[1,1]}_{M,j}+O\left(\epsilon_\text{max}^3\right) \right]^2\\
  & =\sum_j M_j^2 +2\sum_j M_j \Delta^{[1]}_{M,j} + 2\sum_j M_j\left( \Delta^{[2]}_{M,j}+\Delta^{[1,1]}_{M,j}\right)\\
  &\hspace{1.6cm}+\sum_j \left(\Delta^{[1]}_{M,j}\right)^2+O\left(\epsilon_\text{max}^3\right)\\
  &=1 +2\sum_j M_j \Delta^{[1]}_{M,j} + 2\sum_j M_j\left( \Delta^{[2]}_{M,j}+\Delta^{[1,1]}_{M,j}\right)\\
  &\hspace{1.6cm}+\sum_j \left(\Delta^{[1]}_{M,j}\right)^2+O\left(\epsilon_\text{max}^3\right).
 \end{split} 
 \end{equation} 
This holds for any $\epsilon_\text{max}$. Therefore, it follows that both first and second order independently vanish with $\epsilon_\text{max}$ as in 
\begin{equation}
\label{eq:normalization_constraint_first_order}
    \sum_j M_j \Delta^{[1]}_{M,j}=0+O\left(\epsilon_\text{max}^3\right)
\end{equation}
\begin{equation}
\label{eq:normalization_constraint_second_order}
    \sum_j M_j\left(\Delta^{[2]}_{M,j}+\Delta^{[1,1]}_{M,j}\right)=-\frac{1}{2}\sum_j \left(\Delta^{[1]}_{M,j}\right)^2+O\left(\epsilon_\text{max}^3\right) .
\end{equation}
The fidelity of the approximated state and the target one is
\begin{equation}
    \begin{split}
\mathcal{F}&=|\langle v\ket{\tilde v}|^2\\
        &=\left|\sum_j v_j^* \tilde v_j\right|^2\\
       &= \left|\sum_j M_j Z^*_j \tilde M_j \tilde Z_j\right|^2\\
&=\left|\sum_j M_j Z^*_j \tilde M_j  Z(\phi) e^{-\frac{i}{2} \sum_{k=0}^{n-1}(-1)^{j_k}\delta_{j_{[k]}}}\right|^2,
    \end{split}
\end{equation}
where we used \cref{eq:error_phase_first_order}.
In the case of real functions, or more in general whenever 
\begin{equation}
    \sum_{k=0}^{n-1}(-1)^{j_k}\delta_{j_{[k]}}=0
\end{equation}
the fidelity becomes
\begin{equation}
    \begin{split}
        \mathcal{F}&=
        \left(\sum_j M_j\tilde M_j \right)^2\\
        &=\left(\sum_j M_j( M_{j} +  \Delta^{[1]}_{M,j}+\Delta^{[2]}_{M,j}+\Delta^{[1,1]}_{M,j}+O\left(\epsilon_\text{max}^3\right) \right)^2\\
        &=\left(\sum_j M_j^2+  \sum_j M_j\Delta^{[1]}_{M,j}+\sum_j M_j\left(\Delta^{[2]}_{M,j}+\Delta^{[1,1]}_{M,j}\right)\right)^2\\
        &\hspace{3cm}+O\left(\epsilon_\text{max}^3\right) \\
        &=\left(1 -\frac{1}{2}\sum_j M_j\left(\Delta^{[1]}_{M,j}\right)^2\right)^2+O\left(\epsilon_\text{max}^3\right)\\
        &=1-\sum_j M_j\left(\Delta^{[1]}_{M,j}\right)^2+O\left(\epsilon_\text{max}^3\right) ,
    \end{split}
\end{equation}
where we used \cref{eq:normalization_constraint_first_order} and \cref{eq:normalization_constraint_first_order}.
By definition, the infidelity is given by
\begin{equation}
\begin{split}
     &\mathcal{I}:=1-\mathcal{F}=\sum_j M_j\left(\Delta^{[1]}_{M,j}\right)^2+O\left(\epsilon_\text{max}^3\right)\\
     &=\sum_j  \left(\sum_{k=0}^{n-1} (-1)^{j_k+1}\epsilon_{j_{[k]}}G_{j_k+1}(\theta_{j_{[k]}}) \prod_{\substack{l=0\\ l\neq k}}G_{j_l}(\theta_{j_{[l]}})\right)^2\\
     &\hspace{3cm}+O\left(\epsilon_\text{max}^3\right)\\
     &=\sum_j\sum_{k,k'=0}^{n-1} (-1)^{j_k+j_{k'}}\epsilon_{j_{[k]}}\epsilon_{j_{[k']}}G_{j_k+1}G_{j_k'+1}\times\\
     &\hspace{1.2cm}\times \prod_{\substack{l,l'=0\\ l\neq k,l'\neq k'}}G_{j_l}G_{j_l'}+O\left(\epsilon_\text{max}^3\right).
     \end{split}
\end{equation}
We upper bound this sum with the sum of the absolute values and the $|G|$ terms with 1 to write 
\begin{equation}
\begin{split}
     \mathcal{I}&\leq \sum_j\sum_{k,k'=0}^{n-1} \Bigg|(-1)^{j_k+j_{k'}}\epsilon_{j_{[k]}}\epsilon_{j_{[k']}}G_{j_k+1}G_{j_k'+1}\times\\ 
     &\hspace{2.5cm}\times\prod_{\substack{l,l'=0\\ l\neq k,l'\neq k'}}G_{j_l}G_{j_l'}\Bigg|\\
     &=\sum_j\sum_{k,k'=0}^{n-1} \epsilon_{j_{[k]}}\epsilon_{j_{[k']}}\\
     &=\sum_j\sum_{k=k'=0}^{n-1} \epsilon_{j_{[k]}}^2
     + 2 \sum_j\sum_{k>k'}^{n-1} |\epsilon_{j_{[k]}}||\epsilon_{j_{[k']}}|\\
     &\leq \sum_j\sum_{k=k'=0}^{n-1} \epsilon_{j_{[k]}}^2
     + 2 \sum_j\sum_{k>k'}^{n-1} |\epsilon_{j_{[k]}}||\epsilon_{j_{[k']}}|\\
     &\leq \sum_{k=0}^{n-1} \sum_j\epsilon_{j_{[k]}}^2
     + 2 \sum_{k>k'}^{n-1} \sum_j \epsilon_{j_{[k]}}^2\\
      &\leq \mathcal{E}_\mathrm{WHT}\left(\sum_{k=0}^{n-1} \left\|\vc \theta ^{(k)}\right\|_2
     + 2 \sum_{k>k'}^{n-1} \left\|\vc \theta ^{(k)}\right\|_2\right)\\
      &\leq \sum_{k=0}^{n-1} \sum_j\epsilon_{j_{[k]}}^2
     + 2 \sum_{k>k'}^{n-1} \sum_j \epsilon_{j_{[k]}}^2\\
      &\leq \mathcal{E}_\mathrm{WHT}^2\left(\sum_{k=0}^{n-1} \left\|\vc \theta ^{(k)}\right\|^2_2(2k+1)\right)+o\left(\mathcal{E}_\mathrm{WHT}^2\right),
     \end{split}
\end{equation}
where we made use of \cref{eq:relative_error_def} and 
\begin{equation}
    \sum_j \epsilon_{j_{[k]}}^2 =\left\|\vc \theta^{(k)}-\tilde {\vc \theta}^{(k)}\right\|^2_2\leq \mathcal{E}_\mathrm{WHT}\|\vc \theta^{(k)}\|_2^2.
\end{equation}
\end{proof}
\section{The Binary-Tree Boolean-Encoder Algorithm}
\label{app:BTBE}
In section \cref{sec:properties_of_the_multiplexer} we showed how to simplify a uniform multiplexer circuit when the non zero angles are constant. This is obviously not the case in general when dealing with arbitrary data. However, as explained in detail in the next sub-section, it is always possible to rescale the angles to the interval $[0,1]$ and choose a quadrature basis of $L\in \mathbb{N}$ elements in order to approximate each angle with a given precision using Boolean coefficients. This decomposition, together with the addition property of the rotation gates, allows us to reduce an arbitrary uniform multiplexer into a product of $L$ multiplexers at constant angle. The simplification threfore becomes a sequence of independent simplifications where the angle is constant, for which we can use the techniques introduced in \cref{sec:properties_of_the_multiplexer}. Once each group of operations has been simplified, we can merge them back together by summing the arguments of the gates with the same control strings. The result of this process is a product of less controlled and less numerous multi-controlled gates, which can the be decomposed into CX and single qubit rotations using the best known methods in the literature, such as \cite{Claudon_2024,KhattarGidney2025ConditionallyCleanAncillae}. In particular, we utilize Khattar et al.'s methods for decomposing an m-controlled X ($C^mX$) gate into $2n-3$ Toffolis in $O(\log(m))$ depth using two clean ancilla qubits. This decomposition is relevant for us since any $C^mR_y$ gate can be decomposed into two $C^mX$ gates plus single qubit rotations according to the formula 
\begin{align}
\label{MCR_decomposition_into_MCX}
&C^{m}\!\left(R_y(\theta)\right)\nonumber \\
&=
C^{m}(X)
\left(I \otimes R_y\left(-\frac{\theta}{2}\right)\right)
C^{m}(X)
\left(I \otimes R_y\left(\frac{\theta}{2}\right)\right).
\end{align}
An analogous decomposition applies to $C^mZ$.
This appendix provides the implementation details of the \ac{BT-BE}
method summarized in \cref{sec:BT-BE}. We describe the finite-precision
representation of the multiplexer angles, the construction and
simplification of the associated Boolean masks, the collection of
rotations with identical generalized controls, and the synthesis of the
resulting multi-controlled rotations. The algebraic identities that
justify the Boolean simplification are derived separately in
\cref{sec:properties_of_the_multiplexer}, since they are also used by
the \ac{BT-WHT} construction.
A visual scheme of the \ac{BT-BE} is shown in \cref{fig:BT_BE_scheme}.

\subsection{Decomposing angles according to a generalized basis}\label{sec:BT-BE quadrature}

In \cref{sec:BT-BE} we showed how to write a given angle of the multiplexer using an $L$-binary decomposition
\begin{equation}
    \begin{split}
        \theta_{\vc{j}[l]} &\approx \sum_{i=1}^L B_{\vc{j}[l],i}\frac{2\pi}{2^i},
    \end{split}
\end{equation}
it turns out this is only one of infinitely many ways we can approximate a given angle using Boolean coefficients and a partition of the unit. 

More generally, given an angle $\theta \in [0,2 \pi] $, we define the rescaled angle \begin{equation}
    \tilde{\theta} := \frac{\theta}{2 \pi} \in [0,1].
\end{equation}
Now we divide the unit into a finite succession of $L$ real parts $\{a_j\}_\mathbb{R}, j \in [L]$ that fulfils these three constraints 

\begin{subequations}
\label{eq:coefficient-conditions}
\begin{empheq}[left=\empheqlbrace]{align}
    a_j &> 0,
    && \forall j \in [L],
    \\[2pt]
    \sum_{j=1}^{L} a_j &= 1,
    \\[2pt]
    a_j &\leq \sum_{k=j+1}^{L} a_k,
    && \forall j \in [L-1].
    \label{eq:coefficient-bound}
\end{empheq}
\end{subequations}

These conditions ensure that we have a proper basis to represent $\tilde \theta $ up to a precision 
\begin{equation}
    \epsilon \leq  \min_j {a_j}.
\end{equation}
A good method to explicitly parametrize the $\{a_j\}$ is to use the tail sums
\begin{equation}
    r_j:= \sum_{k=j}^L a_k,
\end{equation}
so that $r_1:= 1$ and  $r_{L+1}:= 0$ .

From the definition we derive 
\begin{equation}
   a_j= r_j-r_{j+1}
\end{equation}
and \cref{eq:coefficient-bound} becomes

\begin{equation}
   r_j-r_{j+1}\leq r_{j+1}
\end{equation}
and hence
\begin{equation}
   r_j\geq r_{j+1} \geq \frac{r_j}{2}.
\end{equation}
where we used the fact that all terms must be positive and therefore $\{r_j\}$ is monotonically decreasing.

We now indirectly define the $L-1$ real parameters $\lambda_j$ with the equation \begin{equation}
  r_{j+1} =\lambda_j r_j,
\end{equation}
where $\lambda_j \in [\frac{1}{2} ,1], j \in [L-1]$. 
This allows us to rewrite $a_j$ in terms of $\lambda_j$ as 

\begin{equation}
\label{eq:a_j_in_terms_of_lambdas}
\begin{split}
    a_j &= (1-\lambda_j)
    \prod_{k=1}^{j-1} \lambda_k,\null\quad \forall j \in \{1,\ldots,L-1\},\\
    a_L &= \prod_{k=1}^{L-1} \lambda_k.
\end{split}
\end{equation}
The tuple of $\{\lambda_j\}$ parameters completely defines the basis and if we choose them to be all equal to $\frac{1}{2}$ we recover the binary decomposition, with 
\begin{equation}
    a_j=2^{-j}.
\end{equation}
In order to enforce $\frac{1}{2}\leq \lambda \leq 1$ we can write
\begin{equation}
   \lambda_j(\alpha)= \frac{1}{2}\left(\frac{1+2\alpha_j^2}{1+ \alpha_j^2}\right)
\end{equation}
with $\alpha \in \mathbb{R}$, so that $\lambda(0)=\frac{1}{2}$ and $\underset {\alpha \to \infty}{\lim}\lambda(\alpha) = 1$.
Using this basis we can approximate any scaled angle $\tilde \theta \in [0,1]$ writing it as 
\begin{equation}
\label{eq:generalized_angle_decompositon}
    \tilde \theta \approx \sum_{j=1}^L b_j a_j ,\quad b_j \in \{0,1\},
\end{equation}
with an error $\epsilon \leq a_L =\prod_{j=1}^{L-1}\lambda_j$.
\begin{figure*}[tbph]
\begin{minipage}{\linewidth}
\begin{algorithm}[H]
\caption{\textsc{OptimizeQuadratureBasis}}
\label{alg:optimal-generalized-basis}
\begin{algorithmic}[1]

    \REQUIRE Normalized angle vector
    $\tilde{\boldsymbol{\theta}}^{(l)}
    =
    \left(
    \tilde{\theta}^{(l)}_1,
    \ldots,
    \tilde{\theta}^{(l)}_{2^l}
    \right)^{\mathsf T}$

    \REQUIRE Number of basis elements $L$

    \REQUIRE Reconstruction-error threshold $\varepsilon$

    \REQUIRE Maximum number of candidate evaluations $T$

    \ENSURE Best basis parameters
    $\boldsymbol{\lambda}_{\mathrm{best}}$

    \ENSURE Best basis vector
    $\mathbf{a}_{\mathrm{best}}$

    \ENSURE Best Boolean coefficient matrix
    $B_{\mathrm{best}}$

    \STATE Set $N \gets 2^l$

    \STATE Initialize the best circuit depth
    $D_{\mathrm{best}} \gets +\infty$

    \STATE Initialize the best reconstruction error
    $E_{\mathrm{best}} \gets +\infty$

    \STATE Initialize the evaluation counter
    $t \gets 0$

    \WHILE{$t<T$}

        \STATE Generate candidate parameters
        $\lambda_j\in\left[\frac{1}{2},1\right)$,
        for $j=1,\ldots,L-1$

        \STATE Increment the evaluation counter
        $t \gets t+1$

        \FOR{$j=1,\ldots,L-1$}

            \STATE Compute the $j$-th basis element
            $\displaystyle
            a_j
            \gets
            (1-\lambda_j)
            \prod_{k=1}^{j-1}\lambda_k
            $

        \ENDFOR

        \STATE Compute the final basis element
        $\displaystyle
        a_L
        \gets
        \prod_{k=1}^{L-1}\lambda_k
        $

        \STATE Form the candidate basis vector
        $\displaystyle
        \mathbf{a}(\boldsymbol{\lambda})
        \gets
        \begin{bmatrix}
        a_1 & \cdots & a_L
        \end{bmatrix}^{\mathsf T}
        $

        \FOR{$i=1,\ldots,N$}

            \STATE Compute the optimal Boolean coefficient vector
            $\displaystyle
            \mathbf{b}_i(\boldsymbol{\lambda})
            \gets
            \underset{\mathbf{b}\in\{0,1\}^{L}}
            {\operatorname{argmin}}
            \left|
            \tilde{\theta}^{(l)}_i
            -
            \sum_{j=1}^{L}
            b_j a_j(\boldsymbol{\lambda})
            \right|
            $

        \ENDFOR

        \STATE Form the Boolean coefficient matrix
        $\displaystyle
        B(\boldsymbol{\lambda})
        \gets
        \begin{bmatrix}
        \mathbf{b}_1(\boldsymbol{\lambda})^{\mathsf T}
        \\
        \vdots
        \\
        \mathbf{b}_N(\boldsymbol{\lambda})^{\mathsf T}
        \end{bmatrix}
        $

        \STATE Reconstruct the normalized angle vector
        $\displaystyle
         {\boldsymbol{\theta}}^{(l)}
        \gets
        B(\boldsymbol{\lambda})
        \mathbf{a}(\boldsymbol{\lambda})
        $

        \STATE Compute the reconstruction error
        $\displaystyle
        E(\boldsymbol{\lambda})
        \gets
        \left\|
        \tilde{\boldsymbol{\theta}}^{(l)}
        -
         {\boldsymbol{\theta}}^{(l)}
        \right\|_2
        $

        \IF{$E(\boldsymbol{\lambda})>\varepsilon$}

            \STATE Reject the candidate parameters
            $\boldsymbol{\lambda}$

            \STATE Restart the search with a new candidate

        \ELSE

            \STATE Accept the candidate parameters
            $\boldsymbol{\lambda}$

            \STATE Compute the corresponding circuit depth
            $\displaystyle
            D(\boldsymbol{\lambda})
            \gets
            D\!\left(
            B(\boldsymbol{\lambda}),
            \mathbf{a}(\boldsymbol{\lambda})
            \right)
            $

            \IF{$D(\boldsymbol{\lambda})<D_{\mathrm{best}}$}

                \STATE Update the best circuit depth
                $D_{\mathrm{best}}
                \gets
                D(\boldsymbol{\lambda})$

                \STATE Update the best reconstruction error
                $E_{\mathrm{best}}
                \gets
                E(\boldsymbol{\lambda})$

                \STATE Update the best basis parameters
                $\boldsymbol{\lambda}_{\mathrm{best}}
                \gets
                \boldsymbol{\lambda}$

                \STATE Update the best basis vector
                $\mathbf{a}_{\mathrm{best}}
                \gets
                \mathbf{a}(\boldsymbol{\lambda})$

                \STATE Update the best Boolean coefficient matrix
                $B_{\mathrm{best}}
                \gets
                B(\boldsymbol{\lambda})$

            \ELSIF{
                $D(\boldsymbol{\lambda})=D_{\mathrm{best}}$
                \AND
                $E(\boldsymbol{\lambda})<E_{\mathrm{best}}$
            }

                \STATE Update the best reconstruction error
                $E_{\mathrm{best}}
                \gets
                E(\boldsymbol{\lambda})$

                \STATE Update the best basis parameters
                $\boldsymbol{\lambda}_{\mathrm{best}}
                \gets
                \boldsymbol{\lambda}$

                \STATE Update the best basis vector
                $\mathbf{a}_{\mathrm{best}}
                \gets
                \mathbf{a}(\boldsymbol{\lambda})$

                \STATE Update the best Boolean coefficient matrix
                $B_{\mathrm{best}}
                \gets
                B(\boldsymbol{\lambda})$

            \ENDIF

        \ENDIF

    \ENDWHILE

    \STATE \textbf{return}
    $\boldsymbol{\lambda}_{\mathrm{best}},
    \mathbf{a}_{\mathrm{best}},
    B_{\mathrm{best}},
    D_{\mathrm{best}},
    E_{\mathrm{best}}$

\end{algorithmic}
\end{algorithm}
\end{minipage}
\end{figure*}

\subsection{Decomposing \texorpdfstring{$C^l R$}{controlled-R} into a product of L terms}

Once the $L-1$ real paramters $\{\lambda_j\}$ that define the generalized basis $\{a_j\}$ are chosen , the first step of the algorithm is to use decomposition in \cref{eq:generalized_angle_decompositon} to write any multi-controlled $R_y$ (or equivalently $R_z$) gate as 
\begin{align}
    C^l R_y(\theta) =& C^l R_y(2\pi \tilde \theta) \\
    =& C^l R_y(2\pi \sum_{j=1}^L b_j a_j)\\
    =&\prod_{j=1}^L C^l R_y(2\pi  b_j a_j)\\
    =&\prod_{j=1}^L \left(C^l R_y(2\pi a_j)\right)^{b_j}.
\end{align}
An example of this decomposition for the binary case is given in \cref{fig:binary_decomposition_of_angles}.

\subsection{Grouping of the terms with the same argument and different controls}

Once we have decomposed the each $C^l R $ gate in the multiplexer as a product of at most $L$ gates with fixed rotation angles, we group all of the multi-controlled gates with the same rotation angle but different control strings. This can be done since all of these operations commute within a given multiplexer.

The $l$-th multiplexer consists of a product of $N_l\leq 2^l$ $C^lR_y$ gates with the same target, same control register and different angles from \cref{eq:binarytreepolarangles}.

We can write the multiplexer $M_l$ as
\begin{equation}
M_l=\prod_{\vc{j} \in \{0,1\}^l}  C^lR_y(\vc{j},\theta_{\vc{j}}) 
\end{equation}
 and we decompose it as
 
\begin{align}
   M_l=\prod_{\vc{j} \in \{0,1\}^l} \prod_{k=1}^L  C^lR_y(\vc{j},2 \pi B_{\vc{j},k}^{(l)}a_k),
\end{align}
where we choose the generic basis $\{a_j\}$ and $B_{\vc{j},k}^{(l)}$ are the Boolean coefficients of the decomposition of $\theta^{(l)}_{\vc{j}}$ in that basis.

We then group the terms according to their arguments 
\begin{align}
\label{eq:decomposition_of_grouped_operations}
M_l
=&
\prod_{k=1}^{L}
\prod_{\vc{j}\in\{0,1\}^{l}}
C_{\vc{j}}^{\,l}R_y
\left(
2\pi B^{(l)}_{\vc{j},k}a_k
\right)
\nonumber\\
=&
\prod_{k=1}^{L}
\prod_{\vc{j}\in\{0,1\}^{l}}
\left[
P_{\vc{j}}\otimes
R_y\left(
2\pi B^{(l)}_{\vc{j},k}a_k
\right)
+
\left(\I-P_{\vc{j}}\right)\otimes\I
\right]
\nonumber\\
=&
\prod_{k=1}^{L}
\left[
\left(
\sum_{\vc{j}\in O_{l,k}}
P_{\vc{j}}
\right)
\otimes R_y(2\pi a_k)
+
\left(
\sum_{\vc{j}\notin O^{(l)}_{k} }
P_{\vc{j}}
\right)
\otimes\I
\right]
\nonumber\\
=&
\prod_{k=1}^{L}
\left[
\left(
\sum_{\vc{j}\in O^{(l)}_{k} }
P_{\vc{j}}
\right)
\otimes R_y(2\pi a_k)
\right.
\nonumber\\
&+
\left.
\left(
\I-
\sum_{\vc{j}\in O^{(l)}_{k} }
P_{\vc{j}}
\right)
\otimes\I
\right].
\end{align}
 where 
 \begin{equation}
     O^{(l)}_{k} := \{\vc{j} \in \{0,1\}^l \quad s.t.\quad B^{(l)}_{\vc{j},k}=1 \}
 \end{equation}
 is the support of the k-th column of the binary matrix of coefficients $B^{(l)}$.
 
 \subsection{Simplification of the control strings within each group}

 Once we have grouped the multi-controlled gates based on their argument, we can start simplifying their controls within each group following the procedure in \cref{sec:properties_of_the_multiplexer}, choosing to perform \ac{ESOP} in case of self-inverse operations or \ac{DSOP} otherwise. An example of this is illustrated by \cref{fig:boolean_simplification_colored}.
 In the following we drop the $(l)$ label since the operation is the same regardless of the size. 

 When used as a sub-routine inside \ac{BT-BE}, the decomposition is performed over the $k$-th column of the binary matrix $B_{\vc{j},k}$, whose support is $O_k$. This is evident from \cref{eq:decomposition_of_grouped_operations}.
 
 The output of algorithm is a decomposition consisting of a set of control strings $\mathcal{G}_k, \subset \{0,1,\I\}^l$ with a small amount of terms, whose associated index sets provide a cover of the support $O_k$.
 
In the \ac{DSOP} case, the decomposition is a partition of the support $O_k$ of the k-th Boolean column of $B_{:,k}$ using the subsets of indices $C_{\vc{c}}$ associated to each control string $\vc{c}\in \mathcal{G}_k$
 \begin{equation}
    \bigsqcup_{\vc{c}\in \mathcal{G}_k}^{D_k }C_{\vc{c}} =  O_k,
\end{equation}
where $D_k = \#(\mathcal{G}_k)$ is the length of the decomposition.
In the circuit-logic simplification community, the strings $\vc{c}$ are also referred to as \emph{cubes}, with the symbol $-$ used in place of $\I$.
This terminology comes from a geometric interpretation of the decomposition, which makes the structure of the index sets associated with each control string $\vc{c}$ more intuitive.
As explained in more detail in \cref{sec:appendix geometric interp}, consider the $n$-dimensional hypercube graph $H_n$. The indices in the support $O_k$ correspond to a subset $V_k$ of
$N_k=\#\bigl(O_k\bigr)$
vertices of $H_n$.

The set $V_k$ can be expressed as a disjoint union of a minimal number of lower-dimensional sub-hypercubes. Each sub-hypercube in this partition is then associated bijectively with a control string $\vc{c}$.
 
An example of the control simplification for a group of multi-controlled gates with fixed angles can be seen in \cref{fig:boolean_simplification_colored}.

\subsection{Summing all the terms with the same control string}
Once the simplification has taken place, the resulting multiplexer reads 
\begin{equation}
\begin{aligned}
 {M}_l
=&\prod_{k=1}^L\left(\left (\underset{ \vc{c}\in \mathcal{G}_{l,k}}{\sum }P_j \right)\otimes R_y(2 \pi a_k)\right.\\
&\left. + \left(\I - \underset{ \vc{c}\in \mathcal{G}_{l,k}}{\sum }P_j\right)\otimes \I \right).
  \label{eq:DSOP_simplified_decomposition_of_grouped_operations}
 \end{aligned}
\end{equation}
As the next and final step we consider all of the simplified multi-controlled gates with the same control and replace them with a single multi-controlled gate that has the same control string as the group we removed and whose argument is the sum of all the arguments. In order to do this we need a set containing all the different control strings appearing in the collections $\mathcal{G}_k  $ over all $L$ values of $k$, which we define as 
\begin{equation}
    \mathcal{G} := \bigcup_k \mathcal{G}_k.
\end{equation}
For each unique control string we need the list of values of k, which we call $K_{\vc{c}}$, such that the control string $\vc{c}$ is present in $\mathcal{G}_k  $:
\begin{equation}
    K_{\vc{c}}= \{ k \in [L] \quad s.t. \quad \vc{c}\in \mathcal{G}_k\}.
\end{equation}
Then the grouping and summing step turns the multiplexer into 
\begin{align}
M_l=&\prod_{k=1}^L\left(\left (\underset{ \vc{c}\in \mathcal{G}_{l,k}}{\sum }P_j \right)\otimes R_y(2 \pi a_k)\right.\\
&\left.+ \left(\I - \underset{ \vc{c}\in \mathcal{G}_{l,k}}{\sum }P_j\right)\otimes \I \right)\nonumber\\
=&\prod_{k=1}^L\prod_{\vc{c}\in \mathcal{G}_{l,k}} C^lR_y(\vc{c}, 2\pi a_k) \nonumber\\
=&\prod_{\vc{c}\in \mathcal{G}} \prod_{k\in K_{\vc{c}}} C^lR_y(\vc{c}, 2\pi a_k) \nonumber\\
=&\prod_{\vc{c}\in \mathcal{G}}  C^lR_y(\vc{c}, \sum_{k\in K_{\vc{c}}}  2\pi a_k) 
  \label{eq:resummed_DSOP_simplified_decomposition_of_grouped_operations}
 \end{align}
The final depth of the multiplexer in terms of the simplified multi-controlled operations is equal to the cardinality of $\mathcal{G}$ and it is therefore upper bounded  by the sum of the single cardinalities of $\mathcal{G}_k$ 
\begin{equation}
    \text{D}_\text{BTBE}=D= \#(\mathcal{G}) \leq \sum_k D_k.
\end{equation}

\begin{figure}[tbph]
\begin{algorithm}[H]
\caption{\ac{BT-BE}}
\label{alg:BT-BE}
\begin{algorithmic}[1] 
    \REQUIRE
    \begin{tabular}[t]{@{}l@{\qquad}l@{}}
        $\theta \in [0,2\pi]^{2^l}$ & multiplexer angles \\
        $L \in \mathbb{N}^{+}$            & amount of Boolean digits
    \end{tabular}
    
    \STATE Normalize the angle vector $\tilde{\theta} := \frac{\theta}{2\pi} $
     \STATE Find the optimal basis $a^{(l)}$ and the correspondent matrix of Boolean coefficients $B$ and the other optimization parameters
     \\$\vc{\lambda},\vc{a},B,D,E$ $\gets$ optimizeQuadratureBasis($\tilde{\theta^{(l)}},L,\epsilon,T$)
    
    \FOR{$k=1,\ldots,L$}
        \STATE Compute the \ac{DSOP} of the k-th column $B$ $\mathcal{G}_k^{(l)} \gets \text{\ac{DSOP}}(B_{:,k})$ 
    \ENDFOR
    \STATE Make a set of all the control strings present after simplification\\ 
    $ \mathcal{G}^{(l)}\gets  \bigcup_k \mathcal{G}^{(l)}_k$
    \FOR {$h \in \mathcal{G}^{(l)}$}
    \STATE $K^{(l)}_{\vc{c}}\gets \{ k \in [L] \quad s.t. \quad \vc{c}\in \mathcal{G}^{(l)}_k\}$
    \STATE Sum over the arguments of the gates with the same control string $\vc{c}$
    $\gamma_{\mathcal{G}} \gets  \sum_{k \in K^{(l)}_{\vc{c}}} 2\pi a_k$
    \STATE Define the instruction pair 
    $(\vc{c}, \gamma_{\mathcal{G}})$
    \ENDFOR
    \RETURN $\{(\vc{c}, \gamma_{\mathcal{G}})\}_{\mathcal{H}^{(l)}}$

\end{algorithmic}
\end{algorithm}
\end{figure}

\subsection{Synthesis of multi-controlled rotations}
\label{app:mcr-synthesis}

The Boolean simplification described in \cref{sec:BT-BE} produces a collection of rotations with different numbers of controls. Although these multi-controlled rotations may be retained when they are supported natively by the target hardware, our resource estimates require a decomposition into a common elementary gate basis. In this subsection, we describe the decomposition and cost model used for this purpose.

We denote by $N_{\mathcal A\to\mathcal B}$ the number of gates from the set $\mathcal B$ required to synthesize an operation from the set $\mathcal A$, and by $D_{\mathcal A\to\mathcal B}$ the corresponding circuit depth. When $\mathcal B={\mathrm{CX}}$, these quantities refer specifically to CX count and CX depth; single-qubit gates are accounted for separately.

An $m$-controlled $X$ gate, denoted by $C^mX$, is also referred to as a multi-controlled $X$ (\ac{MCX}) or multi-controlled Toffoli gate. Its action is

\begin{equation}
C^mX
\ket{c_1\cdots c_m}\ket{t}
=
\ket{c_1\cdots c_m}
\ket{t\oplus(c_1\land\cdots\land c_m)}.
\end{equation}

A multi-controlled $R_y$ rotation can be decomposed into two \ac{MCX} gates and two single-qubit rotations. With $\I_{\mathrm c}$ denoting the identity on the control register, we use
\begin{equation}
\begin{aligned}
\label{eq:MCR-decomposition-into-MCX}
&C^mR_y(\theta)\\
=&
C^mX
\left[
\I_{\mathrm c}\otimes
R_y\left(-\frac{\theta}{2}\right)
\right]
C^mX
\left[
\I_{\mathrm c}\otimes
R_y\left(\frac{\theta}{2}\right)
\right].
\end{aligned}
\end{equation}

When at least one control is inactive, both $C^mX$ gates act trivially and the two single-qubit rotations cancel. When all controls are active, the identity

\begin{equation}
XR_y(\vartheta)X=R_y(-\vartheta)
\end{equation}

shows that \cref{eq:MCR-decomposition-into-MCX} applies $R_y(\theta)$ to the target.

Several decompositions of $C^mX$ are available, with different trade-offs between gate count, depth, approximation error, and ancilla requirements~\cite{Claudon_2024,KhattarGidney2025ConditionallyCleanAncillae,Lee2026LogarithmicDepthMCT}. For the resource estimates in this work, we adopt the K2 construction of Khattar and Gidney, which uses two clean ancilla qubits and conditionally clean system qubits~\cite{KhattarGidney2025ConditionallyCleanAncillae}.

Writing the Toffoli gate as $C^2X$, the Toffoli count of this construction is

\begin{equation}
\label{eq:K2-Toffoli-count}
N_{{C^mX}\to{C^2X}}
=
2m-3.
\end{equation}

An explicit upper bound for its Toffoli depth, valid for $m\geq32$, is given in Ref.~\cite{Lee2026LogarithmicDepthMCT} as

\begin{equation}
\label{eq:K2-Toffoli-depth}
D_{{C^mX}\to{C^2X}}
\leq
5\left\lceil\log_2m\right\rceil-5.
\end{equation}

For smaller values of $m$, we use the corresponding directly synthesized circuits rather than extrapolating the bound in \cref{eq:K2-Toffoli-depth}.

To express the cost in terms of CX gates, we replace each Toffoli gate by a standard exact decomposition containing six CX gates~\cite{Shende_2006}. This gives
\begin{align}
N_{{C^mX}\to{\mathrm{CX}}} &\leq 6N_{{C^mX}\to{C^2X}} \nonumber\\
&= 12m-18. \label{eq:MCX-CX-count}
\end{align}

If each Toffoli layer is replaced independently by a decomposition with CX depth at most six, the corresponding CX-depth bound is
\begin{align}
D_{{C^mX}\to{\mathrm{CX}}} &\leq 6D_{{C^mX}\to{C^2X}} \nonumber\\
&\leq 30\left\lceil\log_2m\right\rceil-30, \qquad m\geq32. \label{eq:MCX-CX-depth}
\end{align}

Combining \cref{eq:MCR-decomposition-into-MCX,eq:MCX-CX-count}, the CX count of an $m$-controlled $R_y$ rotation satisfies
\begin{align}
N_{{C^mR_y}\to{\mathrm{CX}}} &\leq 2N_{{C^mX}\to{\mathrm{CX}}} \nonumber\\
&= 24m-36. \label{eq:MCR-CX-count}
\end{align}

The decomposition additionally contains the two single-qubit rotations displayed in \cref{eq:MCR-decomposition-into-MCX}. A conservative depth estimate, in which the two \ac{MCX} decompositions are applied sequentially and each external \mbox{$R_y$-rotation} contributes one layer, is therefore
\begin{align}
D_{{C^mR_y}\to{\mathrm{CX},R_y}} &\leq 2D_{{C^mX}\to{\mathrm{CX}}}+2 \nonumber\\
&\leq 60\left\lceil\log_2m\right\rceil-58, \qquad m\geq32. \label{eq:MCR-CX-depth}
\end{align}

The estimate in \cref{eq:MCR-CX-depth} should be understood as a CX-layer cost model supplemented by the two external $R_y$ layers. The single-qubit gates internal to the selected exact Toffoli decomposition are not counted as additional depth in this model. These bounds are deliberately conservative with respect to the scheduling of different multi-controlled rotations: further cancellations or parallelization may be possible when several simplified gates are compiled jointly.

The low-control cases are treated separately. For $m=0$, the operation is a single-qubit rotation; for $m=1$, a direct controlled-rotation decomposition is used; and for $m=2$, the circuit is synthesized directly rather than through the asymptotic K2 construction. This avoids applying \cref{eq:MCX-CX-count,eq:MCX-CX-depth} outside the regime used in our resource estimates.

Let $\mathcal H$ denote the final collection of multi-controlled rotations produced by \ac{BT-BE} after rotations with identical generalized control strings have been combined, and let $m_g$ be the number of controls of gate $g\in\mathcal H$. The total CX count used in our resource estimates is

\begin{equation}
\label{eq:BTBE-total-CX-count}
N_{{\mathrm{BT\text{-}BE}}\to{\mathrm{CX}}}
=
\sum_{g\in\mathcal H}
N_{{C^{m_g}R_y}\to{\mathrm{CX}}},
\end{equation}

where the direct low-control costs are used for $m_g<32$. A fully serial execution gives the conservative depth bound

\begin{equation}
\label{eq:BTBE-serial-depth-bound}
D_{{\mathrm{BT\text{-}BE}}\to{\mathrm{CX},R_y}}
\leq
\sum_{g\in\mathcal H}
D_{{C^{m_g}R_y}\to{\mathrm{CX},R_y}}.
\end{equation}

When gates act on compatible qubits, circuit scheduling can reduce this depth below the serial bound. Accordingly, \cref{eq:BTBE-serial-depth-bound} is used only as an explicit upper bound, whereas compiled depths are reported separately whenever an actual circuit is constructed.

\begin{figure}[!t]
\centering
\includegraphics[
width=0.8\columnwidth
]{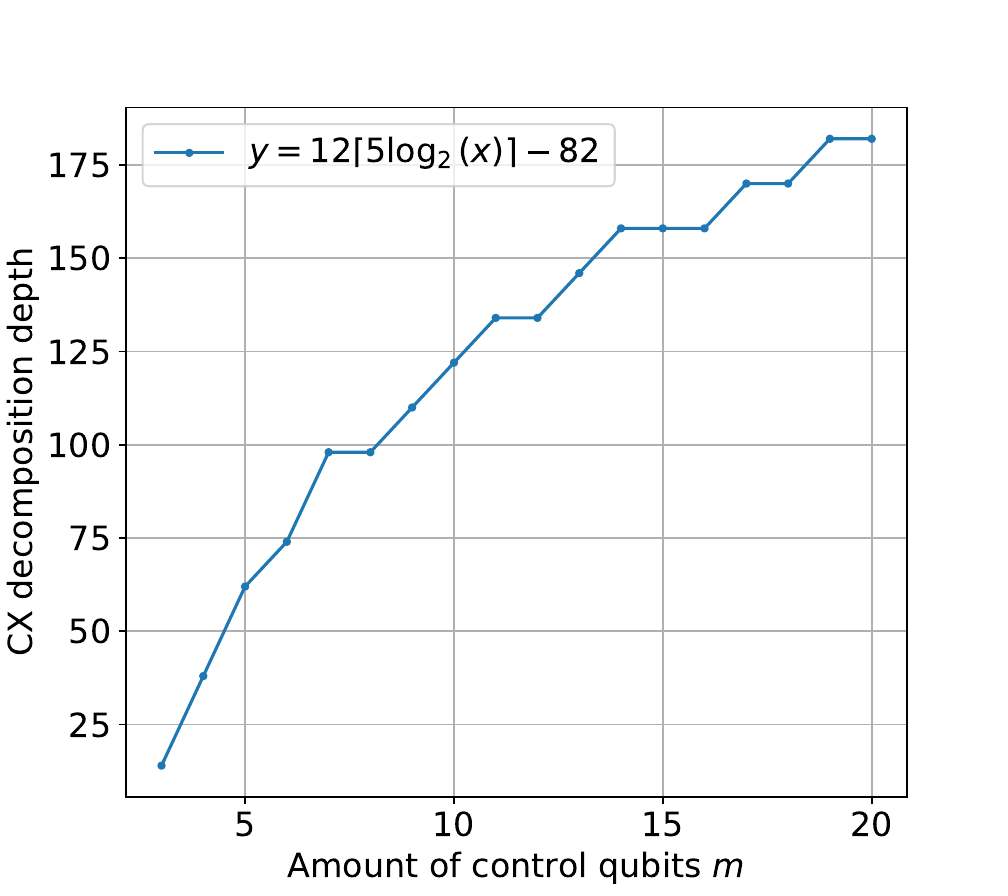}
\caption{Conservative decomposition depth of an $m$-controlled $R_y$ rotation. The estimate uses the two-clean-ancilla K2 construction for each $C^mX$ gate, followed by an exact six-CX decomposition of every Toffoli gate. The plotted quantity is $D_{{C^mR_y}\to{\mathrm{CX},R_y}}$. Small values of $m$ are synthesized directly and are not evaluated using the asymptotic bound.}
\label{fig:MCR_decomposition_into_CX}
\end{figure}

The resource model above is not intrinsic to \ac{BT-BE}. Ancilla-free constructions, alternative space--depth trade-offs, hardware-native multi-controlled operations, or approximate synthesis methods may be substituted without changing the preceding Boolean simplification. The chosen K2 construction provides a concrete and reproducible basis for comparing \ac{BT-BE} with the other encoding methods considered in this work.
\section{Geometric interpretation in terms of cuts of a hypercube graph}
\label{sec:appendix geometric interp}

As laid out in our previous work~\cite{Pagni2025}, the indices of a vector in $\mathbb{C}^{2^n}$ can be identified with the vertices of an $n$-dimensional hypercube graph. More precisely, each index is represented by a binary string $\vc{j}\in\mathbb{F}_2^n$, while a Boolean indicator over all indices is represented by a vector in $\mathbb{F}_2^{2^n}$. In the present work, we additionally associate amplitude information with the vertices of the graph. This information may be regarded as a set of vertex weights and does not modify the underlying combinatorial structure of the hypercube.

We recall the definition of the $n$-dimensional hypercube graph,

\begin{equation}
\begin{split}
H_n&\coloneqq(V_n,E_n),\\
V_n&=\mathbb{F}*2^n,\\
E_n&=
\{
{\vc{i},\vc{j}}
|
\vc{i},\vc{j}\in V_n,
\left|\vc{i}\oplus\vc{j}\right|*{\mathrm H}=1
\},
\end{split}
\end{equation}

where the XOR operation $\oplus$ acts componentwise and
$\left|\vc{k}\right|_{\mathrm H}$ denotes the Hamming weight of $\vc{k}$, i.e., the number of its nonzero entries.

We define a map $H$ that associates a control string $\vc{c}$ containing $n$ symbols from the control alphabet $\{0,1,\I\}$ with an induced sub-hypercube of $H_n$. Let

\begin{equation}
d(\vc{c})\coloneqq\#_{\I}(\vc{c})
\end{equation}

denote the number of occurrences of the symbol $\I$ in $\vc{c}$. The associated sub-hypercube has dimension $d(\vc{c})$ and contains $2^{d(\vc{c})}$ vertices.

This association can be expressed using the intermediate map introduced in \cref{c_string_to_binary}. The positions of the nonzero entries in the Boolean indicator vector $\mathfrak b(\vc{c})\in\mathbb{F}_2^{2^n}$ identify a subset of the $2^n$ vertices of $H_n$, and hence a subset of the entries of the amplitude vector $\vc{v}$. For example,

\begin{equation}
\mathfrak b(0\I)=(1,0)\otimes(1,1)=(1,1,0,0)
\end{equation}

selects the first two entries of a vector $\vc{v}\in\mathbb{C}^4$. The corresponding control string $\vc{c}=0\I$ specifies a zero-control on the first qubit and no control on the second qubit.

The selected vertices form an induced sub-hypercube $H_{\vc{c}}\subseteq H_n$ uniquely determined by the control string $\vc{c}$. We define the map

\begin{equation}
\label{c_string_to_hypercube}
\begin{split}
H:\{0,1,\I\}^n&\longrightarrow
{\text{induced subgraphs of }H_n},\\
\vc{c}&\longmapsto H_{\vc{c}}\coloneqq(V_{\vc{c}},E_{\vc{c}}),
\end{split}
\end{equation}
where
\begin{equation}
\begin{aligned}
V_{\vc{c}}=&\{
\vc{j}\in\mathbb{F}*2^n|b*{\vc{c},\vc{j}}=1\}\\
=&\{
\vc{j}\in\mathbb{F}_2^n
|
j_q=c_q
\text{ whenever }c_q\neq\I
\},
\end{aligned}
\end{equation}

and

\begin{equation}
E_{\vc{c}}=\{{\vc{i},\vc{j}}\in E_n|\vc{i},\vc{j}\in V_{\vc{c}}\}.
\end{equation}

It follows that

\begin{equation}
H_{\vc{c}}\cong H_{d(\vc{c})},
\qquad
|V_{\vc{c}}|=2^{d(\vc{c})}.
\end{equation}

Thus, every occurrence of $\I$ corresponds to a free coordinate of the sub-hypercube, whereas every occurrence of $0$ or $1$ fixes the corresponding coordinate.

In keeping with this formalism, consider the binary angle tree in \cref{fig:angles_tree}. A node at level $m$, with address

\begin{equation}
\vc{j}=(j_0,\ldots,j_{m-1})\in\{0,1\}^m,
\end{equation}

is associated with the sub-hypercube

\begin{equation}
H_{\vc{j}\I^{n-m}}\subseteq H_n.
\end{equation}

This sub-hypercube contains the $2^{n-m}$ computational-basis indices descending from that node. The two children of the node correspond to a bipartition of this sub-hypercube obtained by fixing the next coordinate to either zero or one. We denote this cut by

\begin{equation}
\label{eq:hypercube-tree-cut}
T_{\vc{j}*\I^{n-m-1}}
\coloneqq
\left(
H_{\vc{j}0\I^{n-m-1}},
H_{\vc{j}1\I^{n-m-1}}
\right),
\;
0\leq m\leq n-1.
\end{equation}

For convenience, we use the symbol $*$ to represent the two possible values of the coordinate being cut. At the level of control strings, we therefore write

\begin{equation}
\vc{j}*\I^{n-m-1}
\coloneqq
\left(
\vc{j}0\I^{n-m-1},
\vc{j}1\I^{n-m-1}
\right).
\end{equation}

\begin{figure}[!t]
\centering
\includegraphics[width=5cm]{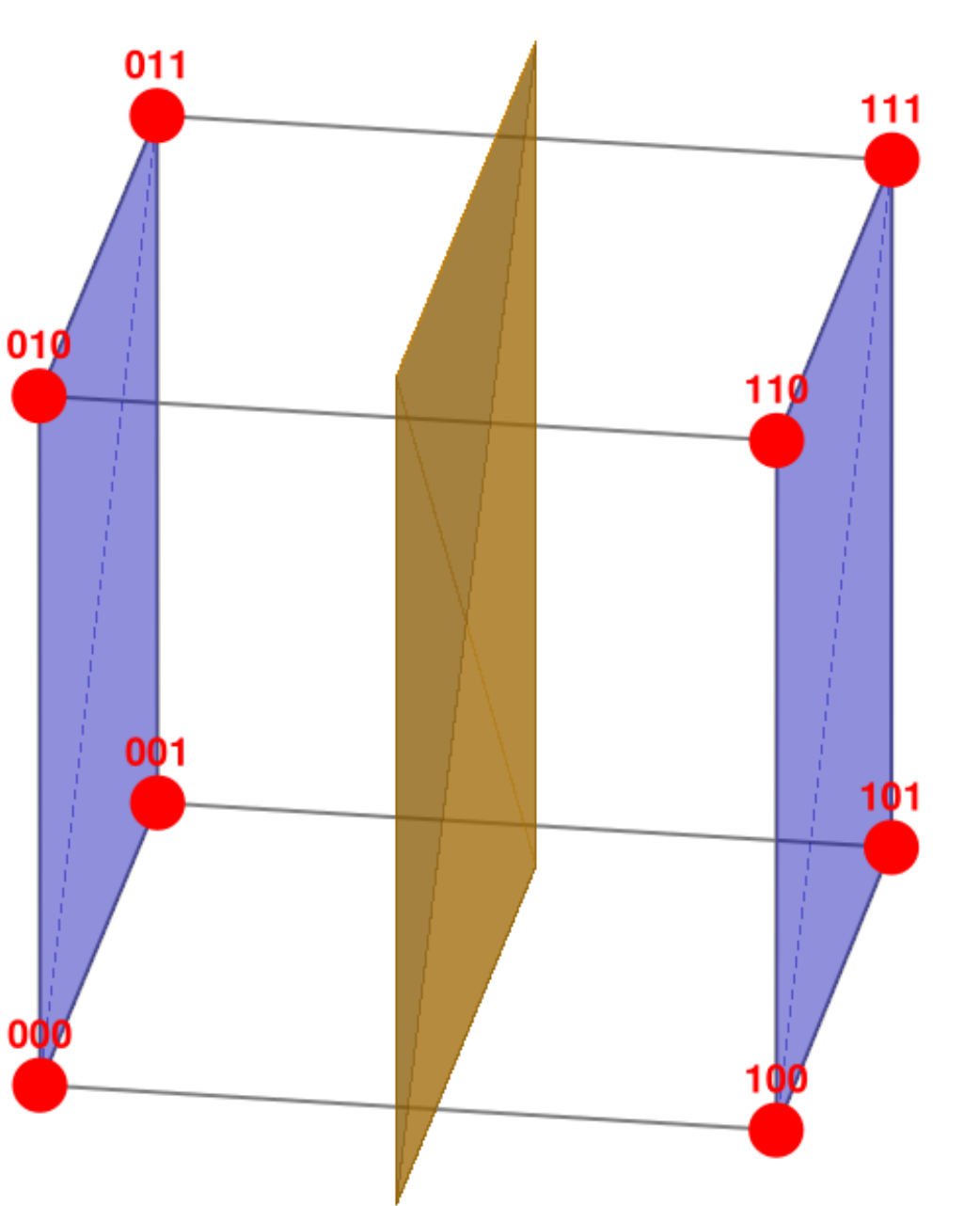}
\caption{Example of the cut $T_{*\I\I}=(H_{0\I\I},H_{1\I\I})$ for $n=3$. The cutting plane, shown in yellow, separates the two faces associated with $0\I\I$ and $1\I\I$, shown in blue.}
\label{fig:cube_cut_example}
\end{figure}

For example, when $n=3$, the cut associated with the root of the tree occurs at level $m=0$ and is labeled by $*\I\I$. It defines the bipartition

\begin{equation}
*\I\I=(0\I\I,1\I\I),
\end{equation}

which divides the cube graph $H_3$ along its first coordinate, as illustrated in \cref{fig:cube_cut_example}.

The usefulness of this notation becomes apparent upon returning to \cref{eq:binarytreepolarangles,eq: phase angle}. The modulus and phase angles associated with a node are determined by the norms and phase information of the two amplitude subtensors supported on the corresponding bipartition. Consequently, every internal node of the angle tree can be interpreted as a cut of a sub-hypercube of $H_n$.

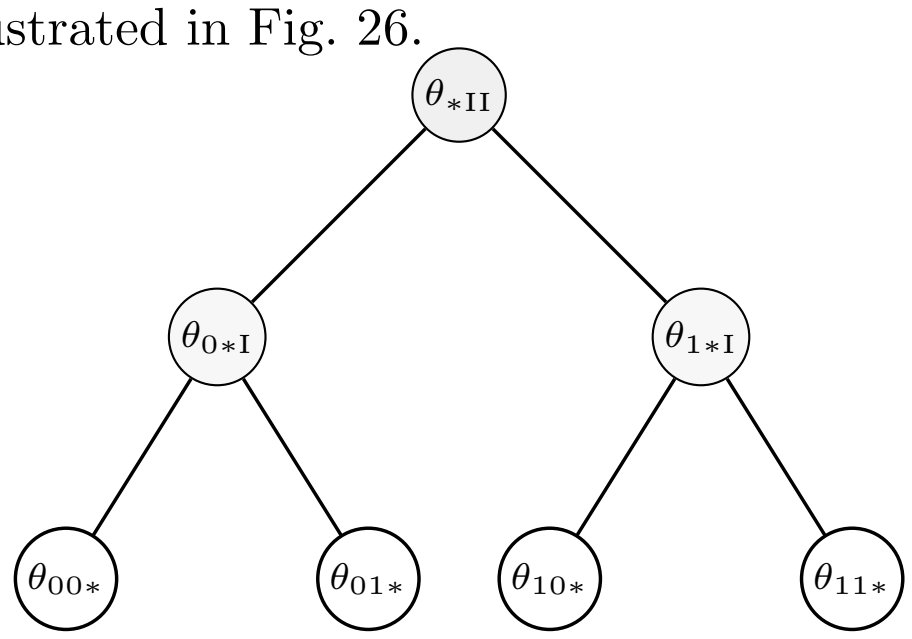
\begin{figure}[!t]
\centering
\begin{tikzpicture}[
    grow=down,
    level distance=16mm,
    level 1/.style={sibling distance=32mm},
    level 2/.style={sibling distance=20mm},
    every node/.style={
        draw,
        circle,
        align=center,
        inner sep=1.2pt,
        font=\scriptsize
    },
    root/.style={
        fill=gray!12,
        font=\scriptsize\bfseries
    },
    internal/.style={
        fill=gray!6
    },
    leaf/.style={
        fill=white
    },
    edge from parent/.style={
        draw,
        line width=0.65pt
    }
]

\node[root] {$\theta_{*\I\I}$}
    child {node[internal] {$\theta_{0*\I}$}
        child {node[leaf] {$\theta_{00*}$}}
        child {node[leaf] {$\theta_{01*}$}}
    }
    child {node[internal] {$\theta_{1*\I}$}
        child {node[leaf] {$\theta_{10*}$}}
        child {node[leaf] {$\theta_{11*}$}}
    };

\end{tikzpicture}
\caption{Binary-tree organization of the $2^n-1$ hypercube cuts for $n=3$. Each internal node corresponds to a bipartition of the sub-hypercube associated with its prefix. The sequence of cuts is ${*\I\I,0*\I,1*\I,00*,01*,10*,11*}$.}
\label{fig:cut_tree}
\end{figure}

For $n=3$, the complete sequence of cuts associated with the binary angle tree is

\begin{equation}
\{
*\I\I,,
0*\I,,
1*\I,,
00*,,
01*,,
10*,,
11*
\},
\end{equation}

as illustrated in \cref{fig:cut_tree}.

The possible cuts of $H_n$ are more numerous than those appearing in the binary angle tree. For example,

\begin{equation}
\left(
H_{\I0\I^{n-2}},
H_{\I1\I^{n-2}}
\right)
\end{equation}

describes a cut of the entire hypercube along its second coordinate. By contrast, the binary angle tree contains only prefix-conditioned cuts obtained according to a fixed ordering of the qubits. Nevertheless, the resulting sequence of $2^n-1$ cuts is sufficient to determine all the modulus and phase angles required to characterize the quantum state. We therefore adopt a fixed register ordering throughout the construction and discuss separately the effect of qubit permutations on the output of the algorithm.

This geometric picture also provides the interpretation used in the Boolean simplification. A generalized control string containing $d$ occurrences of $\I$ represents a $d$-dimensional Boolean subcube containing $2^d$ control addresses. A \ac{DSOP} decomposition therefore corresponds geometrically to a disjoint partition of the selected vertices into such subcubes, whereas an \ac{ESOP} decomposition combines overlapping subcubes according to parity.

\end{document}